%% file: main.tex
\RequirePackage{snapshot}
\documentclass[10pt,twocolumn]{z2-article}

\usepackage[T1]{fontenc}
\usepackage{framed}
\usepackage{mathptmx}
\DeclareMathAlphabet{\mathcal}{OMS}{zplm}{m}{n} % really get the mathptmx's Times/Zapf Chancery look
\usepackage{oneinchmargins}
\usepackage{setspace}
\usepackage{multirow}
\usepackage{bm}
\usepackage{amsmath}
\usepackage{centernot}
\usepackage{amssymb}
\usepackage[hyphens]{url}
\usepackage{alltt}
\usepackage{listings}
\usepackage{cite}
\usepackage{tabularx}
\usepackage{breakurl}
\usepackage{fancyvrb}
\usepackage{relsize}
\usepackage{soul}
\usepackage{array}
\usepackage{booktabs}
\usepackage{enumitem}
\usepackage{currfile}
\usepackage[usenames,dvipsnames,svgnames,table]{xcolor}
\usepackage{color}
\usepackage{tikz}
\usepackage{pbox}
\usepackage{xstring}
\usepackage{xspace}
\usepackage{epstopdf}
\usepackage{graphicx}
\usepackage{float}
\usepackage{makecell}
\usepackage{url}
\usepackage{fixltx2e}
\usepackage{timestamp}
\usepackage{titling}
\usepackage{arydshln} % for wider divider
\usepackage[pdftex,colorlinks,urlcolor=NavyBlue,linkcolor=NavyBlue,citecolor=NavyBlue,hyperfootnotes=false]{hyperref}
\usepackage{pifont}

\usepackage{subcaption}
\usepackage{tcolorbox}
\tcbuselibrary{breakable}

\usepackage{titlesec}
\titlespacing*{\section}
  {0pt}        % left margin
  {3pt}       % space before (vertical)
  {0pt}        % space after (vertical)
\titlespacing*{\subsection}
  {0pt}        % left margin
  {2pt}       % space before (vertical)
  {0pt}        % space after (vertical)
\titlespacing*{\subsubsection}
  {0pt}        % left margin
  {1pt}        % space before (vertical)
  {0pt}        % space after (vertical)

\makeatletter
\def\url@leostyle{%
  \@ifundefined{selectfont}{\def\UrlFont{\sf}}{\def\UrlFont{\small\ttfamily}}}
\makeatother
\floatstyle{plain}
\newfloat{aside}{bth!}{lop}
\floatname{aside}{Aside}

\newlength{\tempa}

\usepackage[frozencache,cachedir=minted-cache]{minted} % code formatting
\setminted{
  fontsize=\footnotesize,
  baselinestretch=1,
  xleftmargin=0.5em,
  linenos,
  numbersep=0.5em,
  frame=none,
  bgcolor=white,
  framesep=0pt,
}
\setmintedinline{
  fontsize=\small,
}

\usepackage[available]{usenixbadges}

\begin{document}
\pagenumbering{gobble} % remove page number
% \pagenumbering{arabic} % OSDI CFP: the page should be numbered
\input{0_0_macro}

\begin{spacing}{1.0}
\input{0_1_title}
\input{0_2_abstract}
\input{1_0_intro}
\input{2_0_bg}

\input{3_0_tunability}
\input{3_5_challenges}
\input{4_0_design}

\input{4_9_impl}

\input{5_0_cases}
\input{6_0_eval}

\input{discussion}
\input{7_0_rel}
\input{8_0_conc}
\input{ack}

\input{99_0_bib}

\newpage
\input{appendix}
\end{spacing}
\end{document}

%% file: 0_0_macro.tex
% Haryadi .. parskip, and par-indent
%\setlength\parindent{0pt}
%\setlength\parskip{5pt}

\newcommand{\beforesec}{\vspace{-.4cm}}
\newcommand{\aftersec}{\vspace{-.3cm}}

\newcommand{\beforesect}{\vspace{-.1cm}}
\newcommand{\aftersect}{\vspace{-.3cm}}

\newcommand{\beforesub}{\vspace{-.3cm}}
\newcommand{\aftersub}{\vspace{-.25cm}}

\newcommand{\beforesubsub}{\vspace{-.2cm}}
\newcommand{\aftersubsub}{\vspace{-.2cm}}

 \newcommand{\zsection}[1]{\section{#1}}
 \newcommand{\zsubsection}[1]{\subsection{#1}}
 \newcommand{\zsubsubsection}[1]{\subsubsection{#1}}

\newcommand{\smush}{0.25in}

\newcommand{\figWidthOne}{3.05in} 
\newcommand{\figWidthHalf}{1.45in} 
\newcommand{\figWidthTwo}{3.05in} 
\newcommand{\figWidthTwop}{1.6in} 
\newcommand{\figWidthThree}{2.2in} 
\newcommand{\figWidthSix}{1.1in} 
\newcommand{\figWidthFour}{1.7in} 
\newcommand{\figHeight}{2.0in}
\newcommand{\captionText}[2]{\textbf{#1} \textit{\small{#2}}}

\newcommand{\zbullet}{\hspace{-0.1cm}$\bullet$}

\newcommand{\eg}{e.g.}
\newcommand{\ie}{i.e.}
\newcommand{\etal}{et al.}
\newcommand{\apriori}{\textit{a priori}}

%-------------------------------------------------------------------
% Haryadi -- new command
\newcommand{\msub}[1]{\vspace{1pt}\noindent{\bf #1}}

\newcommand{\ts}[1]{{\tt{\small#1}}}
\newcommand{\tsb}[1]{{\tt{\small{\bf#1}}}}
\newcommand{\tse}[1]{{\tt{\small{\em#1}}}}
\newcommand{\colorconst}[1]{\textcolor{purple!60}{#1}}
\newcommand{\tss}[1]{{\tt{\footnotesize#1}}}
\newcommand{\exc}{$^{\ddag}$}        % except
\newcommand{\EIO}{\ts{EIO}}
\newcommand{\ENOSPC}{\ts{ENOSPC}}
\newcommand{\EDQUOT}{\ts{EDQUOT}}

%-------------------------------------------------------------------
% For fault injection table
\newcommand{\ip}{bad}
\newcommand{\nullp}{$\emptyset$}

\newcommand{\oops}{o}
\newcommand{\dead}{$\times$}
\newcommand{\alive}{$\surd$}
\newcommand{\unuse}{$\times$}
\newcommand{\use}{$\surd$}
\newcommand{\ic}{$\times$}
\newcommand{\con}{$\surd$}
\newcommand{\gpf}{G}
\newcommand{\npe}{null-pointer}
\newcommand{\usebuta}{$\surd^a$} % unmountable
\newcommand{\hwdetect}{d}

%% TABLE 2
\newcommand{\lateoops}{o$^b$}  % oops happens, but late
\newcommand{\lategpf}{G$^b$}   % gpf, but late
\newcommand{\iop}{i}           % invalid opcode
\newcommand{\detects}{d}       % assertion
\newcommand{\silentret}{s}     % app fails silently
\newcommand{\errorret}{e}      % app returns an error
\newcommand{\appworks}{$\surd$}     % app works!
\newcommand{\usebutar}{$\surd^{ar}$} % read-only and unmountable

\newcommand{\tbls}{\hspace{0.025in}}
\newcommand{\tblss}{\hspace{0.015in}}
\newcommand{\shrinkless}{\vspace{-0.01cm}}

%-------------------------------------------------------------------

%-------------------------------------------------------------------

% axes         
\newcommand{\x}{{\em x}}
\newcommand{\y}{{\em y}}
\newcommand{\xaxis}{x-axis}
\newcommand{\yaxis}{y-axis}

\newcommand{\KB}{~KB}
\newcommand{\KBs}{~KB/s}
\newcommand{\Kbs}{~Kbit/s}
\newcommand{\mbs}{~Mbit/s}
\newcommand{\MB}{~MB}
\newcommand{\GB}{~GB}
\newcommand{\MBs}{~MB/s}
\newcommand{\mus}{\mbox{$\mu s$}}
\newcommand{\ms}{\mbox{$ms$}}

\newcommand{\unix}{{\sc Unix}}

\newcommand{\bquote}{\vspace{-0.25cm} \begin{quote}}
\newcommand{\equote}{\end{quote}\vspace{-0.05cm} }

\newcommand{\zquote}[2]{\begin{quote}
#1 --
{\em``#2'' }
%{\bf -- #1 }
\end{quote}}

\newcommand{\XXX}[1]{{\small {\bf (XXX: #1)}}}

\newcommand{\XXXX}{{\bf XXX}}
\newcommand{\xx}{{\bf XXX}}

% normal
\newcommand{\beforecaption}{\begin{spacing}{0.80}}
\newcommand{\aftercaption}{\end{spacing}}
\newcommand{\mycaption}[3]{{\beforecaption\caption{\label{#1}{\footnotesize \bf #2. } {\em \small #3}}\aftercaption}}

\newcommand{\sref}[1]{\S\ref{#1}}

\newcommand{\xxx}[1]{  \underline{ {\small {\bf (XXX: #1)}}}}

\newenvironment{packed_itemize}{
    \begin{list}{\labelitemi}{\leftmargin=1.0em}
     \setlength{\itemsep}{2.5pt}
     \setlength{\parskip}{0pt}
     \setlength{\parsep}{0pt}
     \setlength{\headsep}{0pt}
     \setlength{\topskip}{0pt}
     \setlength{\topmargin}{0pt}
     \setlength{\topsep}{0pt}
     \setlength{\partopsep}{0pt}
    }{\end{list}}

\newcommand{\smalltt}[1]{\texttt{\fontsize{8.7}{5}\selectfont #1}}
\newcommand{\smallit}[1]{\textit{\scriptsize #1}}
\newcommand{\verysmalltt}[1]{\texttt{\scriptsize #1}}
\newcommand{\verysmall}[1]{\scriptsize #1}
\newcommand*\rot{\rotatebox{90}}

\newcommand{\writeSC}{\smalltt{write()}}
\newcommand{\fsyncSC}{\smalltt{fsync()}}
\newcommand{\msyncSC}{\smalltt{msync()}}
\newcommand{\fdatasyncSC}{\smalltt{x fdatasync()}}
\newcommand{\linkSC}{\smalltt{link()}}
\newcommand{\mkdirSC}{\smalltt{mkdir()}}
\newcommand{\fempty}{$\phi$}
\newcommand{\fexists}{$\surd$}
\newcommand{\creatSC}{{\smalltt{creat()}}}
\newcommand{\unlinkSC}{{\smalltt{unlink()}}}
\newcommand{\renameSC}{{\smalltt{rename()}}}
\newcommand\floor[1]{\lfloor#1\rfloor}
\newcommand\ceil[1]{\lceil#1\rceil}
\newcommand{\totbugs}{60}
\newcommand{\totapps}{11}
\newcommand{\totappsw}{eleven}
\newcommand*{\combination}[2]{{}^{#1}C_{#2}}

\newcommand{\used}{$\surd$}
\newcommand{\usedpar}{$P$}
\newcommand{\useddollar}{$\surd$\textsuperscript{\$}}
\newcommand{\usedadler}{$\surd$\textsuperscript{a}}
\newcommand{\usedparstar}{$P$\textsuperscript{$*$}}
\newcommand{\notused}{}
\newcommand{\numapps}{eight}

\if 0 % colored version
\newcommand{\yes}{$\surd$}
\newcommand{\yesi}{\colorbox{gray!30}{$\surd$\textsubscript{$i$}}}
\newcommand{\nolow}{\colorbox{gray!30}{$\times$\textsubscript{$l$}}}
\newcommand{\no}{\colorbox{gray!85}{$\times$}}
\newcommand{\complower}{$L$}
\newcommand{\compmoder}{$M$}
\newcommand{\comphigher}{\colorbox{gray!85}{$H$}}
\newcommand{\na}{\footnotesize na}
\fi

\newcommand{\yes}{$\surd$}
\newcommand{\yesi}{$\surd$\textsubscript{$i$}}
\newcommand{\nolow}{$\times$\textsubscript{$l$}}
\newcommand{\nomod}{$\times$\textsubscript{$m$}}
\newcommand{\no}{$\times$}
\newcommand{\yeslow}{$\surd$\textsuperscript{$l$}}
\newcommand{\complower}{$L$}
\newcommand{\compmoder}{$M$}
\newcommand{\comphigher}{$H$}
\newcommand{\na}{\footnotesize na}

\newcommand*{\termindex}[2]{$\langle$\textit{epoch}:{#1}, \textit{index}:{#2}$\rangle$}
\newcommand*{\epochindex}[2]{{#1}.{#2}}
\newcommand*{\termindexnovar}{$\langle$\textit{epoch}, \textit{index}$\rangle$}
\newcommand*{\snapid}{$\langle$\textit{snap-index}, \textit{chunk}\#$\rangle$}
\newcommand*{\rafttermindexnovar}{$\langle$\textit{term}, \textit{index}$\rangle$}
\newcommand{\quotes}[1]{``#1''}

\newcommand{\camera}[1]{\textcolor{Black}{#1}}
\newcommand{\addcamera}[1]{\textcolor{Black}{#1}}
\newcommand{\todo}[1]{\textcolor{Red}{#1}}
\definecolor{deepblue}{RGB}{0,0,139}
\newcommand{\deepblue}[1]{\textcolor{deepblue}{#1}}
\newcommand{\rev}[1]{{#1}\xspace}
% \revrm: mark text removed in revision (dark gray + strikethrough)
\newcommand{\revrm}[1]{{\color{darkgray}\st{#1}}}
\newcommand{\hide}[1]{}

%%%
%%% Spacing
%%% 

%\newcommand{\camera}[1]{\textcolor{Black}{#1}}
%\newcommand{\addcamera}[1]{\textcolor{Black}{#1}}
\newcommand{\minisec}[1]{\vspace{0.1em}\noindent\textbf{#1. }}
% \newcommand{\minisec}[1]{\noindent\textbf{#1. }}

% \minisubsec{} gets rid of paragraph indentation and underlines the title
% \newcommand{\minisubsec}[1]{\vspace{0.1em}\noindent\underline{#1. }}
\newcommand{\minisubsec}[1]{\noindent\underline{#1. }}

% the amount of flexible white space ("glue") placed between paragraphs.
\setlength{\parskip}{0.1pt plus 0.2pt minus 0.2pt}

% set space before caption
\setlength{\abovecaptionskip}{2pt plus 2pt minus 2pt}

% % set space for listings
\AtEndEnvironment{listing}{\vspace{-4pt}}
\BeforeBeginEnvironment{minted}{\vspace{-0.4em}}
\AfterEndEnvironment{minted}{\vspace{-0.4em}}

% % change the space between the figure/table and the text
\setlength{\textfloatsep}{0.8pt plus 2pt minus 2pt}

% % change the space among figures/tables
\setlength{\floatsep}{1pt plus 3pt minus 3pt}

%%%
%%% People and comments
%%% 

% \presetkeys{todonotes}{disable}{}
\newcommand{\zhongjie}[1]{\textcolor[rgb]{0.0,0.48,1}{zj:#1}\xspace}

\newcommand{\jing}[1]{\textcolor[rgb]{0.49,0.086,0.443}{Jing: #1}\xspace}
\newcommand{\tianyin}[1]{{\color{red}{ty:#1}}}
\newcommand{\ran}[1]{\todo[author=Ran,color=violet!30,size=\normalsize,inline]{#1}}
\newcommand{\wentao}[1]{{\color{orange}{wt:#1}}}
\newcommand{\yulong}[1]{{\color{teal}{yulongl:#1}}}
\newcommand{\shawn}[1]{\todo[author=Shawn,color=lime!30,size=\normalsize,inline]{#1}}
\newcommand{\TODO}[1]{\todo[author=TODO,color=red!30, size=\normalsize,inline]{#1}}
\newcommand{\tofill}[1]{{\color{red}{?#1}}}
\newcommand{\pie}[1]{{\color{blue}#1}}

%%%%%%%%%%%%%%%%%%%%%%%%%%%%%%%%%%%%%%%%
\newcommand{\TABSEPtiny}{\hspace{0.5em}}
\newcommand{\TABSEPsmall}{\hspace{0.7em}}
\newcommand{\TABSEPlarge}{\hspace{1em}}
\newcommand{\tabincell}[2]{\begin{tabular}{@{}#1@{}}#2\end{tabular}}

\newcommand{\vs}{vs.\xspace}
\newcommand{\etc}{etc.\xspace}
\newcommand{\far}{remote\xspace}

% \providecommand{\para}[1]{\smallskip\noindent\textbf{#1} }
% \newcommand{\para}{\noindent\textbf}

% \newcommand{\paraspace}[1]{\vspace{0.03in}}
% \newcommand{\parab}[1]{\paraspace \noindent {\bf #1} }
% \newcommand{\parae}[1]{\paraspace \noindent  {\em #1} }

% \newcommand{\presec}{\vspace{-0.10in}}
% \newcommand{\postsec}{\vspace{-0.00in}}
% \newcommand{\presub}{\vspace{-0.10in}}
% \newcommand{\postsub}{\vspace{-0.02in}}
% \newcommand{\presubsub}{\vspace{-0.05in}}
% \newcommand{\postsubsub}{\vspace{-0.00in}}
% \newcommand{\prefig}{\vspace{-0.00in}}
% \newcommand{\postfig}{\vspace{-0.10in}}
% \newcommand{\postfigdouble}{\vspace{-0.25in}}
% \newcommand{\postfigcaption}{\vspace{-0.20in}}
% \newcommand{\halfpostfigcaption}{\vspace{-0.00in}}

% Helper command for kernel source references
% #1: file path (after source/), #2: line number, #3: value
\newcommand{\pfconstsrc}[3]{\href{https://elixir.bootlin.com/linux/v6.14/source/#1\##2}{{#1:\allowbreak#2}}, \allowbreak value=#3}

\newcommand{\sysname}{Xkernel\xspace}
\newcommand{\const}{perf-consts\xspace}
\newcommand{\constant}{Performance Constants\xspace}
\newcommand{\perfconst}{{perf-const}\xspace}
\newcommand{\perfconsts}{{perf-consts}\xspace}
\newcommand{\modef}{Mode-Flex\xspace}
\newcommand{\modep}{Mode-Perf\xspace}
\newcommand{\func}{ConstMod\xspace}
% The mechanism
\newcommand{\kprobe}{Kprobe\xspace}
\newcommand{\kprobes}{Kprobes\xspace} % Should be rare.?
% Particular probes 
\newcommand{\kernelProbe}{kprobe\xspace}
\newcommand{\kernelProbes}{kprobes\xspace}
\newcommand{\overwrite}{Inject-Overwrite\xspace}
\newcommand{\sie}{Scoped Indirect Execution\xspace}
\newcommand{\tittb}{TI$\rightarrow$TB\xspace}
\newcommand{\xkgen}{\smalltt{xk-gen}}
\newcommand{\xkcmd}{\smalltt{xk-cmd}}
\newcommand{\xkruntime}{Xk-runtime}
\newcommand{\xkseleton}{XK skeleton}
\newcommand{\xkprog}{XK tuning program}
\newcommand{\xkprogs}{XK tuning programs}
\newcommand{\xktable}{scope table\xspace}
\newcommand{\linuxdevelopers}{Linux developers} % Used in reference author fields

% Codes
% Note: perhaps use minted?
\definecolor{codegreen}{rgb}{0,0.6,0}
\lstdefinestyle{APIDX_C_STYLE}{
    language=C,                            
    basicstyle=\fontfamily{pcr}\selectfont\small,
    keywordstyle=\color[HTML]{1922fb}\bfseries,     
    commentstyle=\color[HTML]{007020}\bfseries,     
    tabsize=4,                             
    breaklines=true,                       
    breakatwhitespace=true,                
    escapeinside={\%*}{*)},                
    morekeywords={uint32_t, uint64_t}
}
\lstset{style=APIDX_C_STYLE}

%%%%%%%%%%%%%%%%%%
% XK numbers
\newcommand{\NumSysctlKnob}{145\xspace}
\newcommand{\NumSysctlUnchanged}{96\xspace}
\newcommand{\NumSysctlChanged}{49\xspace}
\newcommand{\NumSysctlBUG}{20\xspace}
\newcommand{\NumSysctlPBUG}{43\xspace}

\newcommand{\indPos}{location}
\newcommand{\indCode}{update}

%% file: 0_1_title.tex
\setlength{\droptitle}{-0.775cm}
\providecommand{\titlespace}{\quad}
\title{\Large \bf Xkernel: Systematic Tunability of OS Performance Constants\vspace*{-0.5em}}
\title{\Large \bf Xkernel: Rethinking Performance Tunability of Operating System Kernels\vspace*{-0.75em}}
\title{\Large \bf Xkernel: Principled Performance Tunability of Operating System Kernels\vspace*{-0.80em}}

\author {
    Zhongjie Chen$^{\dagger\S}$,\ 
    Wentao Zhang$^{\ddagger}$,\
    Yulong Tang$^{\S}$, \\
    Ran Shu$^{\S}$,\ 
    Fengyuan Ren$^{\dagger}$,\ 
    Tianyin Xu$^{\ddagger}$,\ 
    Jing Liu$^{\S}$
}

\date{\vspace{-0.25em}
    $^{\S}$\textit{Microsoft Research},
    $^{\ddagger}$\textit{University of Illinois Urbana-Champaign}, 
    $^{\dagger}$\textit{Tsinghua University}
    % $^{\P}$\textit{Renmin University of China}
%    \vspace{-0.80em}
}

% \date{\vspace*{-2.5em}}
% \date{}

\maketitle

%% file: 0_2_abstract.tex
\begin{abstract}
The Linux kernel is permeated with constant values that are critical to system performance. 
Many of these constants, referred to as {\it \perfconsts}, are magic numbers with brittle assumptions 
    on hardware and workloads. 
% Such values inevitably become dated, resulting in suboptimal performance of emerging workloads. 
Unfortunately, there is no capability of {\it in-situ} tuning of 
    \perfconst values on deployed kernels.
% ---without a long, disruptive process of rebuilding and redeploying kernel image. 
% In practice, kernel developers are constantly asked to change \perfconsts into tunable knobs 
%    via \texttt{\small sysctl};
% however, runtime updates via \texttt{\small sysctl} is unsafe, 
%    let alone the cost of changing kernel code and deploying it in production.    
This paper rethinks OS performance tunability.
We present \sysname, a system that offers a safe, efficient, and programmable interface for {\it in-situ} tuning 
    of any \perfconsts directly on a running kernel. 
%    towards runtime customization of OS performance. 
% We present \sysname, a system that realizes such a performance tuning 
%    interface for Linux. 
\sysname transforms any \perfconst into a tunable knob on demand using
%     which can be safely and efficiently updated through user-defined policies written in eBPF. 
    a novel approach called Scoped Indirect Execution (SIE). 
SIE captures precise binary boundaries 
    where a \perfconst enters system state and redirects control 
    to synthesized instructions that update the state as if new values were used. 
\sysname goes beyond version atomicity when updating \perfconsts
    to guarantee side-effect safety, a property notably absent in existing kernel update mechanisms. 
Case studies on various OS subsystems demonstrate 
    significant performance benefits of 
    tuning \perfconsts which is made possible by \sysname.
% that 
%    tuning \perfconsts with \sysname yields . 
% For example, \tofill{[put one strongest result]]}
\end{abstract}

%% file: 1_0_intro.tex
\vspace{3.5pt}
\section{Introduction}
\label{sec:intro}
\vspace{3.5pt}

% They exist, they control perf behaviors
Modern operating systems like Linux are permeated with constant values 
    that shape system performance. 
These constants, referred to as {\it performance-critical constants} 
    or {\it \perfconsts}, appear in various source-code forms
    (e.g., macros, literals, and static integers); 
they govern thresholds, time intervals, batch sizes, scaling factors, etc.
Whether to balance latency and throughput or to match batching behavior to
device parallelism, \perfconsts embed design trade-offs and workload/hardware
semantics directly into kernel behavior, forming the kernel's implicit
performance policy.

% However, their value is fixed in the source code, the decision is rather random, and after complication, no way to change it, 
Perf-consts are not tunable in deployed systems 
    without recompiling and rebooting the OS kernels.
Unfortunately, their values are often ``{\it arbitrarily chosen}~\cite{BlkMqReqCnt-2011-Val16}'' by 
    developers based on brittle heuristics, limited testing,
    or assumptions on dated hardware, which ``{\it just happen(ed) to work well}~\cite{ksource:swap_state}.''
However, static magic numbers can hardly serve dynamic workloads or
    diverse hardware configurations, % in modern deployments,
    especially emerging ones that significantly deviate 
    from the time those values were chosen (see~\cite{dukkipati-2010-TcpArgument,BFQHWQUEUETHRESHOLD-2019-Val3,BFQLateStableMerging-2021-Val600,BFQAsyncChargeFactor-2018-Val3, DFLTHROTLSLICE-2018-ValHZ,GSOMAXSIZE-2022-Val65536,ReserveRoot-2022-Val}).

%    and these once-reasonable
%    decisions quickly become outdated as systems evolve

% Tuning potential is HUGE 
Ideally, \perfconsts should be decided at runtime, dynamically adapting to workload
patterns, hardware characteristics, and service-level objectives. 
In practice, the benefits of tuning \perfconsts are substantial. 
In one of our case studies (\S\ref{sec:def}),
    tuning a \perfconst yields 50$\times$ throughput improvement.
Yet such benefits are completely missed as modern OSes 
    provide no mechanism for tuning \perfconsts.
% We define tunability as the ability to adjust a \perfconst dynamically at runtime to fully realize its potential.

% Existing?
Today, tuning a \perfconst typically involves converting it into a runtime knob
    via interfaces such as \smalltt{sysctl}~\cite{SysctlDoc} and
    \smalltt{sysfs}~\cite{SysfsDoc},
    or changing its value in kernel source and
    updating the deployed kernel through live patching~\cite{arnold-2009-ksplice,baumann-atc05,haibo-vee06,cristiano-asplos13,goullon-1978}.
% Today, the practice to tune a \perfconst is either to covert it into a runtime knob
    % through interfaces like 
    % \smalltt{sysctl}~\cite{SysctlDoc} and \smalltt{sysfs}~\cite{SysfsDoc},
    % or to change its value in kernel source code and update the deployed kernel
    % using mechanisms like kernel live patching~\cite{arnold-2009-ksplice,baumann-atc05, haibo-vee06, cristiano-asplos13, goullon-1978}.
However, neither satisfies the needs of performance tuning.
The former (\smalltt{sysctl}/\smalltt{sysfs}) is limited to a small subset 
    of predefined constants with fixed granularity, often system-wide, %values 
    and provides no safety guarantees---the correctness 
    of an update %updating the values 
    depends on manual reasoning.
The latter 
% Existing mechanisms fall short. Converting a constant into a runtime variable
% and exposing it through an interface such as sysctl~\cite{SysctlDoc} is invasive, error-prone,
% and specific to one constant at a time; correctness depends on manual reasoning,
% and updates typically apply system-wide rather than context-specifically. More
    needs to recompile kernel code and apply binary diffs, %~\cite{arnold-2009-ksplice,Kpatch},
    incurring minutes-level delays, which is fundamentally 
    incompatible with fast policy adaptation and online tuning.

% The tunability we want!
% In this work, ...
% I get what 'rethinking' means, it fees like paradigm shifts, still feels abstract (not sharp)
% rethinking is somewhat abused... we do introduce new way of thinking about it, no better solution,
% fine with rethinking; we do need some points like that...
% i'm using machinery rather than mechanism here to distinguish it a bit from SIE
In this paper, we advocate for {\em principled OS performance tunability}---a general
machinery that enables safe, fast value updates 
    for {\it in-situ} performance tuning of {\em any} \perfconsts in deployed OS kernels, 
    {\em without} kernel recompilation or rebooting.
    % add ',' and adjust for better reading rhythm
    % the need of recompiling or rebooting the kernel.  -the need of- is indirect
Such tunability must support expressive, programmable policies, 
    ensure correctness and safety,
    as well as allow millisecond-scale value updates.

% WHY IT IS HARD
% inst diff/instruct gen/safety
Achieving this goal is challenging. First, the system must
    precisely identify all instructions that consume the original constant, 
    in the presence of sophisticated compiler transformations 
    such as constant folding and strength reduction. 
Any missed or incorrectly replaced instructions may leave unsafe remnants in the
    running kernel. 
Second, the system must generate the instructions based on the new value, as 
    well as the tuning policies, without recompilation. 
Finally, the kernel execution may have already produced side effects on runtime state, 
    and any update must not cause conflicts on them.

% SIE % The insight, the uniqieus of a const enables this, the idea 
% The techniques 
We introduce {\em Scoped Indirect Execution (SIE)}, a novel mechanism that addresses
the aforementioned challenges. Our key insight is that 
    a constant---unlike arbitrary code---has
    structural semantics: its influence enters the 
    \rev{machine state (registers or memory)} 
    at a specific point and manifests through a small instruction sequence,
% This property 
    enabling safe, in-situ tuning of the \perfconst 
    without recompilation or rebooting. 
For a \perfconst, the point where its value enters registers or memory, is
{\em well-scoped}: it can be identified by static analysis and be represented by a symbolic state
expression agnostic to compiler optimizations; moreover, it is 
    small enough (typically several instructions) to analyze for side effects.

SIE leverages this structure to identify the precise binary region where the \perfconst
    affects 
    \rev{runtime machine state}
    % runtime system state, 
    termed a {\em critical span}, 
    by deriving the symbolic relationship 
    between the \perfconst and the affected registers or memory. 
Within this span, SIE inserts a set of {\em indirections}: small code snippets tied to
specific kernel address locations. When execution reaches these locations, the
indirections update the \rev{machine state} to reflect the new value as per policy.
Once execution leaves the critical span, the resulting state matches what would
have occurred had the \perfconst been changed directly in the binary. The kernel
binary remains untouched; all modification occurs indirectly and locally.

To ensure safety, SIE analyzes how effects from the critical span propagate to
derive a second region, the {\em safety span}, which encapsulates 
    all consumption of constant-dependent state. 
At runtime, indirections are enabled only when
    the execution is outside the safety span, 
    ensuring that updates never occur while the original effects are still in flight.

% THE SYSTEM
% All the features (policy part)
We build \sysname atop Linux as the first system to realize principled performance tunability
for OS kernels. \sysname implements SIE, reducing update latency from minutes to
milliseconds and supporting programmable tuning policies written in eBPF. 
\sysname does not change the kernel source or require 
    %kernel  just feel redudant (there is a kernel close)
    reboots.
It seamlessly integrates with deployed kernels as its implementation
depends on stable kernel features like \kprobe{}, kernel modules, and eBPF.

% RESULTS
We evaluate \sysname across core Linux subsystems, including CPU scheduling, memory
management, storage, and network. \sysname unlocks previously inaccessible tuning
opportunities, achieving up to 50$\times$ microbenchmark improvement and boosting
real-world application performance, e.g., 1.2$\times$ throughput in RocksDB and 81\%
latency reduction in NGINX. % Beyond individual \perfconsts, 
\sysname enables new capabilities: online exploration of design trade-offs, 
    adaptation to hardware and workload patterns, 
    control of OS-internal maintenance behavior, 
    and coordinated tuning across multiple \perfconsts.

We extensively evaluate \sysname on 140 \perfconsts
% To assess generality and safety, we construct a dataset of 140 \perfconsts
(comparable in scale to the \NumSysctlKnob{} performance knobs exposed by \smalltt{sysctl}). 
We show that SIE applies broadly, introduces negligible 
    runtime overhead (a few hundred cycles per \perfconst update),
    and achieves millisecond-scale policy updates.

% Contributions
In summary, this paper makes the following contributions:
\vspace{-15pt}
\begin{packed_itemize}
    \item Principled OS performance tunability that exposes the
        %to harvest % seems we will harvest
        unexplored performance benefits of pervasive \perfconsts;
    \item Scoped Indirect Execution (SIE), a new approach that enables
        safe, fast performance tuning of any \perfconsts with programmable policies;
    \item \sysname, a practical implementation on Linux,
        which can directly benefit deployed kernels;
    \item Case studies and extensive evaluation demonstrating new
        opportunities and inspiring new tuning techniques;
    \item \sysname is open-sourced at \url{https://github.com/xkernel-org/Xkernel/}.
\end{packed_itemize}
% \vspace{-5pt}

%% file: 2_0_bg.tex
% \vspace{1.5pt}
\section{Background and Motivation}
\label{sec:bg}
\vspace{3.5pt}

\subsection{Perf-Consts in Linux}
\label{sec:def}

A \perfconst is a fixed numeric value used by kernel
  code to control the {\em magnitude} of OS behavior,
  without altering correctness or existence of that behavior.
Table~\ref{tab:perf-const-categories} shows four common types.
Perf-consts are pervasive in Linux, 
  appearing in every kernel subsystem. 
Typical \perfconst representations include 
  macros, literal immediates, and \smalltt{static} \smalltt{const} variables.
% summarizes their tuning regimes with examples.
Perf-consts 
% have significant impacts on OS performance. 
govern core trade-offs
(e.g., latency vs.\ throughput, responsiveness vs.\ utilization);
% that determine application SLOs and system efficiency. 
  they also shape how workloads interact with hardware.
There are strong 
  needs to customize their values based on application SLOs,
  workload behavior, and device characteristics.

\input{tbl-perfconst-class}

\newcommand{\BlkMaxReqCnt}{BLK\_MAX\_REQUEST\_COUNT}
\newcommand{\BlkMaxReqCntTt}{\smalltt{\BlkMaxReqCnt}}

\begin{figure}[t!] 
  \vspace{5pt}
  \centering
  \includegraphics[width=\columnwidth]{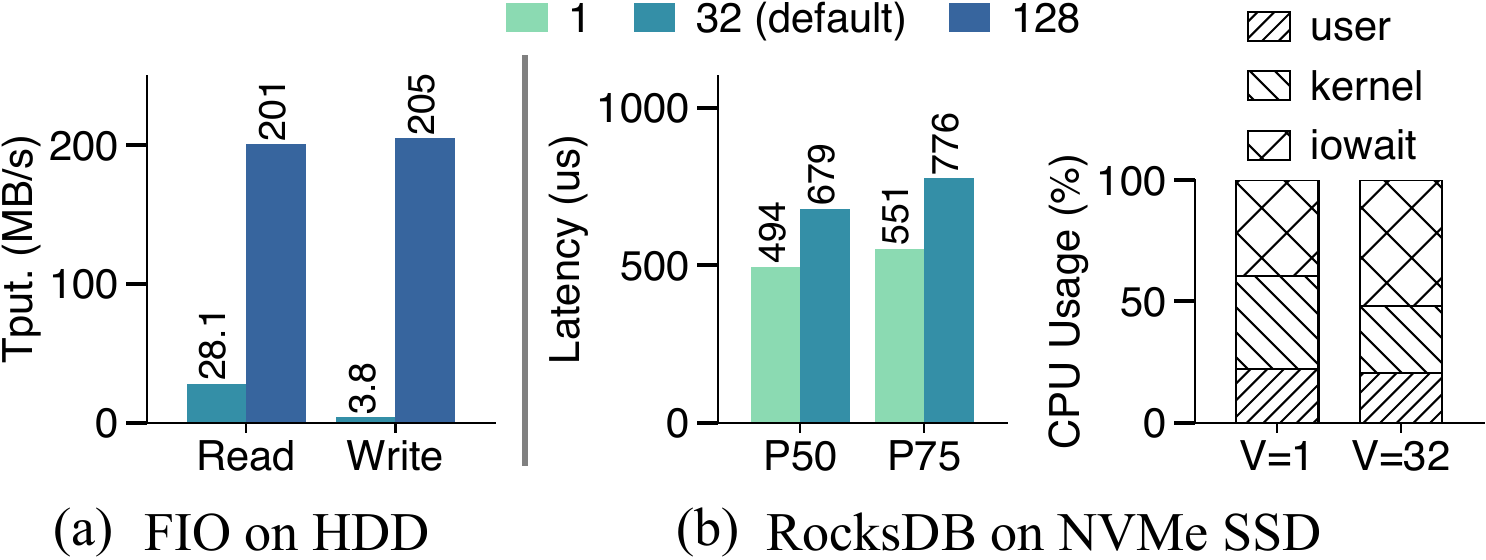}
  \caption{{Performance benefits of changing the value of a 
    \perfconst \BlkMaxReqCntTt{} based 
    on hardware devices and workloads.
    The default value is 32.}}
  \label{fig:cases:blkmq}
  \vspace{5pt}
\end{figure}

\vspace{5pt}
\minisec{A motivating example}
We take
  \BlkMaxReqCntTt{}, a \perfconst
  in Linux's storage subsystem, as an example.
Introduced in 2011~\cite{BlkMqReqCnt-2011-Val16}, 
this \perfconst controls block I/O's plug behavior~\cite{BlockPlug} by delaying
  request submission to merge adjacent requests and reduce device contention. 
Once buffered requests reach the threshold specified by this \perfconst, 
  they are flushed to the device. 
% This behavior sits beneath upper layers (e.g., the page
% cache)\footnote{Complementary to
% \texttt{/sys/block/{device}/queue/nr\_requests}}.
The original commit shows the value was arbitrary: ``{\it $16$ works efficiently to
reduce lock contention... $32$ also works in my tests.}'' 
The choice implicitly reflects then-current hardware---CPU speed, storage performance, and
  lock overhead.
A decade later, in 2021, the value was raised to $32$~\cite{BlkMqReqCnt-2021-Val32}, 
  justified by observed benefits on NVMe devices. 
However, with the wide variety of hardware (HDD, SSD, and NVMe devices) 
  and workload access patterns, a single magic number is inevitably suboptimal.
  
Figure~\ref{fig:cases:blkmq}(a) shows the performance benefits of 
  tuning the \perfconst.
%    with the value 32 (default) and 128.
% Running the FIO workload~\cite{axboe-2014-fio} with 4KB requests 
  % on a 6Gps SAS HDD with 7200 RPM,
\rev{Running a sequential FIO workload~\cite{axboe-2014-fio} with intra-segment
    shuffling (4 KB requests randomly shuffled within each 128 KB segment) 
    on a 7200-RPM SAS HDD,}
    the default value (32) causes the plug to flush too frequently, missing many
    opportunities to merge adjacent requests. 
Increasing the value to 128 allows most requests to be merged, maximizing sequential disk access.
This improves read and write performance by 7$\times$ and 54$\times$, respectively.
% The results are
% Since HDDs perform far better under sequential access than random access, this
    % 7$\times$ and 54$\times$ performance improvements for
    % read and write workloads, respectively.
%   divided into 128 KB segments.
% cannot serve all combinations of devices and workloads.

By contrast, a large value of \BlkMaxReqCntTt\ may not always benefit NVMe SSDs---sometimes request merging has little benefit but adds overhead.
We deploy RocksDB on a 256GB Toshiba XG3 NVMe SSD, and run the \textit{multiread-random} workload 
    from DBbench~\cite{rocksdb} with 32GB dataset (16B keys; 2048B values). 
We use RocksDB's \smalltt{io\_uring}-backed MultiGet API for asynchronous parallelism
  and 
% To stress-test the plugging mechanism at the block layer, we
  use Direct I/O for data transfer between the storage device and user memory.
% \tianyin{It seems that the tuning policy is rather interesting and should be written here}
% The code that implements tuning policies can be found in the appendix.
As shown in Figure~\ref{fig:cases:blkmq}(b),
reducing \BlkMaxReqCntTt\ from 32 to 1 reduces CPU time spent on I/O wait 
    by 12\%, yielding an end-to-end 1.2$\times$ throughput improvement, % (61 KOP/s versus 51 KOP/s), 
    while reducing P50 and P75 latency by 1.37$\times$ and 1.41$\times$, respectively.

\if 0
This case also reveals the slow response cycle of source-level tuning. NVMe
drives were commercially available by 2013 and widely adopted soon after, yet
kernel-level adjustment took until 2021. Static updates and redeployment lag
behind hardware evolution (months) and cannot track workload dynamics
(milliseconds).
\fi

% Finally, the performance space remains unexplored when constants are fixed. As
% shown in~\sref{sec:cases}, tuning this constant for sequential workloads can
% achieve up to a 50$\times$ performance gain -- performance that is impossible to access
% with a hardcoded value.

% say it directly: the performance is foundmentally limited, because the limited tunability of performance consts in OS
% existing efforts
% the needs is real
% not tunable: the problem
% making them tunable: what is the problem

% not tunable
% some are made tunable, but the tunability is still limited (sysctl, sysfs, system_call)
% are fudmentally not systematic, cannot satisfy the need

\vspace{5pt}
\subsection{Limitations of Existing Mechanisms}
\label{sec:limitation}

Despite the strong benefits of tuning \perfconsts,
  modern OSes like Linux provide little support or interface---\perfconsts
  are hardwired into kernel binary once compiled 
  and cannot be changed at runtime once deployed in production.
% We use tunability to describe the degree to which a \perfconst can be adjusted
% at runtime -- allowing its value to adapt to workload patterns, hardware
% characteristics, or evolving system state. Such tuning may rely on a wide range
% of policies, from simple heuristics to dynamic feedback control.

% Despite the substantial performance opportunity, modern kernels provide zero
% tunability.  Many \perfconsts remain hard coded in the kernel source and
% compiled into deployed binaries, preventing runtime adaptation.

% \minisec{Existing mechanisms have narrow scope and are unsafe}
\vspace{5pt}
\minisec{Existing mechanisms are inflexible and unsafe} % avoid collision with scoped
One way to tune a \perfconst{} is to convert a \perfconst into a 
  runtime variable and expose
  it through kernel interfaces like 
  \smalltt{sysctl}~\cite{SysctlDoc}, \smalltt{sysfs}~\cite{SysfsDoc} or system calls.
However, this is not a general mechanism---it requires modifying source code on a
  per \perfconst basis and results in rigid, narrow interfaces.

One fundamental difficulty is to predefine a complete set of \perfconsts{}
  {\it a priori} before deployment; 
  only a very small subset of \perfconsts{} are currently exposed.
  %---performance
  %tuning is often specific to workloads, hardware, and application SLOs.
Linux interfaces like \smalltt{sysctl} and \smalltt{sysfs} are treated 
  as kernel ABIs and therefore prioritize
  stability over flexibility~\cite{LinuxABISymbol}.  
For example, our analysis shows that \smalltt{sysctl} knobs change slowly---among 
  \NumSysctlKnob{} \smalltt{sysctl} knobs, 96 of them have remained unchanged since 2005.
Moreover, decisions to expose
  \perfconsts\ as \smalltt{sysctl} knobs have been largely {\it ad hoc},
  driven by developers' preference and experience (Appendix~\ref{appendix:sysctl-study}).

\rev{Besides the difficulty of upstreaming,
  source-code conversion is fundamentally inflexible: changing the granularity of
  a tuning policy from global (the \smalltt{sysctl} default) to per-process or per-cgroup
  requires modifying source code again, which in turn requires recompilation and rebooting.}

\rev{In addition, \smalltt{sysctl} and \smalltt{sysfs} conversions are known 
  to risk safety~\cite{sysctl-race,sysctl-race-lwn}.
% Extending converted constants to support broader tuning 
  % domains % scopes 
  % (e.g., cgroups) increases kernel complexity and maintenance burden. 
In our study, \NumSysctlBUG of \NumSysctlKnob conversions led to bugs,
largely due to concurrency: for example, the \smalltt{sysctl} setter writes to a shared
global variable from a separate context while core kernel logic reads it
concurrently, easily causing races in practice.
% \jing{@zhongjie, double check numbers?}\zhongjie{43?}
Our inspection shows that \NumSysctlPBUG \smalltt{sysctl} knobs are potentially buggy and 
  lead to races or inconsistent states (see Appendix~\ref{appendix:sysctl-study}).
}

\vspace{5pt}
\minisec{Kernel live patching offers no rescue}
A relevant mechanism is Kernel Live
Patching (KLP)~\cite{LinuxKLP,arnold-2009-ksplice,Kpatch,Kgraft,KernelCare} which
enables 
patching
% code patching of 
deployed kernels without rebooting.
%  and does not require an external setter thread.
% KLP works by modifying kernel source code, recompiling it,
% and applying the binary diff to replace selected regions in the kernel image.
KLP modifies kernel source, recompiles it, and applies the resulting binary diff
to replace selected regions of the kernel image.
% However, KLP is not a solution for \perfconsts\ or performance tuning in general.
However, KLP is not suited for \perfconsts\ tuning.
%or performance tuning in general.
%  are replaced at
% runtime. 
% However, it is a general-purpose code replacement mechanism -- not a
% mechanism for tuning \perfconsts\ -- and exhibits several limitations.

First, KLP is too slow. It takes minutes per patch. Updating a \perfconst value 
  would require recompiling the kernel (with the new value) 
  and patching the binary diff.
By the time a patch is applied, the workload may have changed.

% Second, KLP patches at the function granularity. 
\rev{Second, KLP uses functions as the patching unit.}
When tuning spans multiple
functions (e.g., due to compiler inlining or multiple affected \perfconsts), 
finding safe quiescent points becomes difficult and fragile~\cite{RoughPatch}.
KLP offers no side-effect safety 
  %(hard to track for arbitrary function code) 
  and struggles 
  to support consistent states under multi-threading~\cite{Kpatch,Kgraft}.

%% file: tbl-perfconst-class.tex
\newcommand{\graytxt}[1]{\textcolor{gray}{#1}}

\setlength{\tabcolsep}{1.1pt}
{
\begin{table}[t!]
    \centering
    \footnotesize
    \setlength{\tabcolsep}{4pt}
    \caption{Performance regimes of \perfconsts.}
    \label{tab:perf-const-categories}
    \begin{tabular}{ll}
    \toprule
    \textbf{Category} & \textbf{Explanation and Examples} \\
    \midrule
    \multirow{2}{*}{Threshold} & Triggering a behavior change as a limit or boundary \\
    & \texttt{\graytxt{\#define MAX\_SOFTIRQ\_RESTART} \colorconst{10}} \\
    \midrule
    \multirow{2}{*}{Interval} & Controlling deferred or periodic actions \\
    & \texttt{\graytxt{\#define IPVS\_SYNC\_SEND\_DELAY} \graytxt{(HZ/\colorconst{50})}} \\
    \midrule
    \multirow{2}{*}{Batch Size} & Work processed together per operation to amortize cost \\
    & \texttt{\graytxt{\#define BLK\_MAX\_REQUEST\_COUNT} \colorconst{32}} \\
    \midrule
    \multirow{2}{*}{\shortstack{Scaling Factor}} & A multiplier that adjusts the magnitude or intensity \\
    & \texttt{\graytxt{delta *=} \colorconst{4}; }\\
    \bottomrule
    \end{tabular}
\end{table}
}
\setlength{\tabcolsep}{\tempa}

%% file: 3_0_tunability.tex
\vspace{3.5pt}
\subsection{Our Goal: Principled OS Tunability}
\label{sec:tunability}
\vspace{1.5pt}

Our goal is to address the limitations of 
    existing mechanisms and
    enable safe and fast %efficient, % let's just stick with safe, fast, and time to time in-situ
    % and pervasive 
    tuning
    of any \perfconsts\ in a running OS kernel
    with {\it principled OS tunability}:
%    with following properties:

% To enable safe, effective performance tuning, 
    % the OS must provie principled tunability.
% We present the following goals of principled OS performance tunability.
% In this work, we aim for \emph{systematic tunability} -- a principled and
% general approach to manage the tunability of OS \perfconsts, 
% (tianyin) 
% so that the performance ceiling of modern systems can be explored, experimented with, 
% and realized in production.

% \subsection{Goals}
% dynamic caretaria
% out of box work on any commerial distribution -- no source modification
\minisec{\bf Transparent in-situ tuning on deployed systems}
% Millisecond-scale policy updates
% no reboot (X sysctl, X source modification), no app disruption (result of reboot/interference) -- transparent?
Tuning a \perfconst should not 
    require the slow process of recompilation, redeployment, or rebooting
    the OS kernel. % for each value and policy.
% It must not disrupt any running applications or system services.

% \minisec{Flexible policies and programmability}
\minisec{\rev{Programmability and flexible granularity}}
% easy-to-use (X sysctl, X livepath)
\rev{Effective tuning requires programmable policies
    that can encode sophisticated algorithms and heuristics
    through fine-grained control
    (e.g., TCP flow awareness~\cite{TcpPingpang-2023}).
    % (e.g., flow awareness~\cite{TcpPingpang-2023}).
    % with state, strategy, and kernel observability 
    % to enable fine-grained control (e.g., flow awareness~\cite{TcpPingpang-2023})
    % for diverse application SLOs. 
Policies should apply at flexible granularities
    (e.g., selected cgroups, selected devices, or their combinations)
    instead of fixed ones.}
% Existing global value-based mechanisms cannot meet these
%    needs~\cite{TcpPingpang-2023}.

% Effective performance tuning requires programmable policies that support
%     sophisticated logic, stateful algorithms, and internal kernel observability,
%     enabling workload-aware, hardware-aware, and fine-grained isolation
%     decisions for diverse SLOs. Global value updates, provided by existing
%     mechanisms, cannot accommodate the needs~\cite{TcpPingpang-2023}.
% The system should support programmable tuning policies, enabling 
%    new opportunities for performance improvements beyond
    % {\it when}, {\it what}, and {\it how} to tune (\sref{sec:bg}). 

% \minisec{Millisecond-scale policy update}
% no online recomplication (X livepath)

\minisec{Out-of-box tuning for all \perfconsts}
% not source code modification (upstream time, X sysctl)
% no external dependency
A general tuning mechanism must support {\em any} \perfconsts 
    in the kernel and be fully
    compatible with standard OS kernel distributions.
% , working {\em out of the box}.

% \minisec{Strong correctness and safety guarantees}
% X sysctl, X livepatch, and it is a correct and strong system!
% Challenge only instead?

\minisec{Millisecond-scale policy updates with low overhead}
% kprobe overhead -- sol
% always-on tuning
Performance tuning must be fast and not cause interference on the target OS.
We aim at {\it millisecond}-scale policy updates.
% , and low-overhead tuning for always-on online experimentation.

\minisec{System safety during tuning}
Tuning must be safe at runtime and must not result in inconsistent system states.
Arbitrarily changing a \perfconst value at runtime can be unsafe.

%% file: 3_5_challenges.tex
\vspace{3.5pt}
\subsection{Challenges}
\label{sec:challenge}
\vspace{1.5pt}

Principled tunability introduces new challenges, stemming from the
requirement to avoid recompilation and rebooting.

\minisec{Precisely locating instructions for the \perfconst}
The compiler materializes each \perfconst into instructions in the
kernel binary. However, compiler optimizations can
fold constants, merge lines, %expand a constant into multiple instructions, 
or reorder nearby instructions. A missed instruction leaves part of the original value in
the system state; a misidentified one may overwrite 
    unrelated computation and corrupt the system state.
% semantics.

\minisec{Generating instructions for new values and policies}
When users express new values and policies in a
high-level language, we must generate the required instructions directly.
KLP systems achieve this by recompiling the kernel, which introduces
minutes of delay. 
% \sysname instead synthesizes instructions without recompilation. 
This is challenging because the generated code must interact
correctly with existing kernel instructions and avoid conflicts such as
register clobbering near the original sites.

\minisec{Side-effects of the original value}
The original value may have already propagated, producing
    side effects that kernel execution still depends on. 
A reboot clears these effects but is incompatible with our goal.
We therefore must identify and
detect when these side effects have dissipated so the new \perfconst{} value can be applied
safely at runtime.

%% file: 4_0_design.tex
\vspace{4.5pt}
\section{\sysname}
\label{sec:sys}
\vspace{2pt}

% The last version
\input{4_0_design_afterchallenges}

\vspace{3pt}
\subsection{Design Principles}
\vspace{1.5pt}

% C1,C2,C3 (in the order)
% G1,G2,G3,G4,G5?
% principles are the enduring ``how'' that realizes what we want, either for C, or for G
% also highlights of design choices, in case they are burried in the middle (multi-deminision highlighting than the sec title)

% G2
% what's our main contribution
\minisec{Separating value updates from tuning policies}
% Updating the value requires online instruction generation that interacts with
% runtime kernel state and with the instructions that materialize the original
% value. Value update must follow strict constraints, tuning policies
% themselves may be arbitrary.
\sysname separates mechanism from policy: SIE provides a general mechanism for
safe in-situ value updates, while programmability is
delegated to safe kernel extensions (e.g., eBPF). 
% \sysname remains fully compatible with eBPF programs and leverages the rich tooling and ecosystem already in place.

% G4
\minisec{Synthesizing state update code, not patch instructions}
Directly creating replacement instructions without recompilation is
untenable. Instead, \sysname synthesizes code that updates the system state
described by symbolic state expressions of the \perfconst, turning instruction replacement into a
general state-update function. % applicable to any value. 
% This code is compiled by
% the existing eBPF JIT, fully compatible with kernel infrastructure and
% user-defined policies.

% translate code update into kernel state update
% translate temporary properties to spatial properties

% G1
\minisec{Reusable static processing for fast policy update}
\sysname incurs a one-time offline cost per \perfconst for kernel build,
symbolic execution, and analysis. The resulting artifact (\xktable) enables
millisecond-scale {\it in-situ} updates and is reusable across all future values and
policies. In contrast, live patching or source-code modification requires
kernel rebuilding and rebooting for every change.

\minisec{Decoupling version atomicity and side-effect safety}
% \minisec{Decouple transitioning mechanisms from unit}
% \minisec{Precise CS, and Approximate SS}
% Prior kernel live update systems uses function as the unit to ensure
% version atomicity (as opposed to bottom frame is using the old version code, and
% the top frame is using the new) and side-effect safety, without considering side-effect safety.
\sysname ensures version atomicity %at the granularity of the critical span, 
with correctness guaranteed by symbolic state expressions, when transitioning 
  from the original value to a new one. 
\sysname also offers side-effect safety based on
  safe spans that encapsulate all effects of the original value.
The safety of transition is supported for both per-thread and multi-threading.
% safety independent of transient execution timing.

\minisec{Encapsulated kernel writes; free reads}
To allow user-written policy programs (Xk-tunes) to update \perfconst{} values, \sysname{} provides a
simple, safe API. Writes to kernel state are protected and only done by
\xkruntime{}. % by whitelisting. 
% In contrast, we impose no restriction on reading kernel state.

% \minisec{Minimality}
% fast transition time, easier convergence, easy analysis
% necessary
% catch the perfconst effects
% small so it does not call out, it does not have lock() unlock(), and does not sleep/yield
% however, SS is not minimal -- it is over approximation 
% not sure if CS, SS is concrete yet here to be used in principles
% ok solved

\input{4_1_relation_expression}
\input{4_2_value_enforce}

\input{4_3_runtime}
\input{4_5_program_model}
% \input{4_6_limitations}

%%%%%%%%%%%%%%%%%%%%%%%%%%%%%%%%%%%%%%%%%%%%%%%%%%%%%%%%%%%%%%%%%%%%%%%%%%%%%%%%%%%%%%%%%%%%
% ARXIV
\if 0
\zhongjie{I like the abstraction of the xxx plane. How about Programmable Tuning Plane (PTP) \\Compile once, Modify arbitrarily}
% Great that the plane is liked! We cannot use the abbreviation PTP, at least not
% here; the main reason is that abbreviations become risky when there are too
% many. SIE stays because we need to reference it repeatedly, and it unifies
% both the conceptual novelty and the approach-level innovation.  We might call
% it programmable tuning plane if it works better, but will use the full
% description for better readibility.  still weighing between tuning vs. policy
% -- policy is good to make the point of seperating policy and mechanism --
% highlighting the expressiveness benefit tuning looks a bit charming at first
% glimpse.  another consideration here is the policy plane is not our
% contribution; we enables it (that's the contribution of SIE), but the
% programmability comes from eBPF.  
% Tuning plane somewhat sounds like a feature, rather than an abstraction.
\fi

%% file: 4_0_design_afterchallenges.tex
\sysname is the first system to provide a safe, 
fast,
% efficient, 
and expressive
interface for {\it in-situ} tuning of any \perfconst on a deployed
Linux kernel.
We present the design of \sysname, beginning with {\em Scoped Indirect
Execution} (SIE), a novel mechanism for principled tunability that addresses
the challenges outlined above (\sref{sec:challenge}). We then describe the design
principles and the key techniques that realize SIE and provide programmability
in \sysname.

\vspace{3.5pt}
\subsection{Scoped Indirect Execution (SIE)}
\label{sec:sie}
\vspace{2pt}

% This is the idea/approach
% want people to understand what we do from the high level
% SIE is the key approach of \sysname{} that realizes safe and efficient
%   \perfconst{} tuning.
% the core insight: what's unique about constant
The insight is that \perfconsts exhibit intrinsic
  instruction-level semantics.
% that shed light on in-situ tuning without recompilation or reboot. 
For a \perfconst, the point where its value
enters the \rev{machine state (registers/memory)} forms a {\em well-scoped region}. 
% naturally separating the tuning mechanism from the programmable policy.
% property of the scope (constant's effects)
This region---a {\em critical span}---has 
  three properties: 
  (1) it can be localized by static analysis,
% the constant appears as a literal (possibly transformed) determined entirely by the compiler. 
% Second, 
  (2) it corresponds to a {\em symbolic state expression} that
  captures the relation between the machine state and the \perfconst, 
% despite compiler optimizations that may alter instruction forms. Third, 
(3) the region is {\em small}, involving only a few instructions, 
  which keeps correctness
  reasoning narrow and makes safe runtime updates tractable.

% how the SIE works
% this is the core of SIE, others are used to make it work
SIE identifies the critical span based on the symbolic state expression and
safely redirects execution, at runtime, to a short JIT-compiled instruction
sequence. These instructions execute next to the original code and, when
combined, produce the same effects as if the \perfconst{} were changed
directly. The original binary remains intact; all modifications occur
indirectly, within scoped regions.

\if 0
% the indirection -- source code
The JIT input combines a user policy with synthesized source codes tied to
its invocation site; we call these {\em SIE indirections}. They are synthesized
based on the recovered symbolic state expression, which provides correctness
guarantees for runtime updates without recompilation.
\fi

% second scope: it is analyzed from the expression
SIE ensures side-effect safety by defining a second scope, {\em safe
span}, within which the effects of the \perfconst are encapsulated. 
Transitioning occurs only when execution is outside this
span, converting a temporal coordination problem into a spatial one that can be
analyzed statically. % based on the precisely recovered expression. 

\minisec{SIE in \sysname}
% \sysname{} is a system that realizes the principled performance tunability
%   for OS kernels defined in \sref{sec:tunability}.
% \sysname{} offers a safe, efficient, and programmable interface 
%   for {\it in-situ} tuning of any \perfconsts directly on a deployed Linux kernel.
\sysname implements SIE on Linux.
%  without (re)compiling or rebooting the kernel.
% \sysname essentially transforms a \perfconst into a tunable knob on demand,
%  which can be updated safely at runtime.
% \sysname{} is efficiently for online performance tuning at the microsecond scale.
% Its expressive and programmable interface can support fine-grained tuning policies
%  and can be used by tuning programs or agents.
Figure~\ref{fig:xkernel-arch} gives an overview of \sysname
  and its user-facing workflow.
\sysname{} performs {\it offline} static analysis on kernel code to understand 
  the symbolic state expression, critical span (CS), and safe span (SS) of each \perfconst,
  and maintains them in a global \textit{\xktable}.
%   the scope of each \perfconst{} based on how each \perfconst{} is used in kernel instructions and 
%   affects kernel runtime states; the scope information is 
%   encoded in a data structure termed critical span (CS).
% It then synthesized code to update the kernel states related to the \perfconst
%   using a novel technique termed {\it Scoped Indirect Execution}.
% \sysname{} also captures the transition points in the kernel code 
%   that ensures safe transition from the original \perfconst{} value to a new value;
%   the information is encoded in a data structure termed safe span (SS).
% \sysname{} maintains the CSes and SSes of each \perfconst{} in a global \xktable.
% Each entry maps the constant's ID (i.e., source location) 
%  to its symbolic state expression, critical spans, and safe spans
% (labeled as expression, CS, and SS in Figure~\ref{fig:xkernel-arch}). 
The offline analysis is a one-time effort.
The \xktable only needs to be updated when the deployed kernel is updated.
% \sysname can build this table ahead of time for any set of constants; this cost is paid
% once per constant.

To tune a \perfconst{}, users (or agents) specify the \perfconst{}
  in source code (by source file, line number, and token index)
  and implement tuning policies in eBPF, called {\it Xk-tune}.
An Xk-tune can be loaded at any time in the OS. A lightweight
\xkruntime{} handles attachment and transitions.

\input{fig-tex/fig-xkernel-arch.tex}

% invokes \xkgen\ with the
% constant's ID. \xkgen\ generates a skeleton program exposing a programmable
% interface (e.g., \smalltt{xk\_set()}); the user then edits the skeleton to
% implement their tuning policy.

% When a \xkprog\ is loaded, \xkruntime\ consults the \xktable\ to determine the
% relevant binary spans, synthesizes the required indirections, and feeds both the
% user policy and the synthesized code into the eBPF JIT to produce bytecode.
% These details remain internal and invisible to the user. \xkruntime\ then
% initiates transitioning, with side-effect safety guaranteed by the safe spans.

% Once transitioning completes, any execution that reaches an attached indirection
% point is redirected to the generated instructions. The externally visible
% behavior is ensured to reflects the updated value once execution has passed the
% critical span.

%% file: fig-tex/fig-xkernel-arch.tex
\begin{figure}[t!] 
  \centering
  \includegraphics[width=1\columnwidth]{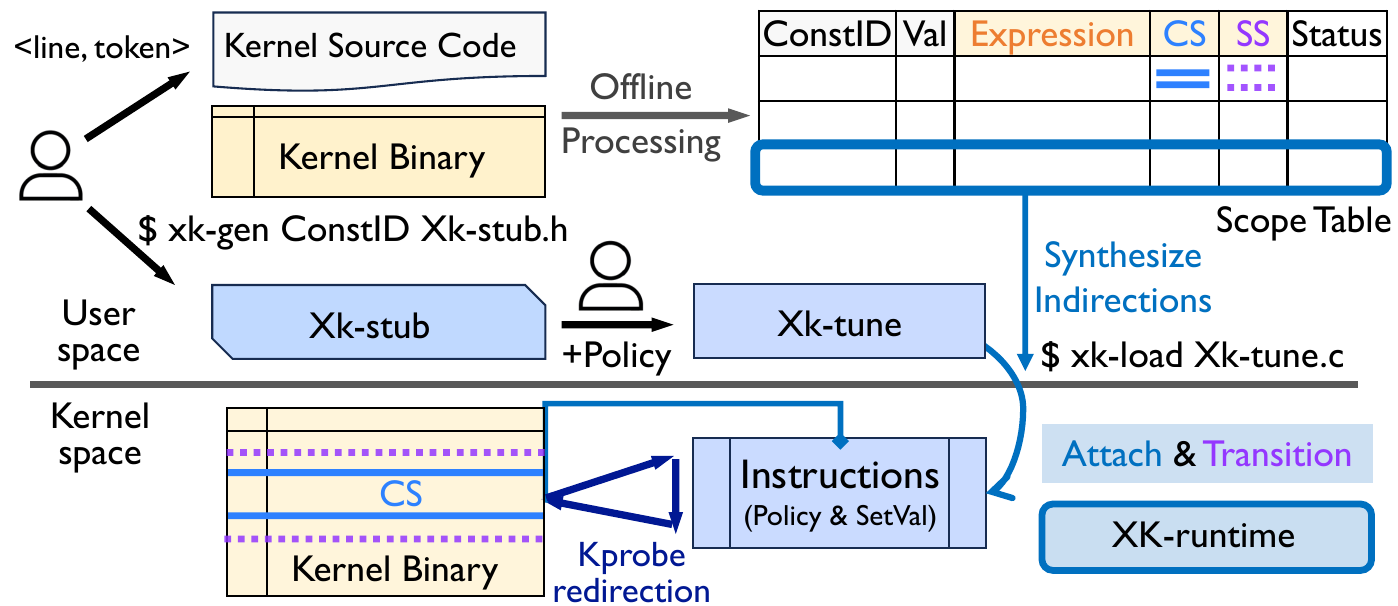}
  \vspace{-6.5pt}
  \caption{{Overview of \sysname}.}
  \label{fig:xkernel-arch}
\end{figure}

%% file: 4_1_relation_expression.tex
% \subsection{The Semantic Scope: Value Enforcement}
% \subsection{Locating and Comprehending Perf-Consts}
\vspace{3pt}
\subsection{Capturing Instruction-level Effect of Perf-Consts}
\label{sec:design:val}
\vspace{1.5pt}

Conceptually, \sysname must replace the effect of instructions tied to 
  the original \perfconst value 
  with the updated effect based on the new value,
  while preserving other runtime state.
A key challenge is to precisely capture the effect, as reversing kernel 
  binaries to original source code is difficult and costly.
Our insight is that the state affected by \perfconsts is tractable to capture.
A \perfconst\ manifests as a numeric value chosen by the compiler 
% independent of \rev{runtime machine state}, 
  and becomes part of runtime state by
  instructions that produce a local numeric effect.
Despite compiler optimizations that obscure its representation, we
    find that {\em symbolic state expression} provides a clean, precise way to
    express how a \perfconst\ affects runtime state.

% We next examine how a \perfconst\ propagates into runtime state, motivating
%    symbolic state expressions (\sref{sec:expression:intuition}), and then describe
%    how we recover such expressions from the binary
%    (\sref{sec:expression:recovery}).

\vspace{2pt}
\subsubsection{How Does a Perf-Const Affect Runtime State?}
\label{sec:expression:intuition}
\vspace{1.5pt}
% What's exactly the problem, give the right definition here (implicitly), make the source -- binary difference clear
% What makes it possible?
% No matter how messy it is, the symoblic expression is clean -- because it refelect a constant, and how a constant can affect the run time state
% the symbolic expression is simple enough to avoid many issues, e.g., handle multi-variable, and match the constant, alias
% the occurance is transient
% Want to show why it works
% Show the idea and intuition to make the further explaintaion easy (i believe reviewers don't need details...)

% What's the problem -- set up the challenge (compiler, it makes decisions)
If a value is a constant and ultimately affects system state, there must be a
well-defined point where it enters \rev{machine state}. This boundary reflects
the compiler's decision to materialize the constant and is determined
statically. Since a constant carries no state before its introduction
(unlike a variable), this entry point is identifiable in principle and is the semantic boundary
we must capture.

\begin{figure}[t!] 
  \centering
  \includegraphics[width=1\columnwidth]{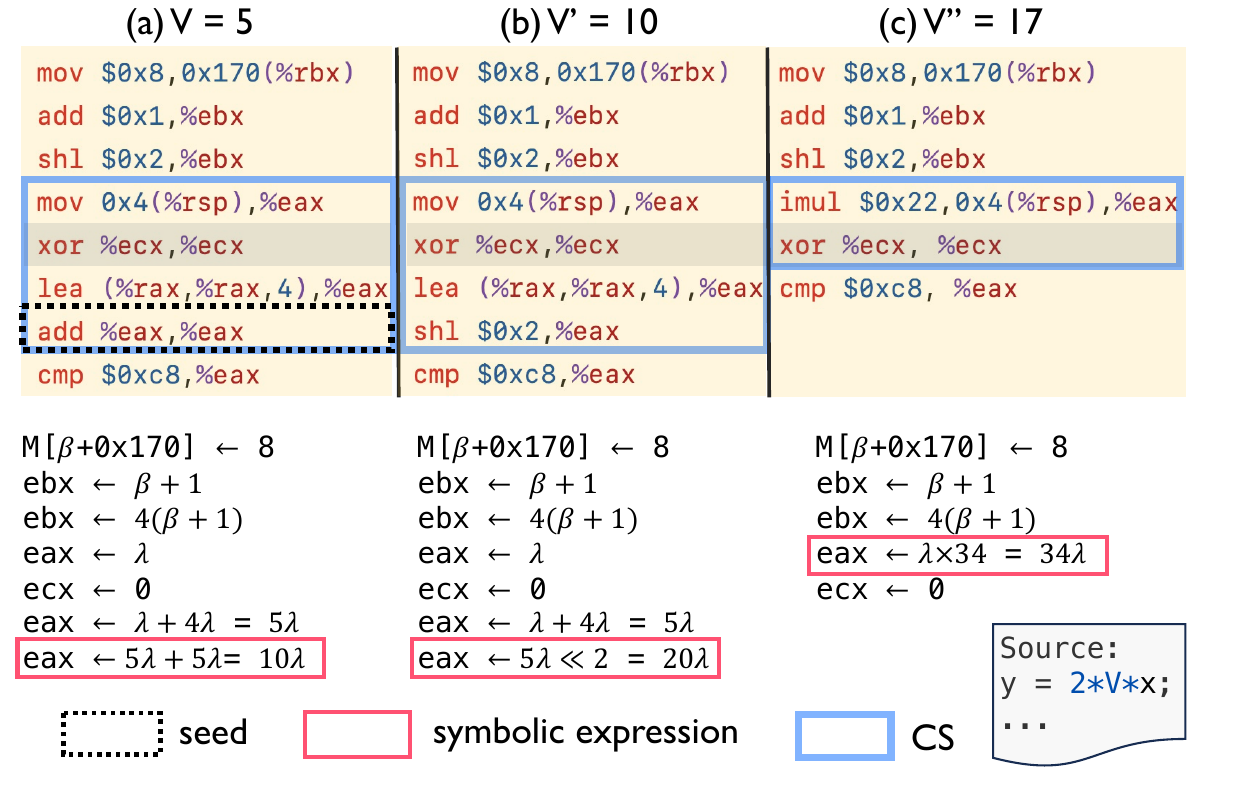}
    \vspace{-8.5pt}
  \caption{{Symbolic expression derived from a \perfconst($V$){}'s seed instructions
    and its critical span.}
    (a) and (b) show instructions from binaries rebuilt with $V'=10$ and $V''=17$.
    The seed is obtained by diffing (a) and (b).
    The $IV$ values in the symbolic expression are 10, 20, and 34.
    }
  \label{fig:recover-expression}
  \vspace{5pt}
\end{figure}

% Describe all the complication
A key challenge is to handle complications due to compiler optimizations. 
A \perfconst\ may be folded,
propagated, or merged with other expressions.
Backend optimizations may further rewrite
it for strength reduction or
% (e.g., using \smalltt{lea}), 
% generate cheaper architectural forms, --> isn't it strength reduction? 
interleave unrelated instructions.
As shown in Figure~\ref{fig:recover-expression}(a), the instruction
sequence for a source-level \perfconst of value 5 contains no immediate value 5 in the
binary. Recompiling the same source with values 10 and 17 produces distinct
instruction sequences (e.g., transforming \smalltt{add} into
\smalltt{shl}, and reducing three instructions to one).
% (e.g., $10 = 2 * 5$ in
% Figure~\ref{fig:recover-expression}). 
Thus, the \perfconst often appears in transformed
form, denoted as $IV$, not its source value~$V$.

% The shape of expression, the simplification we can always try to find R/M <- MathExpression(R/M, V)
% no difference for source code level logic expressions
Despite these transformations, $IV$ can only influence \rev{machine state} by
becoming part of register or memory states, denoted as $R/M$. 
All effects must arise through
instructions expressible in arithmetic and bitwise operations. Even
source-level logical expressions reduce to comparisons such as ``\smalltt{cmp} \smalltt{eax}, \smalltt{edx}'', 
which update condition flags as the arithmetic result of ``\smalltt{eax} \smalltt{-} \smalltt{edx}''.
Thus, regardless of source syntax, the effect of a \perfconst\ in the binary
always reduces to a mathematical relation in the form: $R/M \leftarrow f(R/M, IV)$.
% \tianyin{why R/M inside f?}.
% It's just means some math expression of e.g., eax <- eax*2 is f(eax, 2), just
% means its a math expression f, whose operands include R/Ms, e.g., it could be eax <- edx * 2 + 5
% eax <- edx * 1 + ecx * 2 + 5 (R/M is variable)

% Now, it is the rescue (the idea)
Once expressed symbolically, this relation becomes
explicit. As shown in Figure~\ref{fig:recover-expression}, the same source code 
with $IV$ equal to 10, 20, and 34 produces
% distinct instruction sequences produce 
drastically different binary code, yet their symbolic state expressions all reduce to \smalltt{eax} \textleftarrow\ $IV$ $\times$ \smalltt{eax}.
%  with , respectively. 
Intuitively, compiler transformations preserve the numeric semantics of the constant, 
and those semantics are directly reflected in this symbolic form.

% What we do -- set things up for next
We use symbolic execution to derive a {\em symbolic state expression} that
precisely captures how a \perfconst{} affects runtime state. Symbolic execution
naturally isolates relevant instructions from noise introduced by optimizations
and yields the mathematical relationship between $IV$ and architectural
state $R/M$. Resolving $IV$ back to $V$ yields the final
expression relating~$V$ to \rev{runtime machine state}.

\vspace{3.5pt}
\subsubsection{Deriving Symbolic State Expressions}
\label{sec:expression:recovery}
\vspace{1.5pt}

The first issue is locating the instructions that consume a \perfconst. 
Debug information (e.g., line number) is too coarse-grained---one  
  source line can be compiled into hundreds of instructions, 
  only a small subset of which uses the \perfconst. 
\sysname therefore begins by identifying seed
instructions through binary differencing: we modify the \perfconst\ at the
source level and rebuild the kernel, then use the diff to drive symbolic
execution. 
We run symbolic execution on both binaries and derive the
transformed value ($IV$). % by comparing their symbolic expressions.

\minisec{Finding seeds}
Seed instructions are those that differ when the \perfconst value changes
  in source code.
\sysname finds seeds by assigning the \perfconst a magic value that 
  differs from the original value and rebuilding the kernel binary.
  For example, when $V'=10$, the seed instruction is \smalltt{add eax eax}.
In principle, instructions in the binary diff are the seeds.
% In practice, \sysname addresses noises introduced by 
%   compilation \tianyin{(see \sref{sec:impl})}.

%  Note that such noise leads to false positives, leading to wasted time on
%  symbolic execution, but not correctness, as symoblic expression cannot recover
%  the value.

\minisec{From seeds to symbolic state expressions}
Symbolic execution expands from the seeds to derive a converged symbolic state
expression that reflects their effect. We exhaustively explore instructions
backward and forward until the symbolic state reaches a fixed point while still
incorporating the seed effect.
The two symbolic executions must converge to the
same symbolic form.
In theory, extracting $IV$ from a multivariate symbolic expression may require
multivariate coefficient matching when multiple symbolic variables (registers or
memory locations) appear. However, \perfconsts\ act as simple knobs, and
we did not observe such cases in our experience and our extensive evaluation.

A remaining complication is solver canonicalization: symbolic engines rewrite
expressions in simplified mathematical forms (e.g., \smalltt{shl} \smalltt{2} becomes
$\times 4$). This matters when the source uses shift operations
like \smalltt{\textlangle\textlangle} or \smalltt{\textrangle\textrangle}.
\sysname explores multiple symbolic execution branches and discards
those that cannot preserve a linear relationship with~$V$.

\minisec{Mapping transformed value $IV$ to the original value $V$}
To recover the relation between $IV$ and $V$, \sysname attempts to fit a linear
mapping of the form $IV = a \cdot V + b$. We compile additional binaries using another
modified value $V''$, extract the resulting $(V,IV)$ pairs, and solve for $a$
and $b$ via interpolation. When the recovered mapping is non-linear (e.g., $V
\times V$), \sysname reports the case, allowing users to tune the derived
$V \times V$ rather than~$V$. However, we never encountered such a case in our evaluation.
% \jing{Add some words saying that we don't come across in our dataset?}

\if 0
Seed instructions alone often do not represent the entire CS of a \perfconst,
  as exemplified in Figure~\ref{fig:src-binary-sym}.
\sysname derives the CS by expanding the seed instructions.
It performs both forward slicing to include instructions that directly 
  affect the source operands of each seed (\texttt{eax} and \texttt{edx}
  in Figure~\ref{fig:src-binary-sym})
  and backward slicing to include instructions that are directly affected 
  by the destination operands of each seed (\texttt{eax} in Figure~\ref{fig:src-binary-sym}).
Note that both the forward slicing and the backward slicing 
  are done at the binary level through symbolic execution.
Note that by definition, a CS is within the boundary of a basic block.
\fi

%% file: 4_2_value_enforce.tex
\vspace{3.5pt}
\subsection{Synthesizing Indirections}
\label{sec:synthesis}
\vspace{1.5pt}

The recovered expression precisely captures how the original
value affects runtime state; %, answering what's the origin to be replaced.
the next step is to generate instructions using new value.
% In contrast to seeking those indefinite number of diffs varies for each value in live patching,
\sysname turns the instruction replacement problem into a state-update problem.
% ---the state update must be generic to any new values.
%instead of directly executing instructions correspond to each new value, 
It synthesizes code that overwrites
\rev{machine state}
% \rev{runtime machine state}s 
affected by the original execution.
%  where is correct and also fast.  We call this piece of source code SIE indirections.  
The synthesized indirection is in the form of \{\indPos, \indCode\} pairs;
  when the specified kernel location is reached, 
  the update code ensures that the effects of the system
  states is equivalent to that as if  
%  binary representation for each to using 
  the new value $V'$ of the \perfconst were used in the kernel source.
We introduce a binary representation for each
symbolic state expression termed {\em critical span}
and present our synthesis algorithm.

% (tianyin) This is really overly complex for a simple concept
% Made possible by SIE Indirection's update of states written in source code, we are able
% to leverage existing JIT to compile the SIE indirections and user policies together,
% effectively relying on JIT engine to safely ``synthesize'' instructions for SIE,
% and finally achiving the JIT-syntheized \perfconst update without recompilation.

% We introduce an abstract {\em critical span} (CS) for SIE, as the corresponding
% represeantation of the recovered symoblic state expression in the instruction
% level, which facilites the redirection of control flow, and expressing the
% placement of generated code, formulating correctness invirant of replacement, and
% facilites the overhead optimization.

\vspace{2pt}
\subsubsection{Critical Span}
\label{sec:cs}
\vspace{1.5pt}

% \minisec{Definition}
A critical span represents one occurrence of a \perfconst at the binary level.  
Concretely, a CS is a single-entry, single-exit instruction sequence that begins
  at the first instruction contributing to the \perfconst's symbolic state
  expression and ends either when the resulting state is first consumed or when
  the basic block ends.
% instruction where the value of the \perfconst appears in an arithmetic or bitwise
  %% Jing: it can be logic expression at source level, but only arithmetic or bitwise in instruction level
  %% expression (at the source-code level) and ends at the point
  % where that value % fully (cannot parse the word)
  % becomes part of the runtime state (i.e., influencing registers or memory locations).
% \tianyin{q: is a CS always corresponding to one statement in the source code?}
  % Jing: good question, partial/one/two/many are all possible
%% The CS captures the correctness condition for value update.

% \jing{the more accurate def is: just before the first time the effect (left side
% of the math expression (arithmetic or bitwise)) is consumed, the
% instruction-level behavior is more microscopic and is insightful, we need to
% highlight This!  At the binary level, logical predicates (e.g., x > V) are not
% represented as boolean expressions; instead, the compiler normalizes them into
% arithmetic or bitwise relations over \rev{runtime machine state}. For example, if (x >
% 0) becomes cmp eax, IV, which updates EFLAGS based on the arithmetic relation
% (eax - IV). The branch instruction interprets these flags to realize the
% original logical intent.
% }

\minisubsec{Invariant}
Let $\sigma_{\text{in}}$ denote any machine state, $CS_v$ denote the
critical span executed with value $v$, and $I$ denote the synthesized
instructions that update $v$.
A value update from $v$ to $v'$ is correct when executing the critical span is
observationally equivalent to executing the span with the new value:
\[
\forall \sigma_{\text{in}}:\;
\mathsf{Exec}\bigl(\sigma_{\text{in}},\, CS_{v} \circ I\bigr)
\;\equiv\;
\mathsf{Exec}\bigl(\sigma_{\text{in}},\, CS_{v'}\bigr)
\]
Here, ``$\equiv$'' denotes equivalence of externally visible state at the exit of
the CS.

\minisec{Construction}
CSes can be automatically constructed for a given \perfconst
from the recovered symbolic state expression.
\sysname{} conducts forward and backward slicing 
  to find the start instruction
  that first consumes the value to the last instruction
  that can recover the symbolic state expression. % it is first consumed,
% which is critical for placement decisions. % to enable overhead optimization.
A CS is specified as a pair of [start, end] binary addresses.

\begin{figure}[t!] 
  \centering
  \includegraphics[width=1\columnwidth]{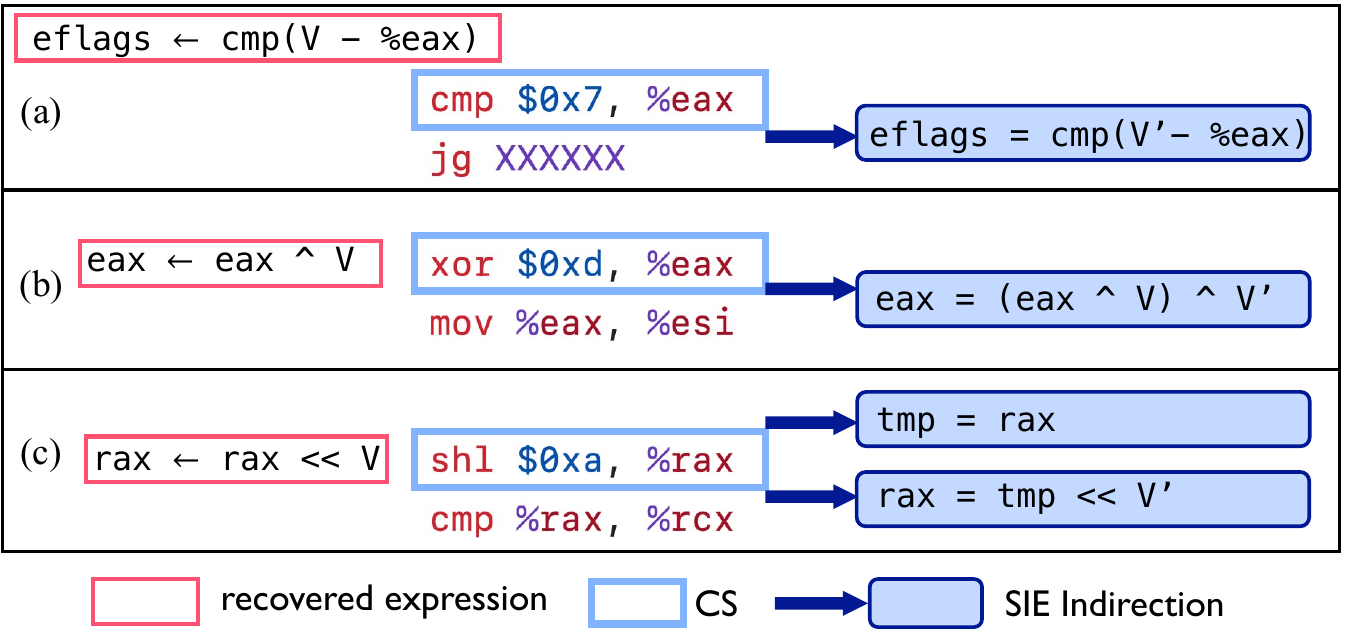}
  \vspace{-9.5pt}
  \caption{{Synthesized SIE indirections for various CSes}. 
  An SIE indirection contains pairs of \{\indPos, \indCode\}. Locations are shown by the blue arrows.}
  \vspace{10pt}
  \label{fig:sie-code-examples}
\end{figure}

% \subsubsection{Kernel State Update Primitive for JIT}
% \label{sec:synchrsis:ptregs}

\vspace{5pt}
\subsubsection{Synthesis Algorithm}
\label{sec:synthesis:algo}
\vspace{1.5pt}

The indirection is synthesized based on the symbolic expression,
  as exemplified by Figure~\ref{fig:sie-code-examples}.
If $R/M$ affected by the \perfconst are not overwritten
within the CS, the indirection consists of a single code insertion located 
after the last instruction in the CS, computing the result using the new value
$V'$ and overwriting R/M (Figure~\ref{fig:sie-code-examples}a).

If $R/M$ are overwritten within the CS, the \indCode{}
  depends on whether the overwriting instructions are reversible.
For reversible arithmetic or logical operations (Figure~\ref{fig:sie-code-examples}b),
\sysname synthesizes inverse logic as part of the \indCode. 

If the modification cannot be reversed (e.g., due to irreversible arithmetic,
bit masking, or information loss), \sysname generates dual-location: 
  the original value is captured before modification and restored
  after the CS (Figure~\ref{fig:sie-code-examples}c).

Note that SIE does not skip the original CS; all instructions
in the CS still execute to preserve state equivalence,
  including instructions that are in the CS but do not 
  depend on the \perfconst, e.g., \smalltt{xor} instructions
  in Figure~\ref{fig:recover-expression}(a). 
By construction of the symbolic state expression, SIE
updates only the states related to the \perfconsts.
% and does not modify any unrelated \rev{runtime machine state}.

SIE relies on the Kprobe mechanism~\cite{kernel-probe} to redirect execution to
JITed \indCode{} code. Because each trigger incurs overhead and different probe
locations have different costs, \sysname adjusts the placement within a CS,
using the symbolic state expression to ensure correctness (details
in~\sref{sec:impl}).
 % by specifying a function and offset. 

% \minisec{Code compilation.}
% The key primitive SIE requires is the ability to update the architectural
% context of a running kernel thread while executing inside a JIT VM (e.g., eBPF).
% Linux already exposes this context through tracing infrastructure such as
% kprobes and ftrace, and eBPF inherits these interfaces. This exposure has become
% part of a stable tracing ABI, as many features, including return-value
% modification and fault injection, rely on it~\cite{BpfTraceRawCtxComment}.

% \sysname extends this access by enabling controlled writes to the architectural
% context. We implement this capability in a kernel module without modifying
% kernel source. Safety guarantees through encapsulation and a restricted
% user-facing API are discussed later in \sref{sec:design:api}.

% \tianyin{Now I see why you need 4.2.2, though I still don't know how things 
%   are put together. There must be one figure to show me step by step.
% Figure 3 alone does not help.}

%% file: 4_3_runtime.tex
\vspace{3.5pt}
\subsection{Safe Transition}
\label{sec:design:transiton}
\vspace{1.5pt}

\input{4_3_zinc_1_ss}

\if 0
we should decompose the problem clearly
[metric] -> decides the scope to apply (e.g., Union of SSes, multiple functions, multiple CSes)
the scope to apply --> the relationship between scope decides the complexity of transition (total amount of time cannot transition and coditions to meet)
* the metric: single code versioning, no side effects
the global consistency solves a orthogonal problem: multi-threading
* it means, even the metric chosen is easier to meet, e.g., CS, multiple threads still need a protocol to ensure the global view of the CSes
granularities -> CS | function | SS
-----------------------------------------------
single thread |     |          |
multi-threads |

Call-chain dependences between granularities
Guarantees:          | version atomicity |    side-effect safety      |
CS                   |   naturally hold  |       X                    |
function             |   stack check     |       X                    |
SS                   |   stack check     |  dependes on SS definition |
\fi

\if 0
\begin{figure}[t!] 
  \centering
  \includegraphics[width=0.95\columnwidth]{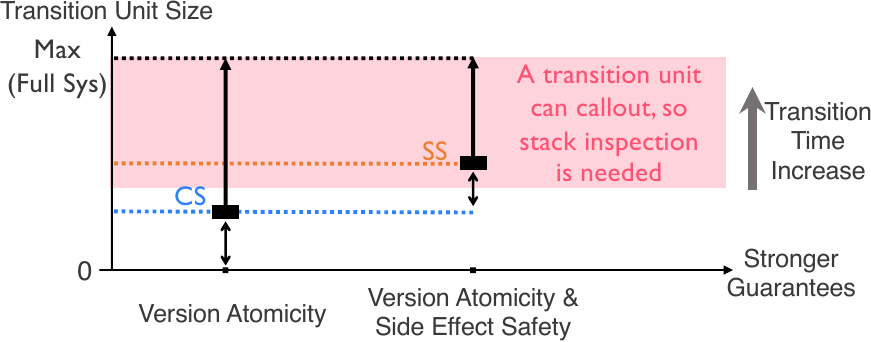}
  \caption{\textbf{The qualitative model of safety properties supported by 
    \sysname}. 
% Single thread is concerned; multi-threading is orthogonal.
    CS is the minimal unit to achieve version atomicity; SS is larger than or equal to CS.
    }
  \label{fig:transition-qual-model}
\end{figure}
\fi

\minisec{Safety properties}
As SSes of a \perfconst{} encapsulate the side effects 
  of its value, \sysname{} enables {\it side-effect safety} 
  when updating \perfconsts.
%  a property notably absent in existing kernel
%  update mechanisms~\cite{arnold-2009-ksplice,Kpatch,Kgraft}.

Existing Kernel live patching (KLP) mechanisms~\cite{arnold-2009-ksplice,Kgraft,Kpatch} 
  offer per-thread {\em version atomicity}---a thread
  must execute either the old or new code, not their mixture~\cite{LinuxKLP}.
KLP does so in the unit of functions, assuming every function
  having a well-defined semantic and being self-contained.
In \sysname, version atomicity is guaranteed through CSes. 
% which are the minimal transition scope of \perfconsts.
% Larger units (e.g., whole functions in kernel live patching) offer the same
% guarantee, but provide no additional benefit and are unnecessarily slow 
%  for \perfconsts.
% \minisubsec{Side-effect Safety}
% Version atomicity does not cover side effects on the system state 
%  created by the old code, 
%  which can affect the new code, leading to unsafe execution.
No existing KLP mechanism provides side-effect safety,
  as side effects of arbitrary function code are difficult to define.
  % as it is difficult
  % hard % Jing: hard is not allowed to use in a paper (Andrea removed it everytime)
  % to define side effects of arbitrary function code.

% The size of the transition unit has two implications.  First, as the unit grows,
% it eventually spans call boundaries. At that point, a stack check is needed to
% ensure that all active frames are free of the transition units. For example, the
% top of the stack may run the new version while a lower frame still runs the old
% one. Using the CS as the transition unit avoids this problem: CSs are small, and
% under \perfconst semantics, they do not call outward.

%Systems such as KLP instead use functions as the transition unit. 
% Function
% boundaries are always larger than CSs and may exceed the SSes. The benefit is
% simplicity: the function entry and return form clear boundaries, and the
% function name symbol is easy to locate and check -- especially when examining
% the stack.

% Second, the size of the transition unit directly affects transition time.
% Larger units reduce the opportunities to find a safe transition point along a
% thread's execution path, increasing the expected time to complete the
% transition.

\sysname{} also guarantees multi-threading safety beyond 
  per-thread safety by enforcing
  {\em global consistency}---all participating 
  threads % tasks (i prefer to use a clear word here and not using their sloopy term, task can mean a call back/coroutine or something, may be put task-... to BG)
  must be
  outside the SSes. 
In effect, the per-thread property must hold for all threads. 
% To safely complete the
% transition, the system must ensure that every thread has exited the transition units.

% Global consistency in terms of version atomicity 
%  is supported by kpatch at the unit of functions~\cite{Kpatch}.
% In \sysname, as a CS is typically much smaller than a function,
%  the transition time is significantly faster (see \S\ref{sec:eval}).
% \sysname also supports global consistency in terms of side-effect safety, while 
%  none existing kernel live patching mechanism does.

% Finally, we argue that systematic tunability requires decoupling the transition
% unit from value enforcement. 
% The CS provides the core code-level correctness
% guarantee and versition atomicity, while the SS captures the required level of
% side-effect safety for runtime state.  Our transitioning mechanism supports
% flexible SS policies for both single- and multi-threading, regardless of
% transition unit size.

% \todo{Need to add how to config the consistency guarantee -- ST or MT somewhere.}

\vspace{2pt}
\subsubsection{Transition Mechanisms}
\vspace{1.5pt}

% Safety spans define the required guarantees and their binary representation as a
%  set of address intervals (e.g., [start, end1/end2]), answering the question of what to
%  check. To enforce these guarantees, we introduce two primitives for 
\sysname{} ensures per-thread and multi-threading
  safety by monitoring the execution at the transition points.
%   transition points and inspecting the stack
%  to check if execution is outside all SSes.
% For multi-threading safety, the monitors use reference counting 
%  to check global consistency.
The monitors are implemented as \kernelProbes{}~\cite{kernel-probe}, 
  inserted at the entry and exit of each SS.
% Finally, we design a transition mechanism that achieves global consistency for any transition unit,
% enabling multi-threading side-effect safety.
These \kernelProbes{} are used only during the transition. 
Once the transition is done, they are removed.
\rev{The same transitioning mechanism is used for rollback and for updating the value of a \perfconst.}

\minisec{Per-thread safety} % Transition Monitoring via Kprobes
% A transition is safe only when execution lies outside all SSes.
% Checks that occurs inside an SS are wasted and must be retried. 
Unlike KLP that implements timer-based polling, 
  \sysname detects safe transition points proactively by monitoring execution at 
  the SS boundary.
% \sysname inserts kprobes at the entry of each SS.
When execution reaches an entry \kernelProbe{}, \sysname checks
  if this SS is deep in the call stack of other SSes (e.g., SS2 in
  Figure~\ref{fig:cs-ss-example})
  by stack inspection.
% If so, the transition has to wait. % Jing: feels this haults system
If so, execution continues and the check is retried upon entering another SS.

\minisec{Global consistency} 
\sysname also supports global consistency for multi-threading safety.
Unlike Kpatch~\cite{Kpatch} that uses a dedicated thread 
  to repeatedly invoke \smalltt{stop\_machine} to
  opportunistically check if all threads happen to be safe for transition. 
% However, \smalltt{stop\_machine} is expensive as it halts all
%  running threads. % -- effectively stopping the world.
\sysname introduces a new mechanism termed {\em self-convergent transition}. 
% to support global consistency.
%  for any transition unit size. 
The idea is to use efficient global reference counts maintained by 
  entry and exit \kernelProbes{} of each SS
  to track safety. 
\sysname invokes \smalltt{stop\_machine} only once to initialize
  the reference count by scanning all running threads 
  and counting the active SS entry boundaries in their stacks.
Afterward, participating threads naturally self-converge when
  crossing SS boundaries and update the reference count.
% Once the kprobes are installed, each thread increments the
% reference count when entering an SS and decrements it when exiting (since an SS may have multiple exits, we attach Kretprobe for unguard). 
When the global reference count is zero, no thread resides in any SS, and the
transition is safe.
% \tianyin{perhaps need a statement say how much benefits it brings}

% Installing Guard/Unguard probes requires a single invocation of
% \smalltt{stop\_machine}. 
% The sum becomes the initial reference value.

% If convergence is never reached, \sysname enforces a timeout 
%  and notifies users. \tianyin{what does timeout mean?}
%  \jing{it's when the system executing for a long but we cannot find any point
%  that all threads are out side of all SS, which is possible in theory; it is monitored
%  by the wall-time and number of triggers, then we cannot do anything about it.}

% Finally, some systems may never reach convergence. For example, a kernel thread
% may loop indefinitely or block within an SS. In such cases, \sysname enforces a
% timeout and notifies the user.

\minisec{\rev{Liveness and timeout}}
\rev{
  \sysname may fail to find a safe transition point for a \perfconst,
causing liveness issues.
This can happen when the instrumented path is not executed after the update
request, or when one or more threads remain inside an SS and prevent
convergence. In either case, \sysname reports the transition as failed after a
configurable timeout.}

% \rev{Rollback is straightforward: \sysname reinstalls the monitor
% \kernelProbes{} to track execution and waits for a safe transition. Once the
% transition succeeds, \sysname marks all SIE \kernelProbes{} as disabled (but not
% removed) to enable the old value; it then safely removes both the SIE and
% monitor \kernelProbes{}. Rollback can encounter the same liveness issue; in that
% case, \sysname reports the rollback as failed.}

\if 0
A single source-level constant may appear multiple times in the binary due to
macro expansion, inlining; and compiler optimizations may obscure its final
form.
\fi

%% file: 4_3_zinc_1_ss.tex
\begin{figure} %[t!] 
  \centering
  \includegraphics[width=1\columnwidth]{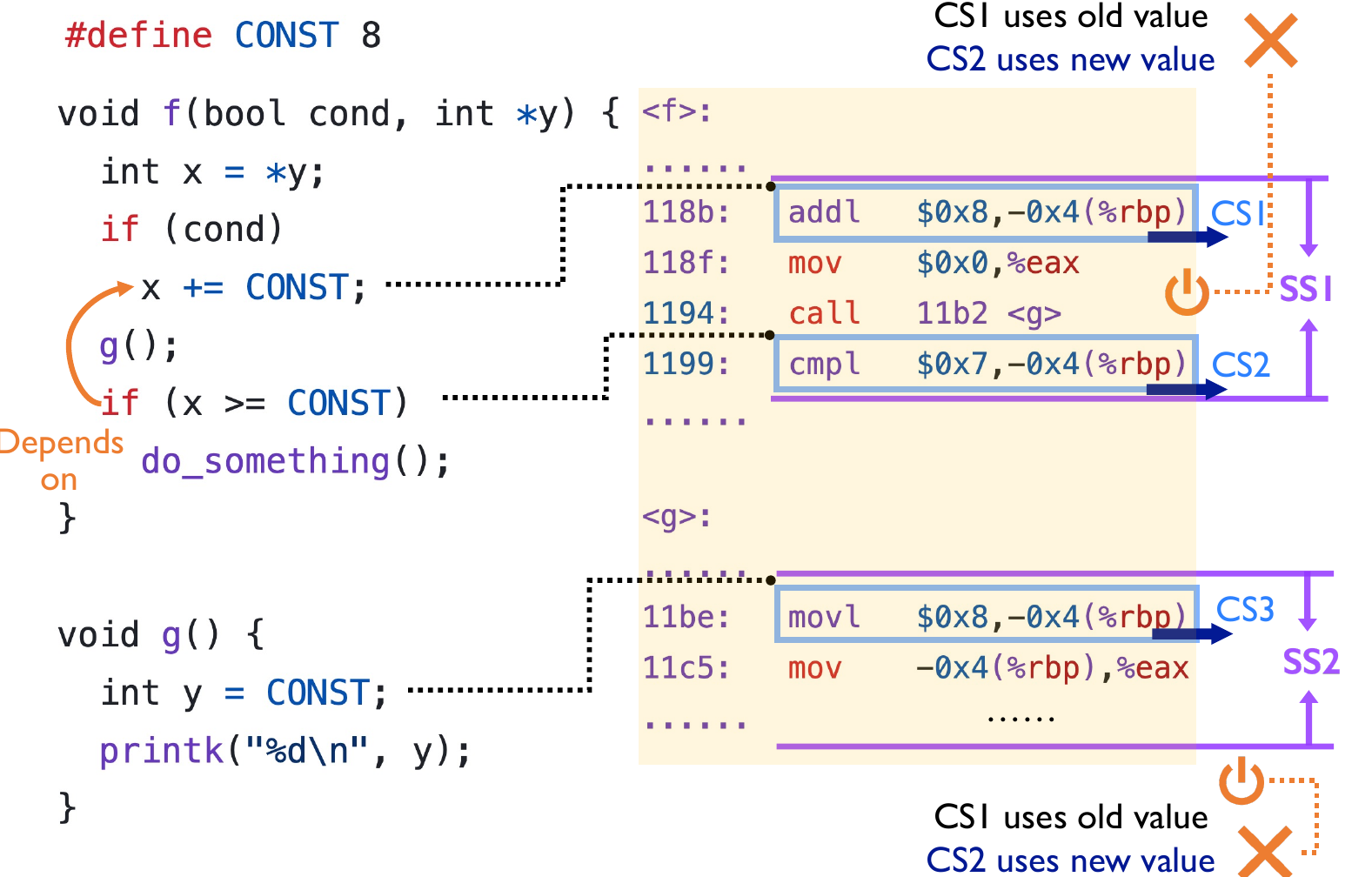}
  \vspace{-7.5pt}
  \caption{An example of a \perfconst's CSes and SSes. 
    Binary representation: SS1[118b, 1199], SS2[11be, 11c8].
    % \tianyin{why you do not keep tracking eax in SS2?}
    % \jing{should include remove printf in SS
    % , + ...}
    }
    \vspace{5pt}
  \label{fig:cs-ss-example}
\end{figure}

A symbolic state expression and critical span (CS) of 
    a \perfconst enables runtime updates of its value. % without recompilation.
However, it does not guarantee a {\it safe} transition of system states
    from the original value to the new.
Figure~\ref{fig:cs-ss-example} gives examples where unsafe
    transition of \smalltt{CONST} could introduce inconsistent system states.
If SIE is enabled when the execution is between CS1 and CS2, the
    system may execute CS1 with the old value and CS2 with the new. 
The program point right before \smalltt{g()}'s return is also unsafe. Although CS3 has no
    dependency on CS1 or CS2, its call stack lies between them.
    % resulting infinite loops.
% In this case, the updates of \perfconsts break 
    % the execution that requires a consistent view of the \perfconst value.
% Without a safe transition point, updates
%    of \perfconsts breaks execution that relies on a consistent view of the \perfconst value.

The core of safe transition is a well-defined transition scope that reflects the
    required safety guarantees; transitions are allowed only when execution
    is outside that scope.
We formalize this scope as a safety span (SS), which
    captures the required safety guarantees and their binary representation 
    (e.g., SS1 and SS2 in Figure~\ref{fig:cs-ss-example}).
%    Combined with our transition mechanisms, 
Safety span provides a concrete
    foundation for safe transition in \sysname.

%To realize safe transition, 
\sysname enforces an update on a \perfconst
    to happen {\it after} all the instructions that consume its 
    value finish their execution, i.e., the lifetimes of all data
    objects derived from the \const have expired upon exiting its safe spans.
\rev{We call this property {\it side-effect safety}. 
    % This is stronger than
    % version atomicity, which only prevents mixing old and new values.
    } 

% In \sysname, safe transition is realized based on the concept of Safe Span or SS in short.

\vspace{3.5pt}
\subsubsection{Safe Span}
\vspace{2pt}
% %. For instance, if the \const defines
% a loop's boundary condition, an ill-timed update could trap the program in a
% non-terminating loop. In another scenario, applying the new value between two
% CSs could violate program semantics by creating an inconsistent state. 

% \minisec{Definition}
A safe span (SS) represents the execution unit in which \emph{transitive
dependencies} of a \perfconst's critical span (CS) are encapsulated (see Figure~\ref{fig:cs-ss-example}).
An SS includes
not only instructions in the CS, but also all subsequent instructions that
consume values derived from the \perfconst.
Concretely, an SS is a single-entry, multi-exit program slice constructed to satisfy
a \emph{confinement} property---any instruction that consumes
    objects that have dependencies with the \perfconst's specific version must execute
within the SS. So, once execution leaves the SS, no thread retains a
dependency on the old value, and switching to a new value is considered safe.
% Formally, an SS has the following invariant.

\minisubsec{Safety Invariant}
Let $T$ denote the set of concerned threads.
A transition is safe only when the system is in a state where no
thread is active in the SS as per program counters (PCs):
\[
  \forall t \in T, PC(t) \notin \bigcup SS
  \Longrightarrow
  \text{transition}(v \rightarrow v') \text{ is safe}.
\]
\sysname{} primarily focuses on data dependencies and the SSes are in the form of 
    thin slices~\cite{sridharan:07}.
Our empirical analysis shows that data dependencies are sufficient in capturing the 
    required safety of updating most \perfconsts. \rev{The SS analysis is designed to be pluggable for future extensions, e.g., to incorporate control-flow analysis if needed.}
% \tianyin{Need a strong statement on control flow; Tianyin will write it later.}

\if 0
\begin{figure} %[!htbp]
    \centering
    \begin{minipage}[b]{0.4\textwidth}
        \includegraphics[width=\textwidth]{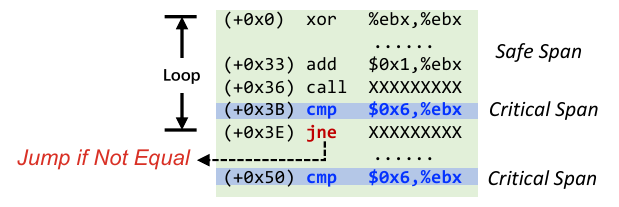}
        \subcaption{Instructions}
    \end{minipage}
    \hfill
    \begin{minipage}[b]{0.49\textwidth}
        \includegraphics[width=\textwidth]{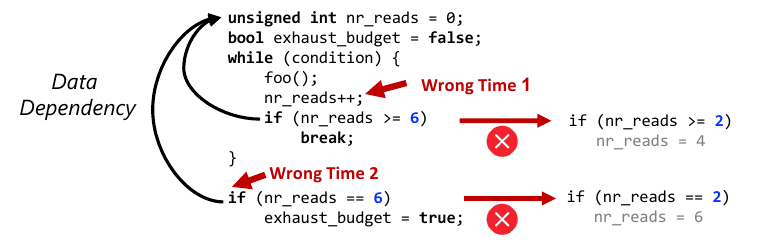}
        \subcaption{Source code}
    \end{minipage}
    \caption{Pitfalls during transition. This example demonstrates two wrong times when applying the new value (2).
    \tianyin{I don't understand this figure at all.}\zhongjie{Replace it with a simpler dataflow case. drop it}}
    \label{fig:need_ss}
\end{figure}
\fi

% Unfortunately, CS is insufficient to provide a safe transition.
% Figure~\ref{fig:need_ss} shows two examples. For instance, if the \const defines
% a loop's boundary condition, an ill-timed update could trap the program in a
% non-terminating loop. In another scenario, applying the new value between two
% CSs could violate program semantics by creating an inconsistent state. 

% The fundamental issue is that CS fails to cover the \textit{external effects} of
% \const. Intuitively, these external effects cease only when the lifetimes of all
% data derived from \const have expired. 

\minisec{Construction}
A safe span of a \perfconst can be automatically constructed 
    from its critical span (CS) using forward thin slicing~\cite{sridharan:07}.
% We introduce the Safe Span (SS), a
% single-entry, multiple-exit instruction sequence, which completely covers a CS's
% external effects by tracking data flow propagation. This abstraction is
% essential for defining safe transition states.
% The high-level idea is to start from the instructions in the CS and 
%    track the data flows of the \perfconst value along all 
%    the data-flow paths.
Any instruction that is transitively flow dependent
    on the \perfconst
    would be included in the SS.
Note that the dependency analysis is inter-procedural,
    because the \perfconst's value can be propagated across 
    function boundaries.
% (tianyin) nahhh no those are implementation tradeoffs; no need to mention them here.
% To balance analysis complexity and accuracy, we bound this recursive
% traversal to a depth of $L$. For example, $L=1$ confines the analysis to direct
% callers and callees. If a data lifetime extends beyond this $L$-depth boundary,
% we deem the CS's effect hard to analyze and report it to the user. 
% The construction logic is as follows: If the dataflow propagates to any qualified
% callers, the SS is computed at the topmost caller(s). Conversely, if all data
% lifetimes influenced by the CS terminate locally (i.e., no qualified callers
% exist), the SS is computed at the current function. If the analysis finds a
% callee that uses the data, the SS is extended to cover that call site. 
If two SSes (constructed from different CSes) overlap, 
    they are merged. 
An SS is specified as a pair of [start, end1/end2/...] binary address intervals 
in the top-level frame once the analysis terminates.

%% file: 4_5_program_model.tex
\vspace{3.5pt}
\subsection{Programmable Policy Plane}
\label{sec:design:api}
\vspace{1.5pt}

% \href{https://netdevconf.info/2.2/papers/brakmo-tcpbpf-talk.pdf}{TCP-BPF: Programmatically tuning TCP behavior through BPF}

\sysname{} provides simple
    APIs to support expressive tuning policies for each \perfconst.
To ensure safety, \sysname{} disallows users from directly
    updating kernel states.
Any user policy code (Xk-tune) must follow the 
    stubs auto-generated by an \sysname\ tool (\xkgen)
    and passed to an in-kernel \xkruntime.
Xk-tune is written in eBPF style
    and has eBPF observability.
\xkruntime\ then compiles Xk-tune into an eBPF \kernelProbe{}
%    in the form of eBPF Kprobe
    which includes SIE-based kernel-state updates.

% while encapsulating the complexity
% of runtime kernel modification. These APIs must remain safe and prevent
% arbitrary or unintended kernel corruption.

% On the user side, we expose simplicity through a command-line tool
% (\xkgen) and lightweight user-space update APIs. 
% Users specify the target \perfconst, and \xkgen\ generates an \xkseleton. The
% user then implements only the policy logic and calls the update APIs. We
% leverage the eBPF ecosystem for observability and tooling, so writing an XK
% tuning program feels similar to writing an eBPF program.

% On the kernel-interaction side, updating kernel state at runtime is inherently
% risky and is normally prohibited from both user space and eBPF programs. 
% We must provide this capability, but strictly control how it can be invoked. In
% \sysname, such privileged updates are only permitted 
% within the generated code for verified SIE indirections.

% To bridge these needs, \sysname introduces a middle layer provided by \xkruntime\ -- 
% the generated backend code. 
% The runtime takes an \xkprog\ and generates the corresponding
% backend eBPF programs, including the SIE-based state updates. These sensitive
% updates are visible only to our runtime and never exposed directly to users;
% users see only safe function interfaces, not their privileged implementations.

\minisec{Usage model and APIs}
To tune a \perfconst, the user first uses a command-line tool \xkgen\ 
    to generate an Xk-tune stub which includes 
    the unique ID of the \perfconst and a header file that declares \sysname{} APIs.
The user is expected to implement the policy in the stub
    using the APIs.
An Xk-tune follows the eBPF event-driven model;
    it is invoked when the \perfconst is used at runtime.
%    \sysname exposes a small set of APIs through a header file included in the 
% \xkprog\ generated by the \xkgen\ utility
% (Figure~\ref{lst:interface}). 
% \xkprogs\ follow an event-driven
% programming model similar to eBPF. Each \smalltt{XK\_PROBE} is invoked whenever
% the associated \perfconst is exercised at runtime. A single structure,
% \smalltt{xk\_ctx}, encapsulates the context required for the runtime and kernel
% interaction.
% Each Xk-tune applies to one \perfconst, regardless the times 
%     it appears in the binary.
One \perfconst may have multiple Xk-tunes according to the number of CS.

% Each \smalltt{XK\_PROBE} corresponds to exactly one \perfconst, regardless of how
% many times it appears in the binary. These details are hidden from users to keep
% the programming model simple. 
% \xkgen\ generates the probe parameters, including the unique identifier specified for the \perfconst. 
% Users must not modify these
% parameters; otherwise, the runtime reports an error (e.g., ``Not Found'').

% // xkernel_helper.h
\begin{figure}[t!]
  \footnotesize
  \vspace{0.75pt}
  \begin{minted}[escapeinside=||,linenos=false]{cpp}
/* handler of xkernel context (xk_ctx) */
typedef const struct xk_ctx * xk_handle_t;
/* User probe definition */
#define XK_TUNE(unique_name, perfconst_id, args...)
/* The set and transition API */
long xk_set(xk_handle_t xk_ctx, u64 val);
bool xk_transition_done(xk_handle_t xk_ctx);
  \end{minted}
  \caption{\sysname API.}
  \vspace{4.25pt}
  \label{lst:api}
\end{figure}

Figure~\ref{lst:api} shows \sysname{} APIs.
Two core APIs are \smalltt{xk\_set} to update a \perfconst\ value 
    and \smalltt{xk\_transition\_done} to check if the transition
    is completed. 
An Xk-tune can read kernel state and invoke \smalltt{xk\_set} to update 
    the \perfconst. 
% The program structure is automatically
% generated as the stub.
% It must run before any update logic; its invocation is generated automatically 
% by \xkgen.
Figure~\ref{fig:xk-program-example-user} shows an example of 
    Xk-tune which is used 
    for the case study on a \perfconst of TCP CUBIC presented in \S\ref{sec:cases}. 
% More policy examples can be found in Appendix~\ref{appendix:policy}.

\begin{figure}[t!]
\footnotesize
\centering
\begin{minted}[escapeinside=||,linenos=false,highlightlines={4-10},highlightcolor=orange!15]{cpp}
XK_TUNE(tcp_hystart, "net/ipv4/tcp_cubic.c:L349:3:0") {
    // 1. Safety guard (mandatory)
    if (!xk_transition_done(xk_ctx)) return 0;
    // 2. User policy logic
    //    - Obtain the current network flow's RTT
    struct sock *sk = (struct sock *)PT_REGS_PARM1(ctx);
    struct bictcp *ca = inet_csk_ca(sk);
    u32 cur_rtt = BPF_CORE_READ(ca, curr_rtt);
    //    - Set the value to 1 only when RTT > 80ms
    if (cur_rtt >= 80000) xk_set(xk_ctx, 1);
    return 0;
} /* my_policy.bpf.c */
\end{minted}
\caption{
    An example of Xk-tune. User-written policy code is highlighted; 
    the rest is from the auto-generated stub.}
    \vspace{5pt}
    \label{fig:xk-program-example-user}
\end{figure}

\minisec{Supporting kernel-state updates}
eBPF programs are restricted to read-only access to kernel memory to prevent 
    arbitrary writes. 
To support controlled updates, \sysname exposes one single kernel-state 
    update API as a BPF kfunc~\cite{BPFKfuncs}. 
% The safety checks of this Kfunc are bypassed
%    by the eBPF verifier and delegated to trusted callers (\xkruntime). 
Kfuncs enable extended functionality 
    through kernel modules. % (as alternatives to BPF helpers).  % without modifying kernel source.
The \sysname{} kfunc, \smalltt{sie\_write\_kernel}, can be invoked only by Xk-tune (see \S\ref{sec:impl}).
%    which is enforced by a whitelist in the kfunc subsystem.
The kfunc calls the SIE indirections generated for the target 
    \perfconst
% \smalltt{long sie\_write\_kernel(void *dst, u32 dst\_size, const void *src);} 
    to modify registers in \smalltt{pt\_regs} or kernel memory.
% why we can modify the previous frame's register in a new frame?
The \smalltt{pt\_regs} holds the CPU register state at the moment the probe is
    triggered and provides a direct reference to the kernel context. Linux already
    exposes this state to tracing infrastructures (e.g., \kprobe{} and ftrace), and it
    is part of a stable tracing ABI~\cite{BpfTraceRawCtxComment}. Our kfunc extends
    this from read access to write access.

\minisec{Transpiling Xk-tune into eBPF}
Each Xk-tune is transpiled into an eBPF \kernelProbe{} (\smalltt{BPF\_KPROBE}), 
    corresponding to a CS for the \perfconst.
% Given an \xkprog, the \xkruntime\ generates a backend program
Figure~\ref{fig:xk-program-example-internal} shows the source-code
    form of the eBPF code 
    of the Xk-tune in Figure~\ref{fig:xk-program-example-user}. 
% This backend is hidden from users for safety. 
\xkruntime{} uses the \xktable to locate the SIE
    indirections for the \perfconst and wraps the update logic into generated functions, e.g., 
    \smalltt{impl\_sie\_logic\_cs1} in Figure~\ref{fig:xk-program-example-internal}. 
A pointer to this function is stored in the \smalltt{xk\_ctx}, effectively implementing 
\smalltt{xk\_set} for that \perfconst.
In each generated Kprobe handler, the user-written policy function (e.g.,
\smalltt{\_\_user\_policy\_tcp\_hystart}) is invoked. 
Hence, the policy code has the kernel context (via \smalltt{ctx})
    and the implementation of SIE indirections is provided as embedded function pointers.
The safety of Xk-tune is checked by the eBPF verifier.
% The resulting backend is an eBPF program that the runtime loads into the eBPF
% subsystem. The verifier checks the safety of the user policy, and the eBPF JIT
% generates binary code for the SIE indirections, which we emit in source form.

\minisec{Transaction semantics}
\sysname supports tuning multiple perf-consts by bundling Xk-tunes 
    in a single file. 
% Alternatively, users can perform tuning in multiple batches by organizing Xk-tunes into separate files. 
All Xk-tunes in the file form an atomic transaction---their tuning logic is loaded or unloaded together. 
Each perf-const may belong to at most one active transaction. 
The \xktable tracks each \perfconst{} status.
If any perf-const is in an active transaction, it cannot be tuned in a new transaction.
\begin{figure}[t!]
\footnotesize
\centering
\vspace{0.75pt}
\begin{minted}[escapeinside=||,linenos=false]{cpp}
// 1. Helper: SIE Indirection
static __always_inline void impl_sie_logic_cs1(
    struct pt_regs *regs, u64 val) {
    // Recovered symbolic state expression:
    //   ebx = (regs->bx) >> 3   (original value: 3)
    u64 new_bx = ((regs->bx) << 3) >> val; 
    // Writing back to pt_regs using the kfunc
    sie_write_kernel(&regs->bx, sizeof(regs->bx), &new_bx);
}
// 2. Kprobe attachment: CS address (function+offset)
SEC("kprobe/tcp_cubic_hystart_check+0x4F")
int BPF_KPROBE(impl_cs_1) {
    // Wrap raw context into safe xk_ctx
    struct xk_ctx xk_ctx = {
        .regs = ctx,
        .set_fn = &impl_sie_logic_cs1,
    };
    // Call user policy (inlined)
    return __user_policy_tcp_hystart(&xk_ctx, ctx); 
} /* my_policy.internal.bpf.c */
\end{minted}
\caption{
    eBPF \kernelProbe{} (in a source form) 
        transpiled from the Xk-tune shown in Figure~\ref{fig:xk-program-example-user}.} 
    \label{fig:xk-program-example-internal}
    \vspace{-0.75pt}
\end{figure}

%% file: 4_9_impl.tex
\vspace{3.5pt}
\section{Implementation}
\label{sec:impl}
\vspace{1.5pt}

\input{tbl-casestudies}

\input{Data/eval_wentao/numbers_auto}

We implement \sysname with $\sim$1.7K lines of kernel C code for \xkruntime.
% including kernel modules and BPF runtime utilities. 
\sysname{}'s toolchain is
implemented in about 11K lines of Python, with $\sim$5K lines for CS and SS
analysis via symbolic execution.

% Functions annotated with \texttt{notrace} can't be attached with a \kprobe and thus can't be tuned.

\minisec{CS and SS analysis}
We implement a symbolic execution engine specific to CS analysis.
%  \tianyin{what tools do you use? Klee? Give some details here.}. 
Upon capturing the CS, it synthesizes all potential SIE indirections (\{\indPos, \indCode\}). 
We implement the SS analysis on kernel bitcode using an LLVM pass. 
By correlating CS and SS with the \texttt{.section} metadata in the binary diff, 
we resolve the precise target function and offset for \kprobe attachment. 
% If there are multiple potential SIE indirections, 
%    we select one that enables Kprobe optimizations. 
% Finally, we store CS, SS, and the best SIE indirection in the \xktable.

% attach to function + offset: how
\minisec{Minimizing \kprobe overhead}
Linux \kprobe may have high overhead if breakpoint or single-stepping traps
    are used ~\cite{jia:atc:24}. 
Linux saves the overhead through boosting (skipping single-step traps when post-handlers are absent) 
    and jump-optimization (replacing breakpoints with direct jumps). 
\sysname exploits these optimizations to minimize runtime overhead.
To enable boosting,
% \minisubsec{Avoid Post-handlers}
\sysname replaces post-handlers with pre-handlers on the next instruction (\smalltt{pc}+1). 
This assumes the next instruction is not a
jump destination; % (which is always the case in our evaluation);
this is almost always the case because \sysname{}
    targets the precise point where $V$ enters runtime states. % ---a transient execution step.
\sysname maximizes jump-optimizations by strategically attaching \kernelProbes{}. 
We implement a new optimization that handles the case when immediate values 
    of \perfconsts are directly used in conditional jumps,
    which Linux Kprobe does not handle. 
We describe the optimization in Appendix~\ref{appendix:branch-optimization}.

\minisec{\xkruntime} The runtime is implemented as a kernel module (\smalltt{xk-sie.ko}). In addition to transpiling Xk-tunes to eBPF \kernelProbes{} and loading them, it registers BPF Kfuncs and manages transitions. \smalltt{sie\_write\_kernel} is registered for SIE internal use, and \smalltt{xk\_transition\_done} is exposed to users to check transition status. It tracks completion via global refcounts maintained in \xkruntime{} (multi-threading) or \smalltt{BPF\_MAP\_TYPE\_TASK\_STORAGE} (per-thread). The kernel module attaches monitoring \kernelProbes at SSes and detaches them with the help of a background monitor kthread. For multi-threading safety, it employs \smalltt{stop\_machine} to initialize refcounts via stack traversal before activating \kernelProbes{}.

\minisec{\sysname tools}
We implement a series of userspace tools.
% \zhongjie{align with Overview 4.2's xk-gen}
% \texttt{xk-tool} orchestrates the user-facing workflow of \sysname. 
% The \texttt{build} command generates the \xktable and Xk-tune stubs
% from the source patch.
\smalltt{xk-build} generates the \xktable from the source patch.
\smalltt{xk-gen} specifies the ConstID in the \xktable and outputs the Xk-stub.
\smalltt{xk-load} invokes \xkruntime{} to transpile the xk-tune code with standard BPF tool-chains, install resulting BPF \kernelProbes{}
and ensure safe transition. % (both internal \smalltt{XK\_PROBE} and user-defined). 
Finally, \smalltt{xk-unload} performs a symmetric, ordered teardown.

%% file: tbl-casestudies.tex
% local macros to ensure no typo of macros!
% io_uring
%% threshold
\newcommand{\IoLocalTwDefaultMax}{IO\_LOCAL\_TW\_DEFAULT\_MAX}
\newcommand{\IoLocalTwDefaultMaxTt}{\smalltt{\IoLocalTwDefaultMax}}
% softirq
%% timing
\newcommand{\MaxSoftirqTime}{MAX\_SOFTIRQ\_TIME}
\newcommand{\MaxSoftirqTimeTt}{\smalltt{\MaxSoftirqTime}}
%% threshold
\newcommand{\MaxSoftirqRestart}{MAX\_SOFTIRQ\_RESTART}
\newcommand{\MaxSoftirqRestartTt}{\smalltt{\MaxSoftirqRestart}}
% block plug
%% threshold
\newcommand{\BlkPlugFlushSize}{BLK\_PLUG\_FLUSH\_SIZE}
\newcommand{\BlkPlugFlushSizeTt}{\smalltt{\BlkPlugFlushSize}}
% shrinker (mem)
%% batch
\newcommand{\ShrinkBatch}{SHRINK\_BATCH}
\newcommand{\ShrinkBatchTt}{\smalltt{\ShrinkBatch}}
\newcommand{\triggerpsec}[1]{{\color{blue}#1}}
\newcommand{\triggeroverhead}[1]{{\color{red}#1}}
% numa
%% batch
\newcommand{\NrMaxBatchedMigration}{NR\_MAX\_BATCHED\_MIGRATION}
\newcommand{\NrMaxBatchedMigrationTt}{\smalltt{\NrMaxBatchedMigration}}
% ca->delay_min >> 3
%% timing
\newcommand{\HystartDelayMax}{HYSTART\_DELAY\_MAX}
\newcommand{\HystartDelayMaxTt}{\smalltt{\HystartDelayMax}}
\newcommand{\HystartDelayTt}{\smalltt{HYSTART\_DELAY\_}}
\newcommand{\HystartDelayMin}{HYSTART\_DELAY\_MIN}
\newcommand{\HystartDelayMinTt}{\smalltt{\HystartDelayMin}}
%% scaling factor
\newcommand{\HystartDelayThreshold}{ca->delay\_min{\scriptsize\textbf{\textgreater\,\textgreater}}3}
\newcommand{\HystartDelayThresholdTt}{\smalltt{\HystartDelayThreshold}}
% https://github.com/zhongjiechen/Xkernel/tree/yltang/Experiment/nr_max_batched_migration

\begin{table*} %[h]
% \small
\footnotesize
\centering
\setlength{\tabcolsep}{4pt}
\caption{Summary of case studies that demonstrate \sysname's features and capabilities.}
\begin{tabular}{cllcl}
\toprule
\textbf{Case \#} & \textbf{Perf-const} & \textbf{Subsystem} & \textbf{Default} & {\bf Performance Tuning Regime} \\
\midrule
Case-1 & \BlkMaxReqCntTt\         & Storage (Block layer)       & 32   & Adapting to hardware devices and workload patterns (see \S\ref{sec:def}) \\
Case-2 & \MaxSoftirqRestartTt\    & CPU (Interrupt)             & 10   & Choosing tradeoffs between tail latency and CPU utilization \\
Case-3 & \ShrinkBatchTt\          & Memory (Reclamation)        & 128  & Controlling kernel internal behavior (e.g., maintenance) \\
Case-4 & \NrMaxBatchedMigrationTt & Memory (Page migration)     &  512 & Tuning and reasoning with kernel and hardware observability \\
Case-5 & \HystartDelayTt[\smalltt{MAX}, \smalltt{MIN}, factor] & Network (TCP CUBIC)  & [16, 4, 3] $\ $ & Collective tuning of interdependent \perfconsts \\
\bottomrule
\end{tabular}
\label{tab:case-study}
\end{table*}

\begin{figure*}[t!]
    \centering
    % First figure (25%)
    \begin{minipage}[b]{0.26\textwidth}
        \centering
        \includegraphics[width=0.99\linewidth]{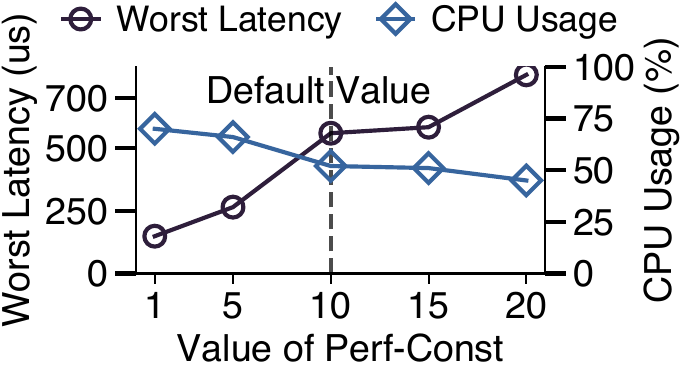}
        \caption{Cost-benefit tradeoff that can be tuned 
            by changing values of \MaxSoftirqRestartTt.}
        \label{fig:cases:softirq}
    \end{minipage}
    \hfill
    % Second figure (29%)
    % \begin{minipage}[b]{0.28\textwidth}
    \begin{minipage}[b]{0.27\textwidth}
        \centering
        \includegraphics[width=0.99\linewidth]{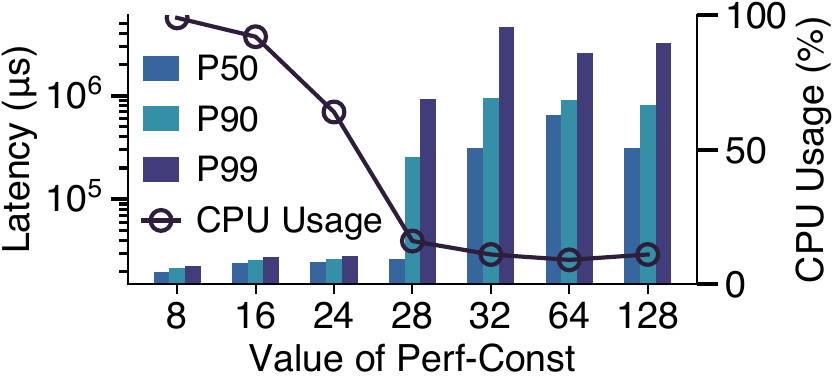}
        \caption{Tuning \ShrinkBatchTt{} can significantly 
            reduce write latency of the workload.}
        \label{fig:cases:shrinker}
    \end{minipage}
    \hfill
    % Third figure (45%)
    % \begin{minipage}[b]{0.45\textwidth}
    \begin{minipage}[b]{0.44\textwidth}
        \centering
        \includegraphics[width=\linewidth]{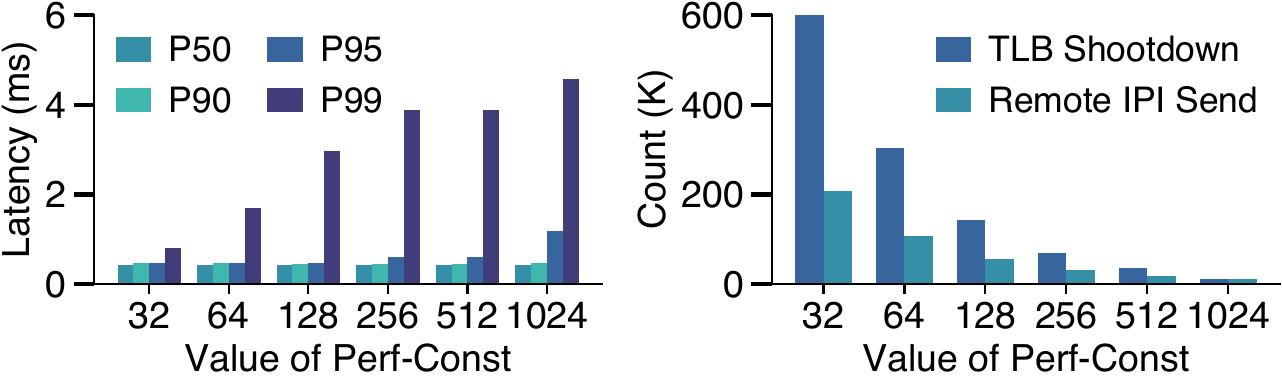}
        \caption{Tuning \NrMaxBatchedMigrationTt{} to reduce tail latency, 
            which requires understanding TLB shootdown behavior.}
% \todo{@tianyin -- CPU TLB shootdown: invalidate TLB entries on other CPU cores; Remote IPI Send: request other CPU cores to invalidate TLB entries.}}
        \label{fig:cases:numa}
    \end{minipage}
    % \caption{\textbf{Performance Improvement of Cases in Table~\ref{tab:case-study}}. (c) CPU TLB shootdown: invalidate TLB entries on other CPU cores; Remote IPI Send: request other CPU cores to invalidate TLB entries.}
    \label{fig:cases}
    \vspace{7.5pt}
\end{figure*}

%% file: Data/eval_wentao/numbers_auto.tex
% Autogenerated. Do not modify by hand.
% https://github.com/zhongjiechen/Xkernel/tree/dataflow/paper-assets

\newcommand{\wentaoNumbersSsAnalysisTimeMean}{11\xspace}
\newcommand{\wentaoNumbersSsAnalysisTimeStddev}{20\xspace}
\newcommand{\wentaoNumbersSsAnalysisTimeMax}{124\xspace}
\newcommand{\wentaoNumbersSsAnalysisTimeMaxMacro}{\texttt{BLK\_MQ\_DISPATCH\_BUSY\_EWMA\_WEIGHT}\xspace}
\newcommand{\wentaoNumbersSsAnalysisTimeMaxInstructions}{5,923\xspace}
\newcommand{\wentaoNumbersSsSizeMedian}{10\xspace}
\newcommand{\wentaoNumbersNumDisjointSsAsm}{371\xspace}
\newcommand{\wentaoNumbersNumDisjointSsIr}{360\xspace}
\newcommand{\wentaoNumbersSsReduction}{66\xspace}
\newcommand{\wentaoNumbersNumPerfConstsWithSsReduction}{26\xspace}
\newcommand{\wentaoNumbersNumLinesLlvmPass}{1,810\xspace}

%% file: 5_0_cases.tex
\vspace{3.5pt}
\section{Case Studies}
\label{sec:cases}
\vspace{1.5pt}

\rev{We present case studies that tune \perfconsts across kernel subsystems,
provide a glimpse at the broader policy space enabled by \sysname,
and discuss the resulting implications for tunability and complexity.}

\subsection{Performance Benefits and Tuning Rationales}
\label{sec:cases:perf}

\rev{Table~\ref{tab:case-study} summarizes the case studies; we then discuss their
    performance benefits and tuning rationales.}
    %to achieve significant performance improvements.
% Table~\ref{tab:case-study} summarizes these case studies.
    % including the one in \S\ref{sec:def} where we 
    % tune \BlkMaxReqCntTt\ based on hardware devices and workload access patterns.

\minisec{Case-1: Adapting to hardware and workload patterns}
\rev{\sysname\ enables users to tailor \perfconst values according to
    specific hardware properties and workload patterns.
    We revisit the \perfconst\ \BlkMaxReqCntTt\ presented in \sref{sec:def}.
% Given a machine with both the SSD and HDD mentioned earlier, 
Users can use
    a single Xk-tune program to simultaneously set the value to 128 for the
    sequential FIO workload on the HDD and to 1 for the RocksDB random workload
    on the NVMe SSD, obtaining the performance benefits shown in
    Figure~\ref{fig:cases:blkmq}.
Meanwhile, other workloads can continue using the default value.}

\minisec{Case-2: Choosing performance tradeoffs}
\rev{\sysname\ enables users to choose points along performance tradeoff curves
to better satisfy SLOs; such tradeoffs are widespread in OS kernels, with examples
such as latency vs. throughput and resource utilization vs. latency.}

We show how \sysname{} enables balancing the tradeoff between CPU efficiency
    and tail latency in the CPU (interrupt) subsystem.
% We show how \sysname{} enables users to balance 
    % cost-benefit tradeoffs. 
The \perfconst, \MaxSoftirqRestartTt\, limits how many times
    the software interrupt (softirq) handler can be rerun in one iteration before yielding,
    preventing indefinite softirq processing.
% It prevents indefinite softirq processing (when new softirqs keep
%     arriving) to starve other tasks.
Increasing its value improves
CPU efficiency but may increase tail latency of running tasks.
Therefore, it has major impacts on latency-critical workloads, especially
    when they are colocated with throughput-oriented workloads~\cite{Prekas-17-ZygOS}.

% The macro resides in softirq path;
% softirq is Linux's software interrupt subsystem, responsible for handling 
% deferred work such as network packet processing, block I/O completions, 
% and timer callbacks.
% Softirq runs in interrupt context and can only be
% preempted by a hardware interrupt. Because softirq handlers often perform heavy
% work (e.g., processing large batches of packets), the kernel must ensure that
% softirq execution does not monopolize the CPU and starve user-space tasks.

% \MaxSoftirqRestartTt\ (\pfconstsrc{kernel/softirq.c}{L483}{10}, threshold)  
% bounds the number of softirq restart loops within a single invocation of
% \_\_do\_softirq() on the same CPU. 
% Each iteration processes all pending softirqs,
% and if additional softirq work is raised during processing, the loop may restart, 
% but only to \MaxSoftirqRestartTt{} times. 
% This bound prevents softirq processing from running indefinitely when new work keeps
% arriving, ensuring that execution eventually yields so other tasks can run.
% The constant was introduced in 2013 as a macro and has remained unchanged
% since then~\cite{SoftirqRestart-2013-Val10}.

% This \perfconst has major impacts on latency-critical workloads when they are
%    colocated with throughput-oriented workloads, a common scenario in modern
%    datacenters~\cite{Prekas-17-ZygOS}. 

% for latency-sensitive tasks because softirq
% handling consumes longer uninterrupted CPU time before yielding.

We run a latency-critical
workload $W_l$ using cyclictest~\cite{cyclictest} and a throughput-oriented workload $W_t$
    on a 4-node cluster with 25 Gbps
    networking, varying the \perfconst values.

Figure~\ref{fig:cases:softirq} shows a clear tradeoff between tail latency of $W_l$
    and CPU utilization.
% T-App's throughput remains stable at 23 Gbps,
The original value (10) provides a reasonable balance with 
    52\% CPU utilization and a worst-case latency of 560 us;
    achieving the optimal tail latency of 149 us requires paying a penalty of 22\% CPU utilization.

\begin{figure}[t!] 
  \centering
  \includegraphics[width=0.85\columnwidth]{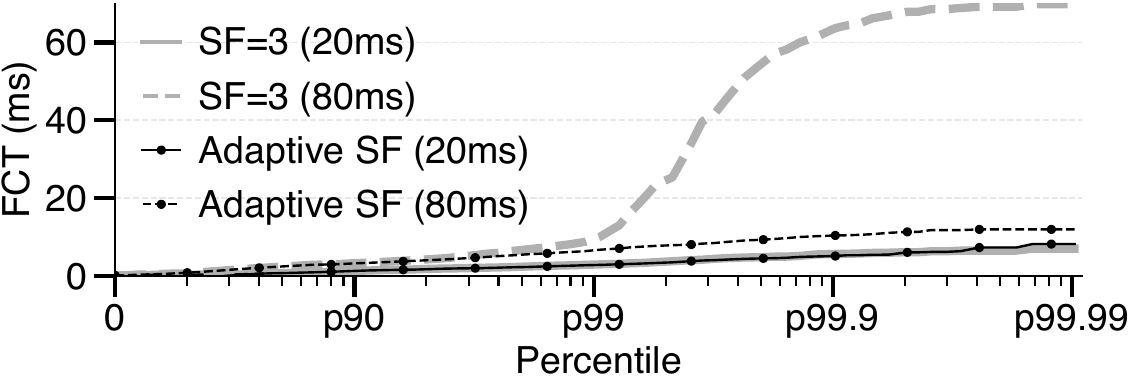}
    \caption{Flow completion time (FCT) of different scaling factors (\perfconsts) 
        under different RTTs running NGINX.}
    \vspace{5pt}
  \label{fig:cases:nginx}
\end{figure}

\minisec{Case-3: Controlling kernel internal behavior}
Oftentimes, application performance is affected by the kernel's 
    internal behavior such as memory reclamation, which is hard to control externally.
With \sysname, kernel {\em internal} behavior can be aligned with the application's 
    access patterns.
%  enabling a workload-aware balance between
%    performance and efficiency.

We demonstrate this capability using \ShrinkBatchTt,
    \rev{a \perfconst for Linux shrinkers, i.e., kernel callbacks that
    reclaim slab objects under memory pressure.
    It controls how many entries a shrinker checks when it scans
    a slab's LRU list for memory reclamation.}
    % which controls how many entries a shrinker checks when it scans
    % a slab's LRU list for memory reclamation.
% The macro is part of the shrinker mechanism to manage object-specific slabs in
% the slab subsystem for memory management. Specific object types for slabs
% include inode, dentry, and zswap\_entry, and all these subsystems rely on
% shrinkers to reclaim memory, and thus mostly utilize the single const value~\footnote{43 shrinkers in total, and very few slabs have their own local value, such as bcachefs.}.
% \ShrinkBatchTt\ (\pfconstsrc{mm/shrinker.c}{L369}{128}, batch size) 
\ShrinkBatchTt\ was introduced before 2005~\cite{LinuxGitInit05} and its value
(128) has remained unchanged since then~\cite{ShrinkBatchBlame}. 
Linux implements 43 shrinkers for different slab objects (zswap entries,
    inode, dentry, etc).
We focus on the zswap-shrinker for the zswap
subsystem~\cite{ZswapLinuxDoc} that is widely used in
datacenters~\cite{Lagar-19-GgFarMem,Tejun-22-IOCost}.

% The constant controls the batch size during reclamation -- an internal OS maintenance task,
% which induces interesting interaction with the user-facing request processing logic
% directly serving workloads, and thus needs to consider workload patterns.

We run a workload that writes data blocks sequentially into a large anonymous
mmaped region and periodically overwrites written blocks,
    which resembles 
% (i.e., reuse distance). Many 
data analytics with a memory budget (constantly triggering zswap). 
% inside a container.
The periodic reuse exposes a classic working-set effect.
When the batch size is small, fewer pages are reclaimed and swapped, increasing
the chance to reuse data in memory. In contrast, an aggressive batch
size may trigger thrashing.
% : pages are reclaimed and swapped excessively, leading to frequent disk access.

As shown in Figure~\ref{fig:cases:shrinker}, the original value of 128, despite the
relatively low CPU utilization (11\%), incurs a large
latency penalty. In fact, any values above 24 cause thrashing with unnecessary disk I/O, slowing
the user workload. In contrast, smaller values ($\leqslant$24) provide much lower latency.

% Moreover, shrinkers across subsystems share the
    % same \perfconst as batch sizes.
% as a side-effect of using the generic
% shrinking infrastructure, which applies a uniform default value. 
% \sysname allows customizing batch sizes for different shrinkers at runtime (see Appendix~\ref{appendix:policy}).
% (Figure~\ref{fig:shrink-code}).

% \begin{figure}[t!] 
%   \centering
%   \includegraphics[width=0.85\columnwidth]{figs/migr_combined.pdf}
%     \caption{\textbf{\NrMaxBatchedMigration}. 
%   Blue bar: CPU TLB shootdown (The current CPU changed a page table entry that
%   other CPUs may have cached.), organze bar: remote ipi send
%   (sending an IPI to another CPU requesting it to invalidate TLB entries.)
%   % yulong added some explanations for latency here
%   % Latency: Perform N consecutive memory accesses—each consisting of one read and one write to the selected hot page—and compute the memory-access latency quantiles over the entire migration process.
%   }
%   \label{fig:cases:numa}
% \end{figure}

% \begin{figure}[t!] 
%   \centering
%   \includegraphics[width=1\columnwidth]{figs/numamigration-crop.pdf}
%   \caption{\textbf{\NrMaxBatchedMigration}. 
%   Blue bar: CPU TLB shootdown (The current CPU changed a page table entry that
%   other CPUs may have cached.), organze bar: remote ipi send
%   (sending an IPI to another CPU requesting it to invalidate TLB entries.)
%   }
%   \label{fig:cases:numa}
% \end{figure}

%\minisec{[NUMA]~{\NrMaxBatchedMigration}}
\minisec{Case-4: Tuning and reasoning with observability}
Written in eBPF, policy code in \sysname{} 
    has observability on kernel and hardware metrics and behavior.
    % which can help performance reasoning.
\rev{These internal metrics are valuable for reasoning about performance,
    confirming tuning effectiveness, and being incorporated directly into
    tuning policies to decide values.}

We use \NrMaxBatchedMigrationTt, a \perfconst\ that specifies the batch size of
    pages migrated across NUMA nodes~\cite{Rayhan-25-DBPageMigration}, to
    demonstrate this feature.
The default value (512) was introduced in 2023~\cite{NrMaxMigration-2023} 
    and has never been updated.
This \perfconst\ controls the amortized costs of TLB shootdowns 
    and page migration. 
Larger batches amortize per-page migration cost but defer TLB shootdowns,
    which could increase stall time for concurrent threads accessing pages under migration. 
Online experiments with \sysname{} help us understand the relationship %tradeoff 
    between local memory access 
    and TLB shootdowns.

We run a workload with one group of threads repeatedly touching
hot data on a NUMA node, while another group continuously
    migrates hot pages to that node to shift  % using the \smalltt{move\_pages()} system
    the hot region over time.  
This workload mimics memory-intensive applications with moving working sets.
% The value should be chosen by weighing benefit of local memory access 
%    against migration stalls and TLB shootdown overhead.

% The macro appears in the generic page migration path in the memory
% management subsystem, which migrates pages between NUMA nodes to improve
% local memory access, a key factor for memory-intensive
% workloads~\cite{Rayhan-25-DBPageMigration}.  

% \NrMaxBatchedMigrationTt\
% (\pfconstsrc{mm/migrate.c}{L1574}{512}, batch size) bounds the total
% number of base pages (via folios) that can be accumulated into a single
% migration batch; it was

% This \perfconst\ controls how the amortized costs of TLB shootdowns and page migration:
% % bookkeeping are amortized: 
%     larger batches amortize per-page cost, but defer TLB invalidations,
%     increasing the latency experienced by active threads. 
    % \tianyin{why?} \zhongjie{@yulong}
    
% \yulong{Jing's description seems somewhat ambiguous; here is my explanation: 

% This parameter balances the granularity of migration against the amortization of TLB shootdown overheads. Smaller batches shorten the duration during which pages remain inaccessible (marked with migration entries), thereby reducing the stall time for concurrent threads that access migrating pages. However, this finer granularity prevents the amortization of TLB shootdowns, leading to a drastic increase in the frequency of system-wide IPIs (Inter-Processor Interrupts).}
% which can negate the gains from migration.

% periodic background migration driven by helper threads is also
% common~\cite{NrMaxMigration-2023}.

We vary \NrMaxBatchedMigrationTt\ values 
    and measure the tradeoff between latency of memory access to hot pages during migration
    and TLB shootdowns, as shown in Figure~\ref{fig:cases:numa}.
A small value can effectively reduce tail latency 
    by prioritizing memory access over migration, with a penalty of increased TLB shootdowns. 
Tuning this \perfconst values requires observability of TLB behavior.
%    which is accessible by eBPF utilities~\cite{bpftrace}. 

%
% In our experiment the default value (512) is
% suboptimal---a value of 1024 strictly outperforms 512 across all metrics.
% \tianyin{this statement is not true as per Figure~\ref{fig:cases:numa}.} \zhongjie{@yulong}

% \yulong{The description here does indeed differ from the image because I used a new machine for testing. The image here comes from new data, demonstrating that the gains previously achieved with `bacth=1024` were due to system jitter.

% I think what we should write about is the impact of increasing the number of TLB shootdowns on the system, and the trade-off between the reduced tail latency we see in this benchmark under these circumstances.}

% highlighting the benefit of online experimentation.

% \takeaway{With \sysname, users can combine kernel and hardware internal metrics
% with their performance goals to reason about and tune \perfconsts and systems
% behavior, enabled by our programmable policy plane and the observability
% provided by the eBPF ecosystems.}

\if 0
\begin{figure}[t!] 
  \centering
  \includegraphics[width=0.75\columnwidth]{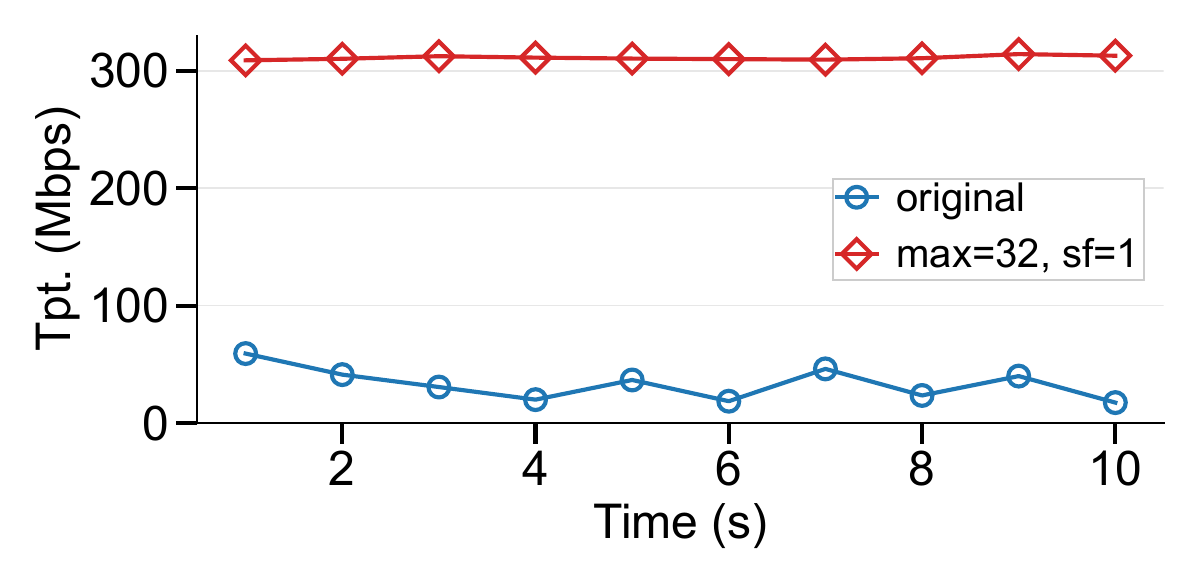}
  \caption{\textbf{\HystartDelayMin}. 
  \jing{Drop, use text}
  }
  \label{fig:cases:cubic}
\end{figure}
\fi

% \minisec{[Network]~{HYSTART\_DELAY\_[]} {\&} {\HystartDelayThreshold}}
\minisec{Case-5: Collective tuning of interdependent \perfconsts}
With \sysname, users can jointly tune multiple \perfconsts that collectively 
    decide fine-grained behavior with transaction semantics (\S\ref{sec:design:api}).

We demonstrate this capability by tuning the behavior of 
    TCP CUBIC's HyStart delay-based congestion window.
When a connection is in slow start, TCP CUBIC checks %(every 8 ACKs) 
    whether the current RTT exceeds a delay threshold computed from three \perfconsts
    (two defined as macros and one used as an operand in a shift operation).
% the minimum RTT and two bounds. 
    If the measured delay exceeds this threshold, CUBIC
    assumes congestion and exits slow start.
% The two macros that clamp the upper and lower bounds are:
% \HystartDelayMaxTt\
% (\pfconstsrc{net/ipv4/tcp\_cubic.c}{L46}{16ms}, tim\-ing) and
% \HystartDelayMinTt\
% (\pfconstsrc{net/ipv4/tcp\_cubic.c}{L45}{4ms}, tim\-ing).  In ad\-di\-tion, a
% scal\-ing fac\-tor (\pfconstsrc{net/ipv4/tcp\_cubic.c}{L439}{3}, scaling factor)
% right-shifts \smalltt{ca->delay\_min} before comparison; if the scaled value
% falls outside the bounds, the fixed timings (16 ms and 4 ms) are used instead.
The optimal threshold is highly dependent on RTT. With high RTTs, the current
\perfconsts are overly sensitive and trigger premature slow-start exit; with
low RTTs, they are too conservative and delay exiting slow start. These
\perfconsts were introduced in 2008~\cite{TcpCubic-2008} and changed three
times~\cite{TcpCubic-2011,TcpCubic-2014,TcpCubic-2019}. 
Prior work reported the importance of tuning them~\cite{Ha-08-CubicHybrid},
    but they remain constants. % and are untunable in practice.

% We run a simple netperf bulk-transfer workload with an RTT of 80 ms. 
% Adjusting \HystartDelayMaxTt\ to 32 ms and setting
% the scaling factor to 1 (relaxing the threshold) yields 5$\times$ to 17$\times$
% throughput improvement (Figure~\ref{fig:cases:cubic}).
% \tianyin{why there are two numbers?}

% With \sysname, users can jointly tune multiple \perfconsts to achieve the desired
% performance. 
% Moreover, \sysname's programmable policy can incorporate flow-level
% information as a runtime tuning criterion to optimize different workloads accordingly.
% across workloads.

% \todo{TODO: add back the per-domain customization experiments here. @zhongjie}

We first ran a microbenchmark to understand scaling factors (SFs) under different RTTs. 
To provide sufficient room for the scaling factor to take effect under high RTT, 
we change \HystartDelayMaxTt\ from 16$ms$ to 32$ms$. 
We find that SF=1 significantly reduces tail latency for slow flows, 
    but increases tail latency for fast flows compared to SF=3 (default). 
Based on this observation, we implement a selective tuning policy 
    that dynamically adjusts the SF of TCP Cubic for flows with 
    long RTT (Figure~\ref{fig:xk-program-example-user}). 

We show % end-to-end 
    performance gains of tuning these \perfconsts\ collectively on NGINX, 
    deployed as a web server for photo and video content~\cite{muralidhar-2014-f4}. 
We run a mixed workload of concurrent 20$ms$ and 80$ms$ flows, representing fast and slow connections. 
% Then, we consider two static configurations of the scaling factor.
% Figure~\ref{fig:cases:nginx}(b) reveals a performance crossover of 
%     Vanilla Cubic (SF=3): lower flow completion time (FCT) for short RTTs, 
%     whereas a reduced value (SF=1) proves superior only for long RTTs. 
As shown in Figure~\ref{fig:cases:nginx}, \sysname achieves a significant 
    FCT reduction (81\% at P99.99) 
    for long-RTT flows while maintaining performance parity for short-RTT flows.

\vspace{3pt}
\subsection{A Glimpse into \sysname-Enabled Policies}
\label{sec:cases:more-policies}
\rev{\sysname\ enables powerful tuning policies and creates opportunities for policy
innovation. 
We briefly discuss three directions that hint at a broader unexplored policy space.}

\vspace{3pt}
\minisec{Flexible granularity}
\rev{One major benefit of \sysname\ is its flexible granularity in applying
    tuning policies.
Unlike hard-coded granularities, such as processes, cgroups, or devices,
    \sysname allows users to define tuning granularity using conditions 
    unforeseen before binary deployment.}

\rev{As demonstrated by the cases in Table~\ref{tab:case-study}, \sysname\
    allows values to be set at the granularity of
    a given application (potentially spanning multiple processes or threads),
    a specific device type,
    or a set of network flows.
Users can update these granularities by changing
    the condition under which \smalltt{xk\_set} is applied in a new Xk-tune.
The zswap shrinker (Case-3) is an interesting example: 43 shrinkers across
    subsystems share the same \perfconst\ as their batch size, yet \sysname
    allows customizing batch sizes for different shrinkers at runtime by checking
    the caller's name (see Appendix~\ref{appendix:policy}).}

\vspace{3pt}
\minisec{Application-informed policies}
\rev{\sysname\ enables a class of application-informed tuning policies that
better coordinate with application-level semantics and knowledge of workload
patterns. This feature relies on eBPF's mature user-kernel communication.
Figure~\ref{fig:policy:app-inform-short} illustrates one such collaboration,
where a workload-pattern hint is passed via an eBPF map.
For example, a RocksDB application can use such hints to distinguish
foreground threads from compaction threads.
}

\begin{figure}[t!]
\footnotesize
\centering
\vspace{0.75pt}
\begin{minted}[escapeinside=||,linenos=false]{c}

XK_TUNE(...) {
    // Get the pid upon invocating this kernel code path
    u64 pid = bpf_get_current_pid_tgid() & 0xFFFFFFFF;
    // Leverage hints from the application process
    int *hint = bpf_map_lookup_elem(&hint_map, &pid);
    if (hint) {  xk_set(xk_ctx, 1);  }
}
\end{minted}
\caption{
    \rev{An example of an application-informed policy.
    The application communicates the hint through an eBPF map (\smalltt{hint\_map}), and
    the Xk-tune uses this hint to apply the tuning value to selected threads.}}
    \label{fig:policy:app-inform-short}
    \vspace{2.25pt}
\end{figure}

\begin{figure}[t!]
\footnotesize
\centering
\begin{minted}[escapeinside=||,linenos=false]{c}
/* Track consecutive failed attempts to merge adjacent 
   blocks and store the count in fail_cnt for use as a
   tuning heuristic */
SEC("kretprobe/blk_attempt_plug_merge")
int BPF_KRETPROBE(blk_attempt_plug_merge_ret, long ret) {
    int *fail_cnt = bpf_task_storage_get(&fail_cnt_map, 
        bpf_get_current_task_btf(), ...);
    if (ret == 0 && fail_cnt) (*fail_cnt)++;
    else *fail_cnt = 0;
}
/* Retrieve the heuristic and tune accordingly */
XK_TUNE(blk_add_rq_to_plug, "block/blk.h:L312:32:0") {
    int *fail_cnt = bpf_task_storage_get(&fail_cnt_map, 
        bpf_get_current_task_btf(), ...);
    // Heuristic: treat the workload as random when many
    // (16, an ad-hoc threshold) consecutive merges fail
    if (fail_cnt && *fail_cnt >= 16) xk_set(xk_ctx, 1);
}
\end{minted}
\caption{
    \rev{An Xk-tune program using ad-hoc workload heuristics. 
    It instruments \smalltt{blk\_attempt\_plug\_merge} with a kretprobe
    to track historical merge failures. 
    % Upon detecting a high failure rate, 
    % a smaller value (1) is used.
    }
    }
    \label{fig:policy:heuristic-short}
    \vspace{3pt}
\end{figure}

\vspace{3pt}
\minisec{Ad-hoc tuning heuristics}
\rev{\sysname\ enables ad-hoc tuning heuristics that are easier to identify
once workloads and deployment scenarios are known, without requiring
such heuristics to be baked into the source code.
Figure~\ref{fig:policy:heuristic-short} illustrates one such heuristic
for Case-1, where kernel-internal observability metrics are used to infer
whether the workload is sequential or random.
Moreover, because an Xk-tune runs on each execution, \sysname\
supports adaptive policies, such as linear or exponential changes
to the tuning value.
}

\vspace{3pt}
\subsection{Tunability and Complexity}
\rev{Choosing values for \perfconsts faces inherent complexity:
the best value depends on the workload, hardware, and policy objective in a
given deployment.
Existing practice hides this complexity behind fixed compile-time choices, but
this can leave significant performance opportunities fundamentally unattainable.
In contrast,
\sysname\ makes \perfconsts\ tunable and programmable in a unified framework,
exposing the broad and powerful policy space discussed above.
}

\rev{Such exposure does not create new complexity; rather, it allows users, for
the first time, to tackle this complexity in a principled manner.
Exposing only a subset of \perfconsts\ is insufficient when multiple constants
jointly affect system behavior; one untunable \perfconst\ can limit the effectiveness
of others. We believe managing the full \perfconsts\ set with \sysname
opens a promising research direction.
}

\if 0
\rev{\sysname exposes a real trade-off between tunability and system complexity.
The complexity is inherent: the best value of a \perfconst depends on the
workload, hardware, and policy objective, and these factors can change over
time. Existing mechanisms hide this complexity behind fixed compile-time choices
or coarse run-time knobs, but doing so also hides optimization opportunities and
can leave substantial performance on the table. \sysname makes this complexity
explicit and programmable, allowing users to express workload- and
hardware-aware policies when the benefit justifies the added policy logic. This
also raises a broader scalability question of how to manage a large number of
tunable constants and their interactions. We view this as a promising direction
opened by \sysname, where higher-level tools can group related knobs, compose
policies, detect conflicting updates, and learn workload-specific
configurations.}
\fi

%% file: 6_0_eval.tex
\vspace{5pt}
\section{Evaluation}
\label{sec:eval}
\vspace{3pt}

\if 0
points to show
- case studies
  - perf gain
  - expressiveness
  - easy to use
- eval
  - it works great
    - safe, general (# of dataset)
    - milisecond-scale policy update
      - we need to show the whole timeline policy update (from load eBPF program)
      - show the formula: code-gen time + loading time + transition time
        - mainly compares with the compilation time vs. livepath
  - the essence of perfconsts (how CS/SSes look like pragmaticlly)
    - CS size, #of CS for each
    - SS size, #of SS for each (how they merge?)
  - online cost
    - fast (low overhead)
    - multi-knobs explination
  - offline cost
    - offline: compliation, symbolic execution
\fi

\if 0
% Testbed
% c6620
28-core Intel Xeon Gold 5512U at 2.1GHz
128GB ECC Memory (8x 16 GB 5600MT/s RDIMMs)
Two 800 GB Mixed-use Gen4 NVMe SSD
Dual-port Intel E810-XXV 25Gb NIC (one port available for experiment use)
Dual-port Intel E810-C 100Gb NIC (one port available for experiment use)
\fi

\begin{figure}[t!] 
  \centering
  \vspace{0.75pt}
  \includegraphics[width=\columnwidth]{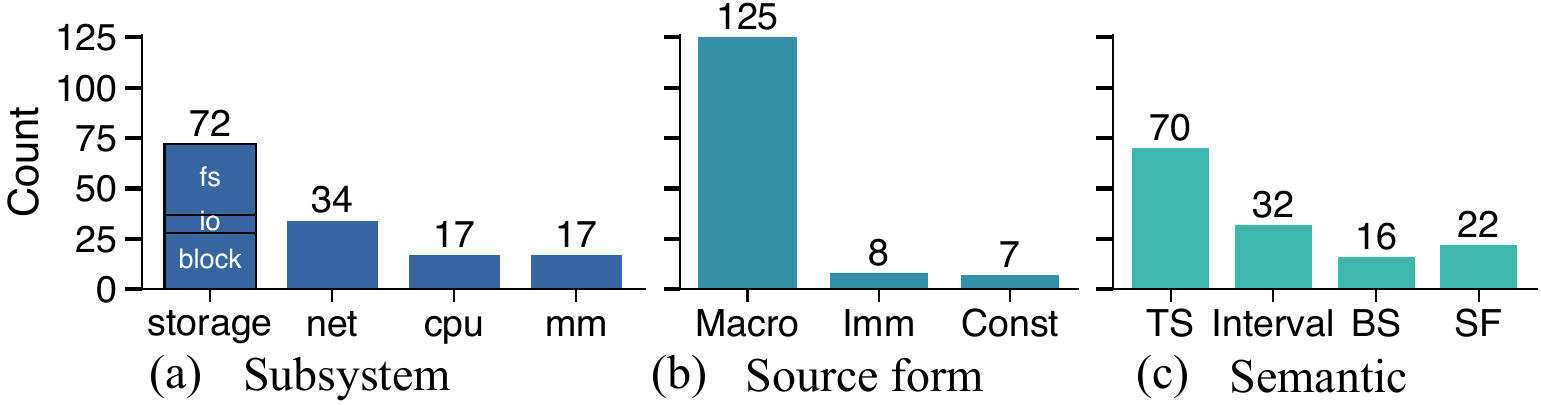}
  \caption{Characteristics of the evaluated \perfconsts.
  The categories are specified in Table~\ref{tab:perf-const-categories}.}
  \label{fig:dataset-src}
  \vspace{6.75pt}
\end{figure}

To evaluate the generality, safety, and efficiency of \sysname, 
  we collected 140 \perfconsts from four Linux subsystems.
These \perfconsts cover common semantics and source-code forms,
  as shown in Figure~\ref{fig:dataset-src}.
Note that the evaluated \perfconsts are a small subset of all \perfconsts
  in Linux, though the number is comparable to that of \smalltt{sysctl} 
  performance-related knobs.
% (details in ~\sref{sec:eval:dataset}).  
Evaluations run on a
  machine with 128\,GB RAM, a 28-core Intel Xeon Gold 5512U CPU (2.1\,GHz), 
  and two 800\,GB NVMe-Gen4 SSDs on CloudLab~\cite{Duplyakin-19-Cloudlab}.

\sysname supports all but one (99.3\%) \perfconsts.
The one \sysname fails to support is due to limitations of \kprobe (see Appendix~\ref{appendix:limitation} for more details). 
  % For example, \smalltt{SEND\_MAX\_EXTENT\_REFS} appears in a function name 
  % with multiple symbol table matches, preventing \kprobe from uniquely attaching to the
  % location (see Appendix~\ref{appendix:limitaion}). 
  % \tianyin{appendix}

We show that:
(1) SIE incurs negligible runtime overhead;
(2) \sysname supports millisecond-scale policy updates and safe transitions;
and (3) the cost of offline, one-time static analysis is about 18 minutes per \perfconst.

\subsection{Characteristics of Critical and Safe Spans}

\begin{figure}[t!] 
  \centering
  \vspace{-1.5pt}
  \includegraphics[width=0.9\columnwidth]{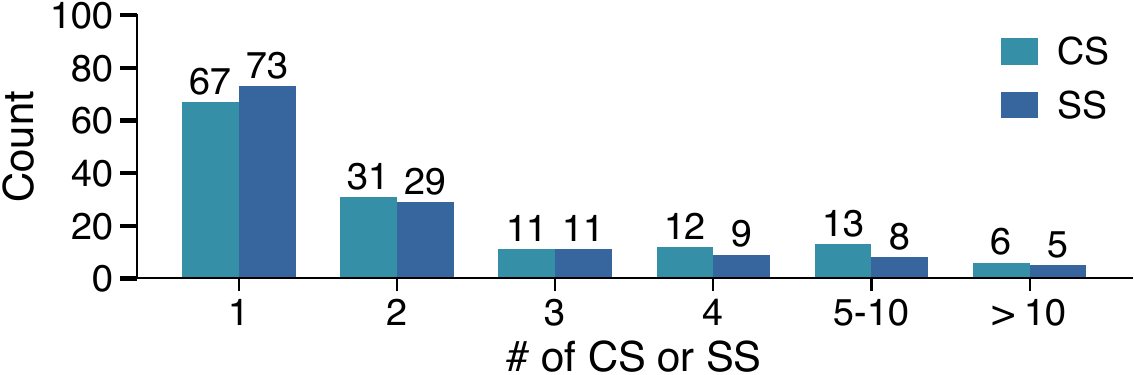}
    \vspace{4.5pt}
    \caption{Number of CS and SS per \perfconst{}.}
    \vspace{5pt}
  \label{fig:dataset-cs-dist}
\end{figure}

We first present the characteristics of critical spans (CSes)
  and safe spans (SSes) of the evaluated \perfconsts. 
The information helps interpret our experiment results.
% We first examine how many CSes each \perfconst produces
As shown in Figure~\ref{fig:dataset-cs-dist}, 
  the 140 \perfconsts have 367 CSes in total.
Most constants are highly localized: 48\% map to a single CS, and 86\% map to fewer 
  than five. 
A long tail also exists: in 4\% of cases, a constant maps to more than ten CSes;
\smalltt{DEF\_PRIORITY} and \smalltt{NFS4\_POLL\_RETRY\_MAX} have the highest
counts (16 and 17) due to aggressive inlining of their enclosing callers.

Out of the 367 CSes, 82 have symbolic values $IV$ 
  that differ from the \perfconst
  value $V$ in the source representation (\sref{sec:expression:recovery}).  
\sysname correctly recovers the symbolic
relations for all cases.  Only three CSes require dual-location indirections to
handle irreversible updates (\sref{sec:synthesis:algo}).

We obtained 300 SSes % 367-\wentaoNumbersSsReduction{}
in total from the 367 CSes.
As shown in Figure~\ref{fig:dataset-cs-dist},
\wentaoNumbersNumPerfConstsWithSsReduction{} of the \perfconsts exhibit CSes that have data dependencies, resulting
in a smaller number of SSes compared to their number of CSes.

% \subsubsection{CS and SS Analysis}

% (\pfconstsrc{include/linux/mmzone.h}{L1293}{12}) 
% (\pfconstsrc{fs/nfs/nfs4proc.c}{L79}{15}) 

\begin{figure}[t!] 
  \centering
  \includegraphics[width=0.8\columnwidth]{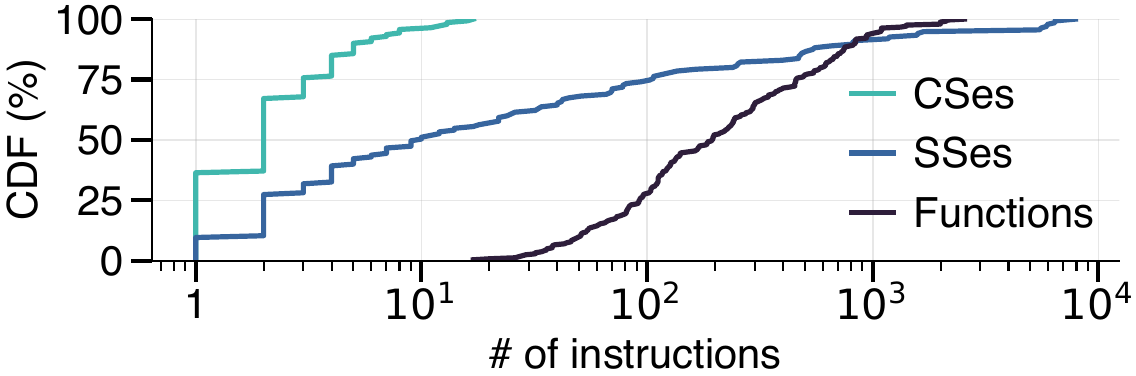}
    \caption{Size of CSes, SSes, and Functions.
    CS size is measured precisely. SS size is approximated
    by summing the instructions in the span, including the callees. 
    For functions, we count only the instructions in the
    function body (excluding callees), so the value is an underestimate.}
    \vspace{7.5pt}
  \label{fig:dataset-ss-cs-func-cdf}
\end{figure}

Figure~\ref{fig:dataset-ss-cs-func-cdf} compares the sizes 
  of CSes, SSes, and kernel functions.
% \minisec{Size comparison: CS, SS, and function}
% To illustrate the compact size of CSes, the mismatch of function-level
% replacement, and how SS expansion captures side effects, we plot the CDF of the
% number of instructions in 
CSes are small, mostly containing a single instruction; only 12 contain two, and
  one each at lengths 3, 4, 5, and 7. This confirms that the
point where a \perfconst first affects runtime state is simple and compact.
The median SS contains only \wentaoNumbersSsSizeMedian{} instructions, showing
that safe transitions are usually confined to a small scope. SS size, however,
has a long tail significantly larger than CS size, reflecting the inherent
complexity of kernel data dependencies. \rev{The largest SS in our dataset is about 8K instructions.} SSes make these dependencies
explicit.
Functions expand the scope unnecessarily
  and are much larger than SS except in
extreme cases, but with weaker safety guarantee than SSes.
% It also provides no natural safety guarantees and leads only to slower transitions.

\vspace{3.5pt}
\subsection{SIE Overhead}
\rev{We evaluate SIE overhead along three dimensions: the per-trigger CPU cost, the frequency at which SIE \kernelProbes{} fire, and the number of SIE \kernelProbes{} enabled at the same time.}

\minisec{\rev{Per-trigger cost (CPU cycles)}}
\rev{An SIE invocation includes the underlying \kernelProbe{} cost plus
SIE-specific work: checking transition state (O1), executing indirection
(O2), and, when needed, reading kernel state (O3) or accessing BPF maps (O4) for
user-space policy interaction. We attach SIE \kernelProbes{} at different
offsets inside a kernel function (\smalltt{vfs\_write}) to compare
jump-optimized and INT3-based probes. We use a single-byte write workload with
an intentionally short kernel path bottlenecked at \smalltt{vfs\_write}.
  % , and
  % measure cycles on a 28-core Intel Xeon Gold 5512U at 2.1GHz using \smalltt{perf}.
Table~\ref{tbl:runtime_cycles} shows that the \kernelProbe{} mechanism dominates:
an empty jump-optimized \kernelProbe{} adds 168 cycles (for 84.2\% of evaluated
CSes), whereas an empty INT3-based \kernelProbe{} adds 1765 cycles. 
O1--O4 add little additional cost.
% On top of this
% fixed \kernelProbe{} cost, 
% Furthermore, O1+O2 raise the total
% cost by only 4\% in the jump-optimized case and 9\% in the INT3 case, while
% O3 and O4 require few additional cycles.
}

\input{tbl-tex/eval-stress-overhead}

\label{sec:eval:runtime_overhead}
\vspace{1.5pt}

% We evaluate the runtime overhead introduced by \sysname. 
% We first present a
% breakdown using stress-test microbenchmarks, then discuss why, in practice, this
% overhead is negligible for real applications.

\begin{figure}[t!] 
  \centering
  \includegraphics[width=\columnwidth]{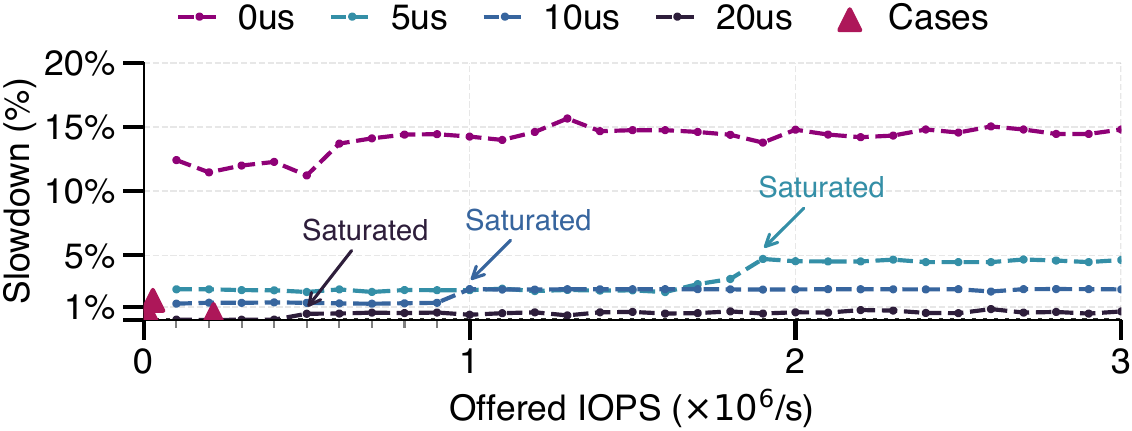}
    \caption{Slowdown of median latency. 
    The trigger rate of SIE \kernelProbes{} is equal to the offered IOPS before 
      saturation and remains unchanged afterwards.}
      \vspace{5pt}
    % \todo{add one line of INT3 push to 1 pct (good to have), + case studies (dots)}
  \label{fig:eval:slowdown}
\end{figure}

\minisec{Trigger frequency}
We evaluate the impact of trigger frequency using a controlled benchmark. We implement a multithreaded workload that uses \smalltt{io\_uring} 
  and performs asynchronous one-byte writes to \smalltt{/dev/null} with an SIE \kernelProbe{}
  attached to \smalltt{io\_issue\_sqe}. 
We vary both the offered IOPS (to control the trigger rate of SIE) 
  and the amount of per-operation computation. 
Figure~\ref{fig:eval:slowdown} shows the slowdown at different trigger rates;
  we also annotate the data points from case studies (\sref{sec:cases})
  by triangles.

When the workload performs essentially no work (0$\mu$s)
  in each operation, the slowdown caused by SIE is 15\%;
the slowdown drops to 5\% and 2\% when each operation includes 5$\mu$s
and 10$\mu$s of computation, respectively. Once per-operation processing reaches
20$\mu$s, the overhead falls below 1\%. The slowdown curve remains relatively
flat as trigger rate increases, indicating that the overhead scales predictably
even under high execution frequency of SIE.
% Even when using \kernelProbe-INT3, applications with modest per-operation cost
% (on the order of 50~$\mu$s) keep the slowdown under 1\%. 
These results show that \sysname introduces negligible overhead, even
when triggered millions of times per second. 
\rev{In practice, applications tend to
execute their own logic for more than 20$\mu$s, making the overhead even more
negligible.}

% Table~\ref{tbl:runtime_cycles} shows the CPU cycles of SIE-related overhead.
% The cycle breakdown explains this trend. 
% For most CSes of evaluated \perfconsts (88.3\%),
%   \sysname{} uses jump-optimized \kernelProbes{} to avoid INT3;
% the fully enabled jump-optimized path adds about 243 cycles, compared with
%   1,858 cycles for the INT3 path on our machine.
% The SIE update instructions themselves (see Figure~\ref{fig:sie-code-examples})
%   take about 20 cycles.

\begin{figure}[t!]
    \centering
    \begin{subfigure}[b]{0.24\textwidth}
        \centering
        \includegraphics[width=\linewidth]{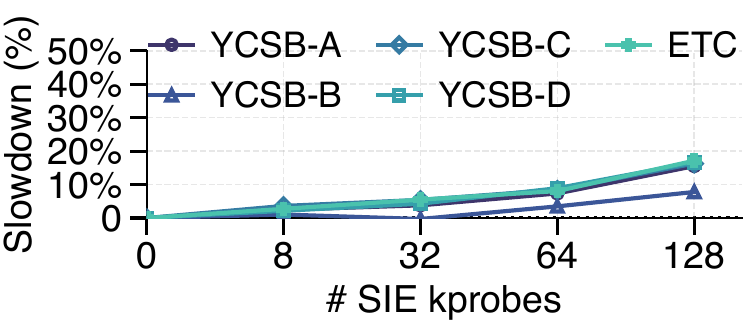}
        \caption{Throughput slowdown}
        \label{fig:app_slowdown_tpt}
    \end{subfigure}
    \begin{subfigure}[b]{0.23\textwidth}
        \centering
        \includegraphics[width=\linewidth]{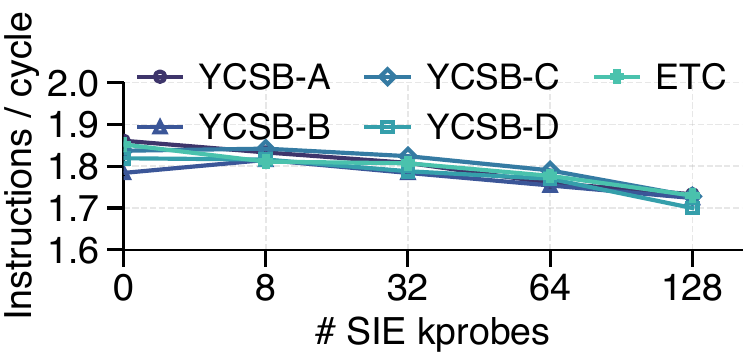}
        \caption{Instructions per cycle (IPC)}
        \label{fig:app_slowdown_ipc}
    \end{subfigure}
    \caption{
      \rev{Redis slowdown with multiple SIE \kernelProbes{}.}}
    \label{fig:app_slowdown}
    \vspace{7.5pt}
\end{figure}

\minisec{\rev{Multiple SIE \kernelProbes{}}}
\rev{We evaluate how the per-trigger cost accumulates
when many tunable constants are active under a real application. We
run Redis~\cite{redis} with YCSB~\cite{YCSB-code} and Facebook ETC
workloads~\cite{FB-ETC}, and 
vary the number of enabled SIE \kernelProbes{};
% enable 0 to 128 SIE \kernelProbes{} at the same.
% For stress testing, 
each SIE \kernelProbe{} is placed on the critical
path. Figure~\ref{fig:app_slowdown} reports throughput and instructions per
cycle (IPC). %measured with \smalltt{perf}. 
With 32 SIE \kernelProbes{},
throughput drops by at most 4\%.  Even with 128 SIE \kernelProbes{}, throughput
slowdown is 7--14\% across workloads, while IPC drops by about 6\%. These
results show that \sysname{} is practical to tune many \perfconsts{} with
modest overhead.}

\noindent\textbf{\rev{Finally,}} \rev{a tuning should be enabled when its
benefit outweighs the pure cost of SIE invocations; in our case studies, the
benefits usually far outweigh this overhead.}

\vspace{3.5pt}
\subsection{Policy-update and Transition Time}
\vspace{1.5pt}

We measure the time to load a policy program (Xk-tune)
and the time to transit from the original value 
  to a new one.
% from loading an XK-tune policy program to when it takes effect.
% The time consists of:
% (1) policy-update time for loading the XK-tune program
% and (2) transition time to the next execution of the code path
%  containing the \perfconst.
%  and completes its transition under SIE).
% The former is inherent to \sysname, whereas the latter depends on runtime behavior.
% execution paths, and transition unit. We evaluate each in turn.

\begin{figure}[t!] 
  \centering
  \includegraphics[width=\columnwidth]{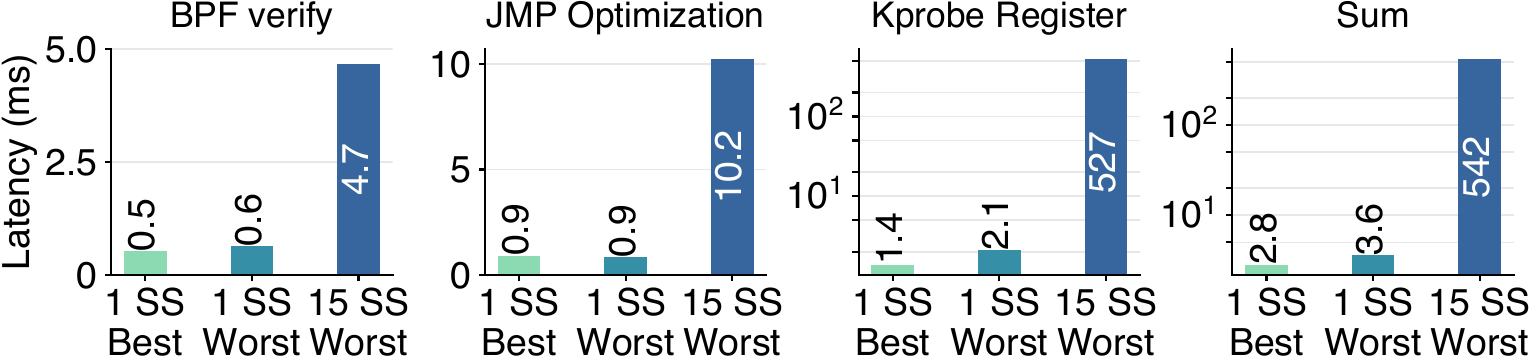}
  \vspace{-5pt}
    \caption{Policy-update time. Best case: only one \kernelProbe{} is required per CS. Worst case: two additional \kernelProbes{} are needed to ensure global consistency.
    % Each \kprobe for SIE consists of all overheads (O1-O4) in Table~\ref{tbl:runtime_cycles}. 
    % We assume the worst-case scenario where each CS maps to a distinct, 
    % non-overlapping SS. For a single CS, there are 2 \kernelProbe and 1 Kretprobe, 
    % for 15 CS, there are 30 \kernelProbe and 15 Kretprobes.
    % The best case means the SS is the same as CS, while the worst case means each CS derives a SS that can't be merged at all.
    }
    \vspace{5pt}
  \label{fig:eval:timeline_latency_bar}
\end{figure}

\minisec{Policy-update time} 
Figure~\ref{fig:eval:timeline_latency_bar} shows the breakdown 
  of loading an Xk-tune program, including
  BPF verification, jump-optimization, and \kprobe registration.
% When an XK-tune is loaded into the XK-runtime, the runtime first performs
% transpiling, which incurs negligible cost. It then loads the resulting eBPF
% program into the eBPF runtime, passing through several steps: BPF verification,
% jump-optimization preparation and application, and finally \kprobe registration,
% as shown in .
The main cost is \kprobe registration, 
  and thus the number of \kernelProbes{} has a major impact
  on policy load time. 
%  affects policy update time. 
Even in the case where the \perfconst has 15 SSes, 
\sysname bounds the policy-update time to 542 milliseconds, meeting our design goal.

\minisec{Transition time}
% Transition time is influenced by factors such as 
%  related code paths and workload intensity.
% code path, and workload intensity.
\sysname provides the near-optimal transition time
  for per-thread version atomicity and side-effect safety, 
  and is scalable with many concurrent threads.
% and finally our global concistency transition mechanisms realizes global consistency
% model without excessive number of \smalltt{stop\_machine}.

\begin{figure}[t!] 
  \centering
  \includegraphics[width=0.95\columnwidth]{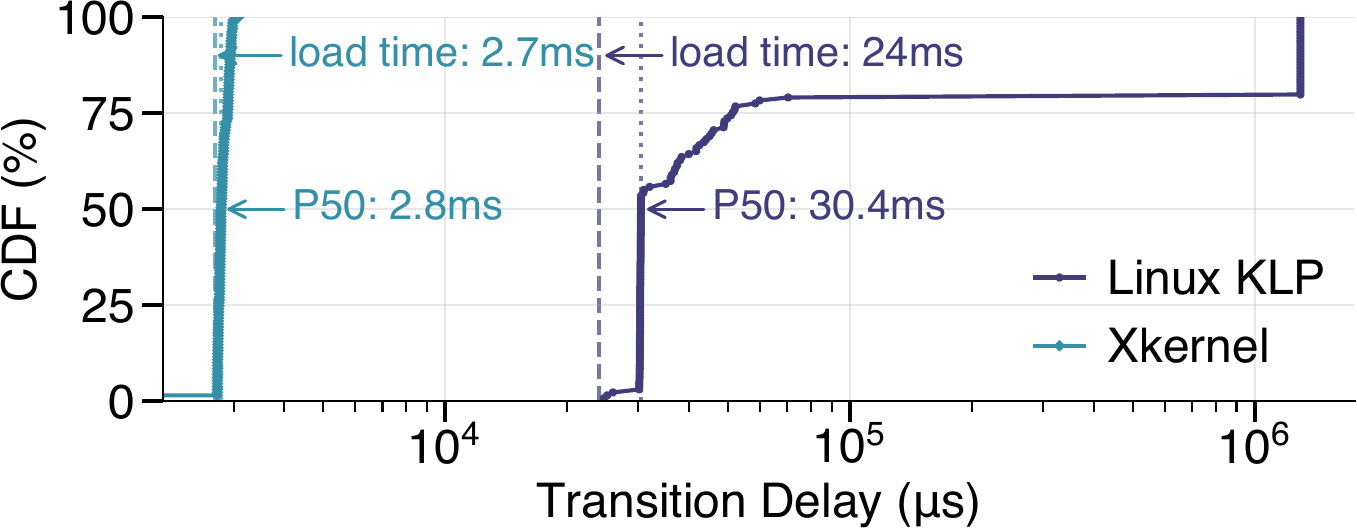}
    \caption{Transition time of \sysname{} using CS versus Linux KLP.
     The target function has 898 instructions.
    }
  \label{fig:per_task_transition_delay}
\end{figure}

\minisubsec{Per-thread version atomicity}
We compare \sysname{} with Linux KLP~\cite{LinuxKLP} on version atomicity;
  both support version atomicity.
% with function-level transition units of in Linux KLP. 
% Function-level units are poorly suited for \perfconst tuning and can lead to long transition tails under
% per-thread version atomicity.
We tune a \perfconst which controls TCP backlog size~\cite{perf-const-tcp-stack}
and measure transition time with an \smalltt{iperf3}~\cite{iperf3} workload 
  of 128 flows (threads).
Figure~\ref{fig:per_task_transition_delay} shows the CDF of the {\it end-to-end}
latency, including the policy-update time (2.7 $ms$ for \sysname
  and 24 $ms$ for KLP modulo 7-minute patch-generation time). The median latency for KLP and \sysname is 2.8ms and 30.4ms, respectively.
% loading excludes 7-minute patch generation required for policy updates.
% \tianyin{@zhongjie: just report the median transition time in the figure. Done},
% The fastest transition delays were 0.74ms (KLP) and 0.64ms (\sysname), while the slowest were 1271.2ms (KLP) and 0.39ms (\sysname).
%
  This shows that CS is a much more efficient transition unit than function.

\input{fig-tex/fig-eval-pertask-ss-transition}

\minisubsec{Per-thread side-effect safety}
% As transition time depends on the triggering rate and the number of SSes, 
Figure~\ref{fig:cases:per_task_ss_transition} shows transition time 
  of four representative cases with different triggering rates
  of Xk-tunes and the number of threads that load Xk-tunes. 
We do not compare with KLP which does not support side-effect safety.
% We run each experiment three times and report the mean and error bars in
In all cases, the transition time is less than 10 milliseconds. 
A per-thread transition completes as soon as the first \kernelProbe{} fires upon entering the
  first SS. %, so the time is largely determined by this interval. 
We see no clear correlation of the triggering rate and the number of the SSes 
  with the transition time---all configurations trigger at least once per millisecond 
  and complete the transition immediately once the entry \kernelProbe{} fires. 

% mechanism scales well and maintains stable delay even with high thread counts
% (16 threads).

\if 0
During the transition phase, each SS entry incurs overhead from the stack
inspector. The complexity of verifying the call stack is $O(M \log N)$, where
$M$ is the stack depth and $N$ is the number of independent SSs. Our
measurements show that for $M=10$ and $N=15$ (the worst case in our dataset),
the average inspection latency is 4 $\mu$s.
\jing{Not sure if we want to reserve this, stack check never fails...}
\fi

% \minisubsec{Per-thread transioning scalability}

% IR to assumble address, manual efforts, not very accurate is fine
% what we want to show here
% support different transition unit
% how fast it takes to converge
% \todo{show the results of huge win using CS for global consistency (optional), need to make the kpatch work}
\minisubsec{Global consistency}
Figure~\ref{fig:cases:global_ss_transition} shows transition time 
  with global consistency on side-effect safety with multiple threads.
The time increases compared to per-thread safety.
We do not compare with Kpatch~\cite{Kpatch} because % it adopted KLP's per-thread
  % transitioning model by default in 2015, and 
  its global-consistency mode was
  later deprecated~\cite{kpatch-bugs} and it does not support side-effect safety.
% cite: https://github.com/dynup/kpatch/pull/1355

\begin{figure}[t!] 
  \centering
  \includegraphics[width=0.75\columnwidth]{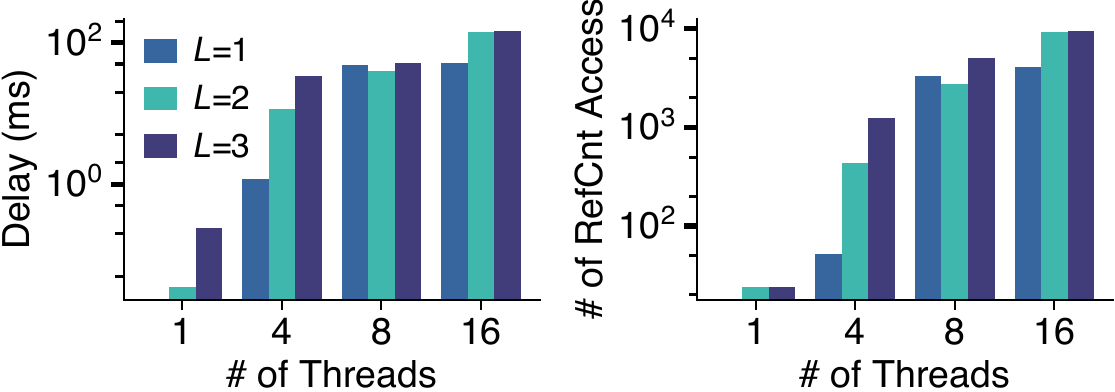}
    \caption{The impact of SS size with global consistency. 
    $L$ is the length of function calls to reach the CS of a \perfconst; 
    it is used to control the SS size.}
  % indicates the call relationship of SS's enclosing function relative to
  %  CS L=1: current; L=2: caller; L=3: caller's caller.
    \label{fig:eval:global_converge}
\end{figure}

To understand the time in more depth, we run a controlled experiment using the
microbenchmark in \sref{sec:eval:runtime_overhead}, 
  varying the SS size (controlled by $L$ levels of call stacks) and the number of threads.
Figure~\ref{fig:eval:global_converge} shows that transition time increases with
  higher concurrency (fewer safe points) and larger SS sizes. 
The number of reference-count updates shows the same pattern: a larger number of updates
indicates a lower likelihood of leaving the SS.
Overall, the transition time is low (144$ms$) even under high concurrency (16 threads). 
% The global consistency model provides a
%  foundation for policies that require coordination across domains.

% (tianyin) this comes out of nowhere
% to avoid pitfalls such as the deadlock introduced when write-back thresholds are tuned per device~\cite{MemSyctlDeadlock}. 
% a capability absent in current general live-update systems.

\if 0
  \minisec{Appication Interference}
  \jing{less important}
  \minisubsec{Per-Thread}
  \todo{Does KLP have app interference (per-task)? I feel so?}
  \minisubsec{Multi-Threading}
\fi

\vspace{3.5pt}
\subsection{Offline Static Analysis Time}
\label{sec:eval:dataset}
\vspace{1.5pt}

% Our dataset of 140 \perfconsts spans major Linux subsystems (CPU, storage, and memory management), 
% three syntactic representations, and the four common
% semantic types (Table~\ref{tab:perf-const-categories}).
% Figure~\ref{fig:dataset-src} summarizes the distribution.

% Our selection is intentionally conservative. We exclude
% many immediate values because their effects are strictly local -- and therefore
% trivial for SIE. The final size (140) is comparable to the number of existing
% \smalltt{sysctl} performance knobs (149), and \sysname successfully supports all of them.

% We constructed the dataset through manual inspection of kernel source code,
% reviewing comments and execution paths to determine whether a constant affects
% performance and to classify its role. This process required substantial effort
% and subsystem understanding, and therefore reflects our familiarity rather than
% the true prevalence or distribution of \perfconsts across the kernel.  The
% dataset is intended to demonstrate the generality of the approach, not to
% exhaustively capture its full performance potential.

\sysname incurs a one-time cost per \perfconst for static analysis;
  after that, the \perfconst can be tuned online anytime.
%  to generate an entry in the \xktable. 
% Each entry includes the symbolic state expression and the
% binary intervals corresponding to the CSes and SSes. 
This offline process consists of
two kernel compilations, symbolic execution to recover symbolic state expressions and
CSes, and analysis to construct SSes. On average, the process takes 18 minutes per \perfconst.
The analysis of different \perfconsts can be done in parallel using spot instances.
Compilation takes about seven minutes (two
compilation runs executed in parallel across 56 threads). 
The remaining time is dominated by SS construction, which
varies significantly based on the dependency structure. On average
\wentaoNumbersSsAnalysisTimeMean{} $\pm$ \wentaoNumbersSsAnalysisTimeStddev{}
minutes and up to \wentaoNumbersSsAnalysisTimeMax{} for the most complex case.
%  such as
% \smalltt{\wentaoNumbersSsAnalysisTimeMaxMacro{}} with
% \wentaoNumbersSsAnalysisTimeMaxInstructions{} instructions in its SS. 
% Symbolic execution is negligible (\todo{xx} ms), as CSes contain only a few instructions.

% The \xktable\ allows \sysname to support millisecond-scale updates at runtime---{\it process once offline, reuse anytime}.
% The cost composes linearly, yielding orders-of-magnitude savings compared to KLP: 
%  for $N$ \perfconst with $M$ values each, \sysname needs only $2N$ recompilations, vs.
% $M^N$ patches (recomplication) for live patching.
% \tianyin{???}

% The primary offline cost of \sysname is to construct the \xktable.
% For CS, the process requires compiling the target
% twice in parallel, locating seed instructions via binary diff, and running
% symbolic execution. 
% Recompilation is the bottleneck. 
  % Using a 56-thread Intel
  % Xeon Gold 5512U, we executed two parallel rebuilds (splitting CPU resources
  % equally) on Linux 6.14, which took 7 minutes.
% (assuming a pre-built kernel source). 
\rev{Overall, \sysname's offline cost is linear in the number of
  \perfconsts in the kernel source code.
% but is independent of the values explored at runtime. 
For $N$ constants, constructing the \xktable requires $2N$ recompilations.
In contrast, KLP or source-level recompilation requires one binary per target
value; with $M$ possible values per constant, all combinations can require up to
$M^N$ builds.
% In addition, the compilation, symbolic execution, and SS construction of
% each \perfconst can be parallelized to save time. 
The \xktable is reusable across machines running the same kernel
binary, allowing cost amortization across a deployment.
%one offline-analysis result to serve an entire deployment.
}

%% file: tbl-tex/eval-stress-overhead.tex
%AMD
% \begin{table}[t!]
% \centering
% \resizebox{0.95\linewidth}{!}{
% \begin{tabular}{lcc}
% \toprule 
% \textbf{Action}
%  & \textbf{CPU Cycles Per Op} & \textbf{Overhead}\\
% \midrule
% Baseline & 2296 & - \\
% Empty Kprobe (INT3) & 4440 & +93\% \\
% \sysname (INT3+T) & 4590 & +\textbf{100\%} \\
% \sysname (INT3+T+S) & 4667 & +\textbf{103\%} \\
% Empty Kprobe (JMP) & 2831 & +23\% \\
% \sysname (JMP+T) & 2976 & +\textbf{29\%} \\
% \sysname (JMP+T+S) & 3063 & +\textbf{33\%} \\
% \bottomrule
% \end{tabular}
% }
% \caption{\textbf{Runtime Overhead of Transition Model and SIE in \sysname. \textbf{T}: Transition Model; \textbf{S}: SIE.}}
% \label{tbl:runtime_cycles}
% \end{table}

%INTEL
\setlength{\tabcolsep}{1.1pt}
{
\begin{table}[t!]
\centering
% \scriptsize
\footnotesize
\begin{tabular}{p{38mm}p{13mm}p{17mm}!{\vrule width 0.5pt}>{\centering\arraybackslash}p{12mm}}
    \hline
    \textbf{~Operation}
    & \textbf{Cycles/Op~} & \textbf{Overhead(\%)} & \textbf{Cases \#}\\
    \hline
    Baseline & 946 &  &  \\
    \hline
    Empty BPF Kprobe (JMP) & 1114 & +18\% &  \\
    \sysname (JMP+O1) & 1146 & +\textbf{21}\% &  \\
    \sysname (JMP+O1+O2) & 1167 & +\textbf{22\%} & 309 \\
    \sysname (JMP+O1+O2+O3) & 1186 & +\textbf{25\%} & (84.2)\% \\
    \sysname (JMP+O1+O2+O3+O4) & 1187 & +\textbf{25\%} &  \\
    \hline
    Empty BPF Kprobe (INT3) & 2711 & +187\% &  \\
    \sysname (INT3+O1) & 2752 & +\textbf{191\%} &  \\
    \sysname (INT3+O1+O2) & 2801 & +\textbf{196\%} & 58 \\
    \sysname (INT3+O1+O2+O3) & 2795 & +\textbf{195\%} & (15.8)\% \\
    \sysname (INT3+O1+O2+O3+O4) & 2799 & +\textbf{196\%} &  \\
    \hline
\end{tabular}
\caption{
    \rev{CPU cycles breakdown of one SIE invocation. 
    O1: Transition state check; O2: Indirection's execution; O3: Kernel
    Memory Read; O4: BPF MAP Access 
    % (INTEL(R) XEON(R) GOLD 5512U).
    }
    }
\label{tbl:runtime_cycles}
\vspace{5pt}
\end{table}
}
\setlength{\tabcolsep}{\tempa}

%% file: fig-tex/fig-eval-pertask-ss-transition.tex
\begin{figure}[t!]
    \centering
    \begin{subfigure}[b]{0.24\textwidth}
        \centering
        \includegraphics[width=\linewidth]{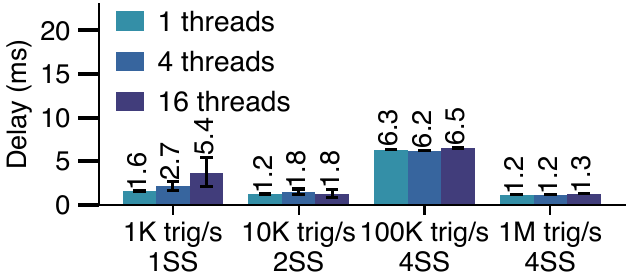}
        \caption{Per-thread}
        \label{fig:cases:per_task_ss_transition}
    \end{subfigure}
    \begin{subfigure}[b]{0.23\textwidth}
        \centering
        \includegraphics[width=\linewidth]{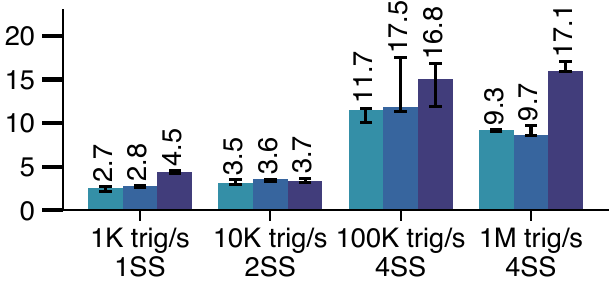}
        \caption{Global consistency}
        \label{fig:cases:global_ss_transition}
    \end{subfigure}
    \caption{Transition time of side-effect safety. 
    % \tianyin{put the case in the axis}
    % Case1: 1K triggers/s, 1 SS; Case2: 10K triggers/s, 2 SS; Case3: 100K triggers/s, 4 SS; Case4: 1M triggers/s, 4 SS.
    }
    \label{fig:transition}
\end{figure}

%% file: discussion.tex
\vspace{5pt}
\section{Discussion}
\label{sec:discussion}
\vspace{2pt}

\if 0
\jing{@tianyin, do we need to mention they relies on \smalltt{pt\_regs},
we can do that because kprobe handler can get a direct reference of the pt-regs,
kernel can potentially only make a copy in theory. (not practical in practice as kprobes etc dependes on it)}

\tianyin{Clarify what we do (and do well) and what we don't do.
We don't focus on policy or auto-tuning; a ton of work there already
(refer to the related work)}

\tianyin{Give a summary on the safety model. 
One thing we do not do is the safety of the value --- what if 
    the person changed a bad value? There is a threat model 
    we probably should clarify.}
\tianyin{Another thing we do not do is control-flow for SS;
    the abstraction is general, but static control flow 
    is expensive and unnecessary. We basically give up 
    when we find there are complicated control flow dependences}
\fi

% \minisec{Safety model}

% A CS is inherently concurrency-safe. 
% When a constant is being materialized into runtime state, 
%     its instructions do not interleave with lock-related operations 
%     (required for locking correctness).
% % so the original concurrency semantics are preserved.
% % : either the lock is held or it is not. 
% CSes also has no instructions that may sleep or handle
% interrupts. An XK-tune preserves this concurrency
% semantics and introduces no races, unlike existing mechanisms (e.g., \smalltt{sysctl}) 
%     that can create reader-writer races.

\minisec{Applicable constants of \sysname}
\rev{\sysname prefers safety over completeness, while still supporting a broad
range of \perfconsts.}
\sysname does not target constants which can directly change kernel memory layout (e.g., array
sizes or struct padding), as such changes can introduce
dangerous pointer arithmetic and are challenging to make safe.
\rev{\sysname rejects the constants it cannot handle upon detection.}
% To detect these cases, \sysname compares the DWARF debug information of the resulting binaries.
% Any change to data structure size or member offsets signals a layout dependency,
% and \sysname considers such constants as untunable. 
\rev{The layout changes can
be reliably detected by reading debug information such as DWARF (e.g., \smalltt{pahole}). 
\sysname can also detect and reject dead-code elimination cases where the offline
analysis finds the constant absent from one binary.
% \smalltt{pahole} to extract a canonical layout signature from the original and
% modified kernel builds; \sysname rejects modifying consts if any type's size,
% alignment, or field offsets change.
}

\if 0
To maximize flexibility, \sysname does not impose built-in bounds on new values
(except when architectural operand-width constraints are violated). We view such
bounds as part of the performance semantics of each \perfconst, which should be
defined on a per-constant basis rather than embedded in the mechanism. Users may
freely add these checks in their XK-tune programs; \sysname treats them simply
as policy, keeping mechanism and policy cleanly separated.
\fi

\vspace{0.5pt}
\minisec{\rev{Concurrency safety}}
\rev{A CS is a very short instruction sequence
(Figure~\ref{fig:dataset-ss-cs-func-cdf}) that materializes a constant into
runtime state; it typically contains no lock operations, blocking primitives, or
memory barriers. An XK-tune only changes this materialization, so it preserves
the original concurrency semantics and does not introduce the risk of races from
breaking locks.
The kprobe mechanism also provides a solid foundation, as probe insertion is atomic.
% unlike existing
% mechanisms (e.g., \smalltt{sysctl}) that can create reader-writer races.
}

\vspace{0.5pt}
\minisec{\rev{Choosing a good value}}
To maximize flexibility, \sysname{} does {\em not} impose any built-in bounds on
new values (except for architectural constraints such as register width limits).
Such bounds belong to the performance semantics of each \perfconst, not the SIE
mechanism. \rev{Users can encode range checks, device-specific constraints,
specifications defined by RFCs, and any other value checks in XK-tune. \sysname
treats them as policy, keeping mechanism and policy separate.}

% (tianyin) think again, I don't think it's worthwhile to dig a hole here
\if 0
The safety-span abstraction and transitioning mechanism in \sysname are general
and can support any desired side-effect safety property, provided that the
analysis produces the corresponding safety spans. While \sysname currently uses
data-dependency analysis to construct safety spans,
a control-flow analysis pass could be incorporated to cover
control-dependent behaviors as well.  However, we posit that full control-flow
analysis is unnecessary and too costly for the relatively simple performance
constants we target.
\fi 

The focus of this work is the mechanism and interface to realize 
    principled OS tunability (\sref{sec:tunability}), not the policies on deciding the optimal values of each \perfconst.
We expect the new opportunities and flexibility enabled by
    \sysname to inspire many new tuning techniques (e.g., with AI~\cite{Liargkovas-25-OsTuningAgent}).

\vspace{0.5pt}
\minisec{\rev{Maintenance of offline results}}
\rev{The \xktable is tied to the exact kernel source, compiler version, and
build configurations to match the running binary. Xkernel assumes that the
compiler version and build configurations of the running kernel are available.
This holds because (1) they are inherently known for customized kernels, and (2)
for out-of-the-box kernels, they can be retrieved from official sources. 
% We
% expect these to exactly match the kernel binary managed by the Xkernel runtime.
}
% \rev{Spurious instruction differences caused by non-semantic binary changes
% (\eg, instruction reordering or offset changes) can make binary differencing
% less robust. \sysname allows users to provide hints of source lines to filter
% these differences using debug information (\ie, \smalltt{debug\_line}),
% retaining only instructions mapped to source lines that reference the target
% \const. If the debug information is unavailable or unreliable, \sysname
% explicitly reports the case for manual inspection, which is rare in practice.}

% \rev{\sysname's current SS analysis is based on data dependencies, which is
% sufficient for most \perfconsts. If users require control-dependency reasoning
% or domain-specific lifetime rules, \sysname allows them to provide custom
% analyses through our pluggable interface. We leave more advanced SS analysis for
% future work.}

% \minisec{\rev{Trade-off between tunability and complexity}}

\vspace{0.5pt}
\minisec{\rev{Out-of-tree modules and drivers}}
\rev{\sysname is not limited to in-tree kernel code. SIE also supports
loadable kernel modules and vendor drivers with available source code. Each
\xktable entry is tied to a specific module or driver version and must be
regenerated after it is rebuilt or updated.}

%% file: 7_0_rel.tex
\vspace{1pt}
\section{Related Work}
\label{sec:rel}
\vspace{1.5pt}

% \href{https://arxiv.org/pdf/2203.12132}{Runtime Software Patching: Taxonomy, Survey and Future Directions}

% 1. Kernel live patch, Ksplice, Kpatch, kGraft, KernelCare, Linux KLP.

% 2. bpf kernel extension, sock\_ops, bpf tuning

% \minisec{Kernel Live Patching}
\minisec{OS and system performance tuning}
Tuning has long been essential for OS
performance~\cite{gregg2014systems,Infokernel-sosp03,Saltzer-09-PCSD}. Existing
practice, however, relies on the limited set of knobs exposed through
\smalltt{sysctl} and \smalltt{sysfs},
    leaving substantial performance potential unexplored. 
Kernel configuration systems (e.g., Kconfig~\cite{KconfigDoc}) customize the OS
    but are designed for feature
    selection~\cite{She-11-RE-Feature-Model,Tartler-11-Linux-10k-feature-problem,Tartler-14-90K-ifdefs,Kuo-20-Cozart},
    not performance tuning, since they operate purely at compile time.
Recent techniques using AI agents show promise in improving
    search strategies~\cite{Liargkovas-25-OsTuningAgent,Zero-Mod}, 
    but they are limited to a fundamentally constrained tuning space. 
We view \sysname as a new foundation to expand this space 
    and advance tuning techniques. % to realize workload-optimal performance in situ.
Our techniques and principled tunability can potentially benefit many other
    systems that actively explore tuning
    policies~\cite{LlamaTune,StorageXTune,RightChord,LLMKV-storage24}.

\if 0
Kernel configuration system like Kconfig~\cite{KconfigDoc} provides a way to customize the
    OS~\cite{Gazzillo-17-Kmax,Mahmud-23-ConfD,Kuo-20-Cozart}.
But, as Kconfig controls static compile-time configuration,
    it does not support in-situ performance tuning.
% is possible, but
% like source conversion, it leads to long recompilation times and high cost, not
% to mention the long-standing challenges of kernel configuration
% management~\cite{many}. 
After all, Kconfig is designed for feature
    selection~\cite{She-11-RE-Feature-Model,Tartler-11-Linux-10k-feature-problem,Tartler-14-90K-ifdefs},
    not performance tuning.

Our principle is not limited to OS kernels, but can be generalized 
    to any system software.
Existing performance tuning approaches of userspace systems such as
    databases~\cite{LlamaTune}, 
    storage systems~\cite{StorageXTune}, key-value
    stores~\cite{LLMKV-storage24}, and memory management~\cite{RightChord}
    share the same limitations---the tuning is limited to only exposed configuration 
    knobs.
\fi
% These efforts, however, rely on exposed configuration knobs and thus inherit the
% same limitations as source-conversion-based tuning in OS kernels, where adding
% new knobs typically requires recompilation and rebooting.
% We posit that they face the same constraints and can benefit from principled
% tunability.

\vspace{2pt}
\minisec{Programmable kernel extensions}
% they mostly need kernel modification or read-only, or target on a specific problem customization
% we are general but focus on perfconsts
Kernel extensibility has long been a central topic in OS
research~\cite{ExtensibleAstray97}. 
This area has seen renewed momentum with the rise of eBPF,
% a modern Linux mechanism that enables user-space code to run
% safely inside the kernel. Beyond its early use in packet
% filtering~\cite{vieira-2020-fast} and observability~\cite{bpftrace}, 
 which has been used to customize kernel behavior across many
subsystems~\cite{XRP,xdp,etran-nsdi25,PageFlex-atc24,CacheExt-sosp25,SchedExt-linux, syrup-sosp21}.
These efforts share our motivation of adapting kernel behavior to diverse
workloads and hardware. 
% However, they target specific policies and require
% kernel source modifications, which creates deployment challenges.
Our work focuses on \perfconsts and principled tunability.
While \sysname uses eBPF as its policy interface, its core contribution is SIE,
which enables principled in-situ tuning. 
% (tianyin) this is untrue -- we need a runtime inside the kernel
% \sysname can also be deployed on
% unmodified Linux kernels, requiring no changes to kernel source.
TCP-BPF~\cite{Brakmo-17-TcpBpf} provides a similar capability by allowing TCP
parameters to be tuned through an eBPF program. However, it is limited to the
TCP subsystem and requires substantial engineering effort to restructure kernel
code paths. In contrast, \sysname is general to any \perfconsts.

\if 0
eBPF allows for dynamically and safely injecting user-defined handlers into the
kernel. It has been widely used for packet filtering~\cite{vieira-2020-fast},
tracing~\cite{cilium}, or even application offloading~\cite{ghigoff-2021-bmc}.
Recently, TCP-BPF~\cite{Brakmo-17-TcpBpf} allows tuning TCP parameters with
limited programmability using eBPF. However, this approach is limited to TCP and
difficult to extend to all other kernel subsystems. \sysname is the first work
that brings safe, programmable tuning for perf-const to the entire OS kernel.
Sche-ext, cache-ext, eTran, Rex, 
\fi
% Page-Flex:  608 lines of changes to kernel
% cache-ext
% Implementing cache_ext required adding ∼2000 lines to the
% kernel. Only a fraction of these lines modified the core kernel:
% 210 lines in the page cache (mostly the eBPF hooks an
% eTran: change kernel source

\vspace{2pt}
\minisec{OS and program live update}
SIE shares attributes with live-update systems, but is designed
    specifically for \perfconsts{}---users (or agents) specify a value and \sysname
synthesizes the update instructions. OS live update has been widely
studied~\cite{baumann-atc05, haibo-vee06, cristiano-asplos13, goullon-1978} and has led to products such as
Ksplice~\cite{arnold-2009-ksplice}, Kpatch~\cite{Kpatch}, and
KLP~\cite{LinuxKLP}. As discussed, these systems treat functions as the unit of
version atomicity, making them a poor match for \perfconsts. User-space
live-update systems also operate at function-level 
unit~\cite{Rommel-20-Wfpatch,Kitsune-oospla12}.  A recent effort,
PlugSched~\cite{ma-2023-efficient} provides live updates for CPU scheduling but depends
on policy-specific understanding of kernel scheduler data structures to
reconstruct state.  In contrast, \sysname provides a general and principled
mechanism without relying on specific kernel data structures.

% Put them into all.bib, use the same format and name convention (sorted by year)
% That gives us a consistent view of bib items

\if 0
PageFlex: Flexible and Efficient User-space Delegation of Linux Paging Policies with eBPF. {PageFlex-atc24}
Transforming Policies into Mechanisms with Infokernel {Infokernel-sosp03}
Set It and Forget It: Zero-Mod ML Magic for Linux Tuning {Zero-Mod}
Can Modern LLMs Tune and Configure LSM-based Key-Value Stores? {LLMKV-storage24}
Striking the Right Chord: Parameter Tuning in Memory Tiering {RightChord}
Kitsune: efficient, general-purpose dynamic software updating for C {Kitsune-oopsla12}
\fi

%% file: 8_0_conc.tex
\vspace{4pt}
\section{Conclusion}
\label{sec:conc}
\vspace{1.5pt}

% Some sense of the summary of the main to start conclude? feel free to roll back
We presented \sysname{} which uses Scoped Indirect Execution (SIE), the first mechanism
to achieve principled OS tunability and enable safe, fast in-situ
tuning of {\em any} \perfconsts in OS kernels.
\sysname offers a new path for OS performance
    and opens untapped performance opportunities.

\if 0
Principled tunability provides the foundation 
and the first step toward
achieving truly optimal performance. 
This capability enables a wide range of
policies, including optimization and machine-learning techniques, as well as
emerging approaches such as LLM-driven OS
tuning~\cite{Liargkovas-25-OsTuningAgent}, all supported by our programmable
policy plane.
\fi

%% file: ack.tex
\vspace{4pt}
\section*{Acknowledgement}
\vspace{1.5pt}

We thank our shepherd and the anonymous reviewers for their insightful comments,
and Shawn Zhong for his help at an early stage.
% We thank our shepherd and the anonymous reviewers for their insightful comments. 
% We thank Cloudlab for the development and evaluation infrastructure. 
% We thank Shawn Zhong for his help in the early stage of this work. 
This work is supported  by National Natural Science Foundation of China (NSFC) 
    under Grant No. 62132007 and No. 62221003, and by NSF CNS-2145295.

%% file: 99_0_bib.tex
{\small
   \bibliographystyle{acm}
   \bibliography{all-defs,all,urls,xk-urls,all-confs}
}

%% file: appendix.tex
\appendix

\input{appendix_sysctl}

\vspace{1.5pt}
\section{Optimization for Conditional Branches}
\label{appendix:branch-optimization}
\vspace{1.5pt}

When the immediate value of a \perfconst{} is used in a conditional branch 
    (e.g., Figure~\ref{fig:sie-code-examples}(a)), 
    the location target of SIE is the corresponding \smalltt{jmp} instruction. 
However, a conditional jump cannot be jump-optimized by Kprobe~\cite{jia:atc:24}.

\begin{figure}[H] 
  \centering
  \includegraphics[width=1\columnwidth]{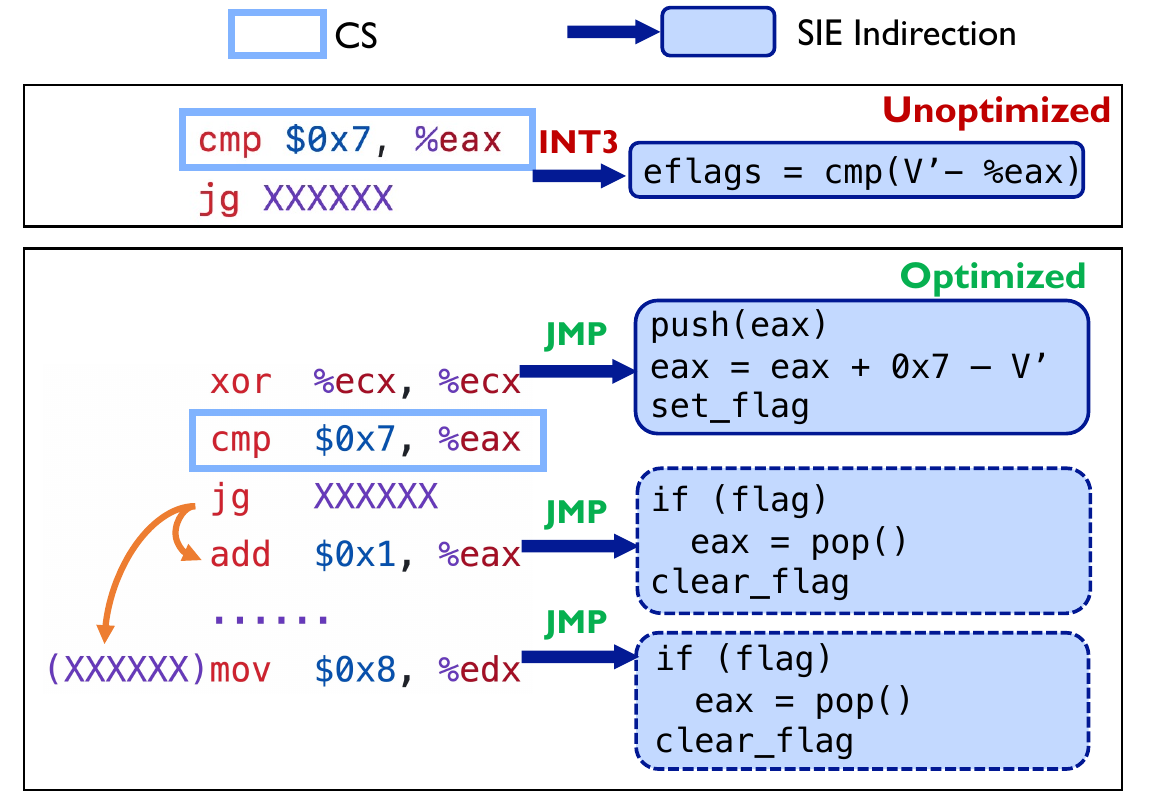}
  \caption{Optimization for conditional branches. 
  The \smalltt{cmp} instruction is undesired because its length is less than 5 bytes, 
    which prevents Kprobe jump-optimization. 
    \sysname{} attempts to attach the probe to a preceding instruction to 
    enable Kprobe jump-optimization.}
  \vspace{-5pt}
  \label{fig:appendix:branch-opt}
\end{figure}

For such cases, we implement a new optimization that allows jump-optimized kprobes.
The key idea is to modify $R/M$ in the \texttt{cmp} in advance 
    (at a position that jump-optimization is possible), 
    and restore it along the control flow via extra jump-optimized kprobes, 
    synchronized by a task-local flag, as shown in Figure~\ref{fig:appendix:branch-opt}.
In this way, the same effect as synthesizing an SIE indirection to modify \smalltt{eflags} 
    can be achieved by proactively adjusting $R/M$ used in the conditional jump. 
% \tianyin{The reason Kprobe jump-optimization is enabled is that ...} \zhongjie{i think we don't care why it can but care whether it can. If we really need to explain the reason, i think we need 5-6 lines.}
For example, in Figure~\ref{fig:sie-code-examples}(a), changing the immediate value 
    from $7$ to $4$ is functionally equivalent to adding $3$ to \smalltt{eax} before the comparison. 
More generally, if we want to replace $V$ with $V'$, we can adjust the $R/M$ 
    operand by $\Delta V = V - V'$ in advance, 
    and later restore its original value along all relevant control-flow paths.
This optimization increases the fraction of jump-optimized kprobes from 66.6\%
    to 88.3\% in our evaluation.

To implement this optimization, \sysname{} inserts two extra \kernelProbes{}
    at the jump target and the subsequent instruction. % immediately after the jump instruction. 
These \kernelProbes{} query the task-local storage and restore the operand. 
To ensure restoration occurs only when control flow originates from the CS, and not from unrelated execution paths, 
    we employ a task-local flag. 
This flag is set upon entry to the first \kernelProbe{} and cleared when the two extra \kernelProbes{} trigger.
If the first \kernelProbe{} and either of these two \kernelProbes{} fail to enable jump-optimization, 
    we fall back to the original SIE location.
When successful, this optimization replaces one INT3-based \kernelProbe{} with two jump-optimized \kernelProbes{}. 
Despite introducing extra \kernelProbes{}, this approach is beneficial: a jump-optimized \kernelProbe{} is an order of magnitude cheaper than an INT3-based \kernelProbe{}.
% Figure~\ref{fig:appendix:branch-opt} illustrates how to apply this optimization for case 1 in Figure~\ref{fig:sie-code-examples}.

\vspace{1.5pt}
\section{Kprobe Limitation}
\label{appendix:limitation}
\vspace{1.5pt}

In our evaluation (\S\ref{sec:eval}), \sysname{} failed to handle one \perfconst,
    where the target kernel functions could not be attached.
The \perfconst, \smalltt{SEND\_MAX\_EXTENT\_REFS}, is located in \smalltt{fs/btrfs/send.c} 
    within the \smalltt{check\_extent\_items()} function. 
In Linux v6.14, the symbol \smalltt{check\_extent\_item()} has multiple definitions 
    (e.g., appearing simultaneously in the kernel core and a loadable module). 
When registering a kprobe using the syntax \smalltt{[MOD:]SYM[+offs]}, the kernel's 
    \smalltt{trace\_kprobe} mechanism resolves the symbol name. 
However, to prevent ambiguity, the kernel rejects the registration if multiple matches are found, 
    causing the kprobe creation to fail with \smalltt{EINVAL} or \smalltt{-EADDRNOTAVAIL}.

\vspace{1.5pt}
\section{Policy Code with \sysname}
\label{appendix:policy}
\vspace{1.5pt}
% \sysname enables users to write expressive policies for tuning, 
% unlocking substantial design and optimization space. 
We present three Xk-tune programs that illustrate the policies enabled by
    \sysname (\sref{sec:cases:more-policies}): object-level tuning granularity,
    application-informed policy, and ad-hoc heuristics.
These examples show how users express such policies in \sysname.

\begin{figure}[t!]
\footnotesize
\centering
\vspace{2.5pt}
\begin{minted}[escapeinside=||,linenos=false]{c}
XK_TUNE(do_shrink_slab, "mm/shrinker.c:L381:128:0") {
    // 1. Safety Guard (Mandatory)
    if (!xk_transition_done(xk_ctx)) return 0;
    struct shrinker *s = (struct shrinker *)
        PT_REGS_PARM2(ctx);
    char name[32];
    if (bpf_probe_read_kernel(name, sizeof(name), 
        &s->name) < 0) return 0;
    // 2. Tune SHRINK_BATCH only for zswap
    if (bpf_strncmp(name, 12, "zswap-shrink") == 0)
        xk_set(xk_ctx, 64);
    return 0;
} /* SHRINK_BATCH.bpf.c */
\end{minted}
\caption{
    An Xk-tune for zswap shrinker; it identifies
    zswap shrinkers by checking the name.} 
    \label{fig:policy:shrinker}
    \vspace{-0.9pt}
\end{figure}

\begin{figure}[t!]
\footnotesize
\centering
\begin{minted}[escapeinside=||,linenos=false]{c}
#define MAX_ROCKSDB_THREADS 16
struct { // Filled by RocksDB threads
    __uint(type, BPF_MAP_TYPE_HASH);
    __uint(max_entries, MAX_ROCKSDB_THREADS);
    __type(key, u32);
    __type(value, int);          
} hint_map SEC(".maps");
XK_TUNE(blk_add_rq_to_plug, "block/blk.h:L312:32:0") {
    // 1. Safety Guard (Mandatory)
    if (!xk_transition_done(xk_ctx)) return 0;
    // Get the pid upon invocating this kernel code path
    u64 pid = bpf_get_current_pid_tgid() & 0xFFFFFFFF;
    // 2. Leverage hints from the application
    int *hint = bpf_map_lookup_elem(&hint_map, &pid);
    if (hint) xk_set(xk_ctx, 1);
    return 0;
} /* BLK_MAX_REQUEST_COUNT.bpf.c */
\end{minted}
\caption{
    An application-informed policy for RocksDB. 
    RocksDB characterizes threads with random-read patterns
    and exposes their IDs through a BPF map. 
    Xk-tune leverages this hint to enable per-thread tuning.} 
    \label{fig:policy:app-inform}
    \vspace{5pt}
\end{figure}

Figure~\ref{fig:policy:shrinker} shows an Xk-tune program
    that customizes \perfconsts for different objects.
    Written in eBPF, Xk-tunes can incorporate many features
    % such as call-stack checks, 
    to customize tuning granularity.
    The \perfconst\ is \ShrinkBatchTt\ (Case-3 in \sref{sec:cases:perf}).
Figure~\ref{fig:policy:app-inform} shows an Xk-tune program for
    an application-informed policy for a RocksDB application.
    The \perfconst\ is \BlkMaxReqCntTt\ (Case-1 in \sref{sec:cases:perf}).
Figure~\ref{fig:policy:heuristic} shows an Xk-tune program
    that coordinates with other \kernelProbes{} and uses ad-hoc workload heuristics.
    The \perfconst\ is \BlkMaxReqCntTt\ (Case-1 in \sref{sec:cases:perf}).

\begin{figure}[H]
\footnotesize
\centering
\begin{minted}[escapeinside=||,linenos=false]{c}
#define MERGE_FAIL_THRESHOLD 16
struct { // Filled by blk_attempt_plug_merge()
    __uint(type, BPF_MAP_TYPE_TASK_STORAGE);
    __uint(map_flags, BPF_F_NO_PREALLOC);
    __type(key, int);
    __type(value, int);          
} hint_map SEC(".maps");
/* Track consecutive failed attempts to merge adjacent 
   blocks and store the count in fail_cnt for use as a
   tuning heuristic */
SEC("kretprobe/blk_attempt_plug_merge")
int BPF_KRETPROBE(blk_attempt_plug_merge_ret, long ret) {
    struct task_struct *task = bpf_get_current_task_btf();
    int *fail_cnt = bpf_task_storage_get(&hint_map, 
        task, NULL, BPF_LOCAL_STORAGE_GET_F_CREATE);
    if (!fail_cnt) return 0;

    if (ret == 0) (*fail_cnt)++;
    else *fail_cnt = 0;

    return 0;
}
XK_TUNE(blk_add_rq_to_plug, "block/blk.h:L312:32:0") {
    // 1. Safety guard (mandatory)
    if (!xk_transition_done(xk_ctx)) return 0;
    struct task_struct *task = bpf_get_current_task_btf();
    // 2. Leverage hints from blk_attempt_plug_merge()
    // Heuristic: treat the workload as random when many
    // (16, an ad-hoc threshold) consecutive merges fail
    int *fail_cnt = bpf_task_storage_get(&hint_map, 
        task, NULL, 0);
    if (fail_cnt && *fail_cnt >= MERGE_FAIL_THRESHOLD)
        xk_set(xk_ctx, 1);
    return 0;
} /* BLK_MAX_REQUEST_COUNT.bpf.c */
\end{minted}
\caption{
    An Xk-tune program using ad-hoc workload heuristics for RocksDB. 
    The Xk-tune instruments \smalltt{blk\_attempt\_plug\_merge} with a kretprobe
    to track historical merge failures. 
    Upon detecting a high failure rate, the Xk-tune adjusts the threshold to a smaller value (1).} 
    \label{fig:policy:heuristic}
\end{figure}

\vspace{-10pt}
\section{\sysname Tool Commands}
\label{appendix:cli}
\vspace{1.5pt}

We show a few commands that use \sysname tools (\S\ref{sec:impl}).

\begin{minted}[linenos=false, framesep=2mm, bgcolor=white!5]{shell}
xk-build /usr/src/linux tune.patch # > ConstID
xk-gen ConstID xk-stub.h
xk-load [global/task/imm] xk-tune.c
xk-unload ConstID/all
\end{minted}

%% file: appendix_sysctl.tex
\section{\texttt{\bf Sysctl} Performance Knobs}
\label{appendix:sysctl-study}
\vspace{1pt}

We did a pilot study of \smalltt{sysctl} knobs on 
    Ubuntu 24.04 LTS which uses Linux v6.14. 
The number of \smalltt{sysctl} knobs may vary slightly across distributions.
%    but should not affect our conclusions.
We found 702 \smalltt{sysctl} knobs in total (deduplicated due to NIC name prefix) 
  and inspected all of them.
Among them, 145 \smalltt{sysctl} knobs are related to performance tuning. 
The others are for debugging, observability, and feature toggles.

We study the evolution of these 145 \smalltt{sysctl} knobs,
  and summarize the results in Figure~\ref{fig:sysctl-history}.
We observe that the \smalltt{sysctl} knobs evolve slowly.
Among the \NumSysctlKnob{} \smalltt{sysctl} knobs, 
  \NumSysctlUnchanged{} have not changed since 2005 (earlier
  commit history is unavailable). 
Only \NumSysctlChanged{} knobs were indeed evolved from constants---\perfconsts
  studied in this paper.
Among them, 19 (columns 1--19 in
Figure~\ref{fig:sysctl-history}) were historically converted from a \perfconst. 
The other 30 were extended from
  system-wide values to per-namespace (cgroup) values.
Reading the commit messages and the related discussions,
  it is clear that converting a \perfconst{}
  into a \smalltt{sysctl} knob is very slow due to the conservative upstream practice in Linux. 
% Source conversion and upstreaming slows deployment and adoption. 
For example, the TCP ping-pong threshold was set to 1 originally, 
  raised to 3 in 2019, reverted to 1 in 2022, and only recently 
  exposed as a \smalltt{sysctl} in response to application demands (e.g., SQL workloads) in 2023~\cite{TcpPingpang-2023}.
Therefore, waiting for a \perfconst{} to be converted to a \smalltt{sysctl} knob 
  may be unwise.

\input{fig-tex/fig-sysctl.tex}

Note that converting a \perfconst into a \smalltt{sysctl} knob is non-trivial 
  and error-prone. % frequently introduces bugs.
As shown in Figure~\ref{fig:sysctl-history}, \NumSysctlBUG knobs required bug fixes after
their conversion, taking an average of 12 years to resolve. Most issues
stem from race conditions introduced by the new global variable. 
We check the bug pattern of all the 149 \smalltt{sysctl} knobs,
  and identify \NumSysctlPBUG additional knobs with likely unresolved bugs.
Basically, converting a \perfconst{} into a writable interface adds
complexity: external writers (i.e., callers of the interface) can update the
value while kernel threads may still read the old one. 
Fine-grained control adds further complexity.
% Once a value is customized on a per-entity basis, complexity
% grows further, especially when entities must coordinate or share a global
% budget. 
For example, changing the dirty-ratio knob to operate per device
  introduced a deadlock because the \smalltt{sysctl} configuration path was not
  aware of the new semantics~\cite{MemSyctlDeadlock}. 
\sysname{} addresses these problems by safely updating \perfconst values through SIE (see \S\ref{sec:discussion}).

%% file: fig-tex/fig-sysctl.tex
\begin{figure}[H]
  \centering
  \vspace{-2.5pt}
  \includegraphics[width=1\columnwidth]{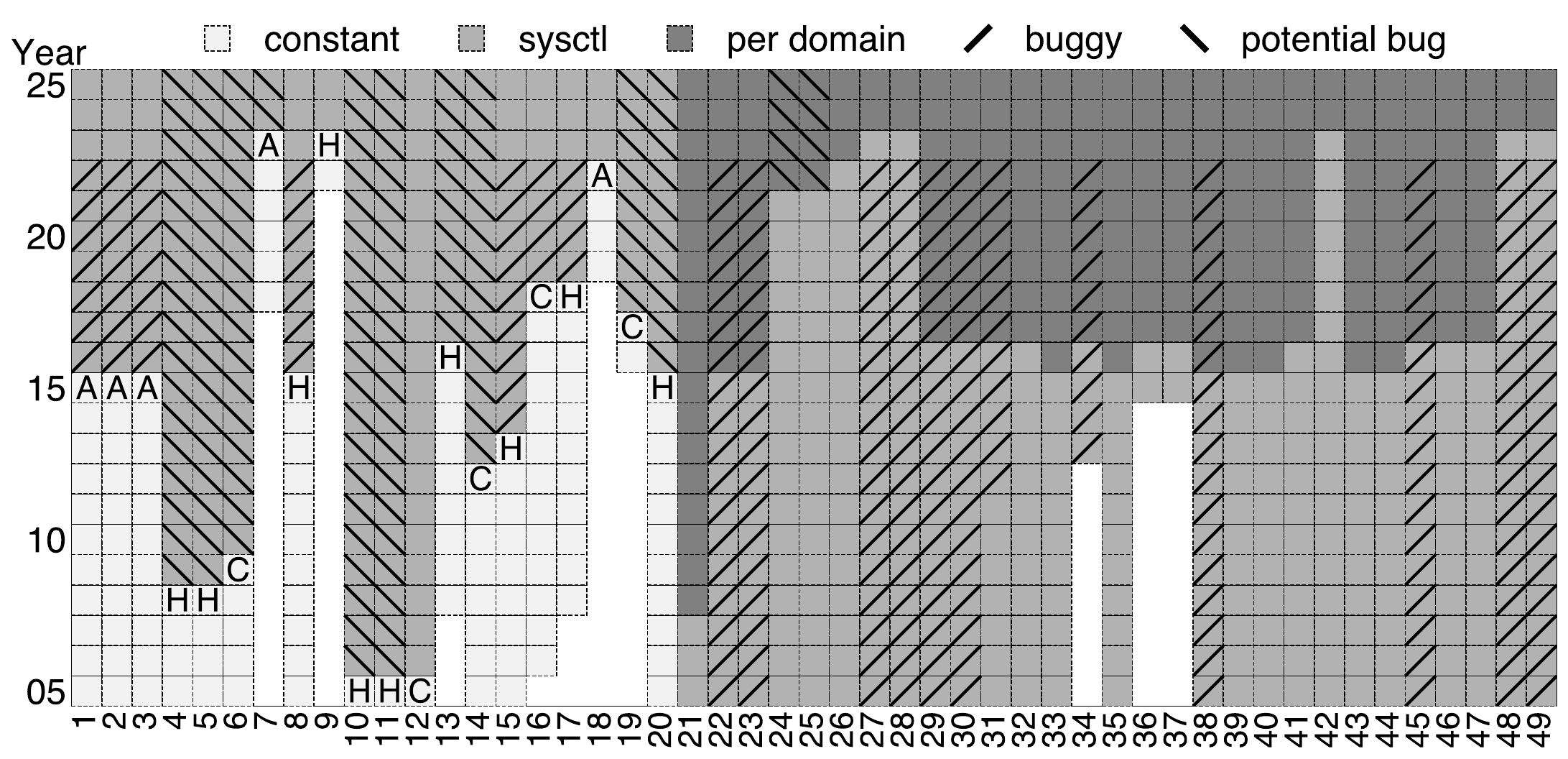}
  \caption{Evolution of \smalltt{sysctl} knobs.
    Each box shows the status of a knob.
    constant: a fixed \perfconst in the source code;
    \smalltt{sysctl}: exposed as a \smalltt{sysctl} knob.
    per-domain: the \smalltt{sysctl} knob is made per namespace.
    buggy: time between the introduction of a bug and the commit that fixes it.
    potential bug: a likely bug identified by our inspection.
    A/H/C: reason for the change (A: application-driven, H: hardware-driven, C: exposing control of an inherent trade-off).
  }
  \vspace{-5pt}
  \label{fig:sysctl-history}
\end{figure}